\documentclass[a4paper, 11pt]{article}
\pdfoutput=1
\usepackage{jcappub}
\usepackage[T1]{fontenc}
\usepackage{natbib}
\usepackage{txfonts,makecell}
\usepackage{hhline}
\usepackage{hyperref}
\usepackage{cleveref}
\usepackage{array}
\usepackage[numbers]{natbib}
\crefname{equation}{eq.}{eqs.}
\newcounter{mysfig}
\renewcommand\themysfig{(\alph{mysfig})}
\makeatletter
\newcommand\Scaption[1]{%
\refstepcounter{mysfig}%
\vskip.5\abovecaptionskip
  \sbox\@tempboxa{\small\themysfig~#1}%
  \ifdim \wd\@tempboxa >\hsize
    \small\themysfig~#1\par
  \else
    \global \@minipagefalse
    \hb@xt@\hsize{\hfil\box\@tempboxa\hfil}%
  \fi
  \vskip\belowcaptionskip}
\makeatother
\newcommand{\diff}{\mathrm{d}}

\title{An improved model-independent assessment of the late-time cosmic expansion }

  \author[a,b,c]{Balakrishna S. Haridasu}
  \author[b,c]{Vladimir V. Lukovi\'{c}}
  \author[d,e]{Michele Moresco}
  \author[b,c]{Nicola Vittorio}

  \affiliation[a]{Gran Sasso Science Institute , Viale Francesco Crispi 7, I-67100 L'Aquila, Italy}
  \affiliation[b]{Dipartimento di Fisica, Universit\`{a} di Roma "Tor Vergata", Via della Ricerca Scientifica 1, I-00133, Roma, Italy}
  \affiliation[c]{Sezione INFN, Universit\`{a} di Roma "Tor Vergata", Via della Ricerca Scientifica 1, I-00133, Roma, Italy}
  \affiliation[d]{Dipartimento di Fisica e Astronomia, Universit\`{a} di Bologna, Via Gobetti 93/2, I-40129, Bologna, Italy}
  \affiliation[e]{INAF - Osservatorio Astronomico di Bologna, Via Gobetti 93/3, I-40129, Bologna, Italy}

  \emailAdd{sandeep.haridasu@gssi.it}
  \emailAdd{vladimir.lukovic@roma2.infn.it}
  \emailAdd{michele.moresco@unibo.it}
  \emailAdd{nicola.vittorio@roma2.infn.it}
  \abstract{
  In the current work, we have implemented an extension of the standard Gaussian Process formalism, namely the Multi-Task Gaussian Process with the ability to perform a joint learning of several cosmological data simultaneously. We have utilised the ``low-redshift'' expansion rate data from Supernovae Type-Ia (SN), Baryon Acoustic Oscillations (BAO) and Cosmic Chronometers (CC) data in a joint analysis. We have tested several possible models of covariance functions and find very consistent estimates for cosmologically relevant parameters. In the current formalism, we also find provisions for heuristic arguments which allow us to select the best-suited kernel for the reconstruction of expansion rate data. We also utilised our method to account for systematics in CC data and find an estimate of $H_0 = 68.52^{+0.94 + 2.51 (sys)}_{-0.94} $ $\textrm{km/s Mpc}^{-1}$ and a corresponding $r_d = 145.61^{+2.82}_{ - 2.82 - 4.3 (sys)} $ Mpc as our primary result. Subsequently, we find constraints on the present deceleration parameter $q_0 = -0.52 \pm 0.06$ and the transition redshift $z_T = 0.64^{+0.12}_{-0.09}$. All the estimated cosmological parameters are found to be in good agreement with the standard $\Lambda$CDM scenario. Including the local model-independent $H_0$ estimate to the analysis we find $H_0 = 71.40^{ + 0.30 + 1.65 (sys)}_{- 0.30 } $ $\textrm{km/s Mpc}^{-1}$ and the corresponding $r_d = 141.29^{ + 1.31 }_{-1.31-2.63 (sys)}$ Mpc. Also, the constraints on $r_d H_0$ remain consistent throughout our analysis and also with the model-dependent CMB estimate. Using the $\mathcal{O}m(z)$ diagnostic, we find that the concordance model is very consistent within the redshift range $z \lesssim 2$ and mildly discrepant for $z \gtrsim 2$.
  }

\begin{document}
\maketitle
\flushbottom
\section{Introduction}
\label{sec:Introduction}

The current era of precision cosmology has allowed us to constrain cosmological models to a very high degree of precision \cite{Collaboration16b, Alam16, Joudaki17a, Riess18}. However lacking explanations for several physical aspects within these models \cite{Mortonson11}, this creates a need for alternative approaches for analysing the data and making cosmological inferences. Such a need and the ease to conduct phenomenological analysis has led to the use of several model-independent methods, in the last decade.

Amongst several model-independent approaches, Principal Component Analysis \cite{Feng16, Zhao17, Miranda17, Mortonson10b, Mortonson09a, Vanderveld12}, smoothing methods \cite{Shafieloo12a, Shafieloo06, LHuillier17}, cosmographic and polynomial regression methods \cite{Cattoen08, Moertsell09, Gruber14, Gomez-Valent18, Capozziello18}, Gaussian process (GP) formalism \cite{seikel12, Seikel13, Shafieloo12} etc., are a few. See \cite{Vitenti15} for an overview of some of these methods. GP has been a popular ``model-independent'' method that has been utilised in the past years to reconstruct the low-redshift cosmic expansion history, henceforth the dynamics of the late-time evolution \cite{Seikel13, Seikel13a}. GP has been implemented to draw inferences on acceleration \cite{Zhang16, Bilicki12}, estimates of $H_0$ \cite{Busti14a, Busti14, Yu17, Gomez-Valent18}, curvature of the universe \cite{Wei17}, to perform model falsification by estimating several diagnostics \cite{Yu17, Seikel12a, Sahni14, Nair14}, and to study the dynamics of the Dark Energy (DE) / scalar field equation of state (EoS) \cite{Holsclaw10, Holsclaw11, Seikel13, Nair14}. More recently, it has also been implemented to study model selection in \cite{Bilicki12, Yennapureddy17, Melia18} and estimation of linear anisotropic stress in \cite{MartaPinho18}.

In this work we implement an extended GP formalism, which allows one to perform a Joint-analysis of ``low-redshift'' observations such as Supernova Type Ia (SN) \cite{Riess18}, Cosmic chronometers (CC) \cite{Jimenez02, Moresco16a} and the anisotropic Baryon Acoustic Oscillations (BAO) \cite{Eisenstein05} datasets. To conduct a joint analysis using the cosmological data in an independent/single GP often requires assumptions of the several cosmological parameters such as present expansion rate ($H_0$), sound horizon at drag epoch ($r_d$) etc. Such assumptions, in turn, make the overall analysis and thereby any inferences mildly model-dependent (if not significantly). In this respect, we implement a model-independent joint learning method called Multi-Task Gaussian Process (MTGP) \cite{Caruana98, Bonilla07, Bonilla08}, which provides a provision to model the covariance between different tasks (essentially datasets), thereby allowing various tasks to complement each other and hence improve the overall reconstruction/prediction for each of the datasets, in comparison to the independent GPs of individual tasks. In particular for a cosmological dataset, one can make better predictions in the redshift range with less significant observations, by complementing the primary dataset with the secondary dataset(s) which provide better measurements in that range of redshifts. Apart from providing improved predictions, MTGP also allows one to infer the parameters (e.g., $r_d$) from the reconstructions, in contrast to assuming them a priori.

The tension in the local direct estimate of $H_0 = (73.48 \pm 1.66)$ \text{km/s Mpc$^{-1} $} (hereafter R18) in \citep{Riess18a} and indirect estimate of $H_0 = (66.93 \pm 0.62)$ \text{km/s Mpc$^{-1} $} (hereafter P16) from the ``high-redshift'' Planck collaboration \citep{Collaboration16b} (See Table 8 therein) has been a well established problem of precision cosmology. GP formalism has been implemented many a time to estimate $H_0$ as an intercept ($z = 0$) of the extrapolated predictive reconstruction of CC data in several works \cite{Busti14a, Busti14, Yu17}. \cite{Yu17}, have reported $H_0 = 67.42 \pm 4.80$ \text{km/s Mpc$^{-1} $}, preferring a lower $H_0$ value in comparison to R18. More recently, \cite{Gomez-Valent18} have combined the expansion rate data (E(z)) from supernova \cite{Riess18} and CC data to obtain $H_0 = 67.06 \pm 1.68$ \text{km/s Mpc$^{-1} $}. These estimates are clearly very consistent with ``high-redshift'' P16 and retain the well-known tension with R18. As is also mentioned in \cite{Gomez-Valent18}, the GP formalism suffers from the caveat of having to choose the specific covariance function, essentially a prior (\textit{of sorts}) on the reconstruction. In turn, this choice could lead to a bias in the estimate of $H_0$. Correcting for this effect would require better ways to model the GP, such as, learning the covariance function instead of assuming a fixed model, or to implement a joint analysis of different datasets. While the earlier option is a statistically rigorous procedure (see e.g., \cite{Abdessalem17}), it is not suitable for the current compilation of limited cosmological data. On the contrary, there lie prospects in the latter option for the same, which we implement here (partly due to the newer E(z) data). It is also important that the joint analysis is complemented with a proper kernel selection criteria to overcome any possible biases/specificities of the assumed covariance functions. However, we anticipate that given the smooth nature of the cosmic expansion rate data, one might not require a kernel selection but on the contrary find very consistent results with each of the kernels implemented.

In this work, we focus on the estimation of constant parameters $H_0$, $r_d$ from the MTGP formalism for several suitable combinations of data. As the GP formalism is also able to predict derivatives of the reconstruction regions, evaluation of deceleration parameter ($q(z)$) \cite{Zhang16, seikel12} and estimation of deceleration/acceleration transition redshift ($z_T$) \cite{Leon06, Lima12} have been of key interest. The SN distance data was able to provide a reliable reconstruction of $q(z)$ up to $z \sim 0.8$ (see e.g., \cite{seikel12}), however, the CC data is unable to do the same unless otherwise a prior on $H_0$ is assumed \cite{Zhang16}. In \cite{Yu17}, it has been reported that no good estimate of $z_T$ is available from the CC data, which we try to revise in this work.  We also reconstruct relevant cosmological diagnostics to comment on their estimation and hence the inferences. In particular we reconstruct the $\mathcal{O}m(z)$ diagnostic \cite{Sahni08, Sahni14, Zunckel08} and its derivative. These diagnostics and their variations have been implemented on several occasions to assess the validity of $\Lambda$CDM \cite{Sahni08}. The CC dataset usually implemented in the literature are derived based on an assumption of a particular model for the stellar evolution \cite{Moresco12b, Moresco16a, Moresco15}. We discuss the differences in the $H_0$ estimates and hence the systematics, obtained using the alternate data (this issue was also addressed in \cite{Gomez-Valent18}). We also illustrate several subtleties of the newer method utilised in this work by implementing it with several combinations of the data.

The paper is organised as follows: In \Cref{sec:Method} we describe the newer MTGP method, in \Cref{sec:Data} we then briefly present the datasets, corresponding modelling and also introduce the theoretical framework used to derive inferences. We have also included a brief discussion on the issue of kernel selection in this section. In \Cref{sec:Results} we present results for cosmological inferences from our analysis, along with a discussion on systematic effects and finally summarise our conclusions in \Cref{sec:Conclusions}.

\section{Method}
\label{sec:Method}
In this section, we describe the basic formalism of the Multi-Task Gaussian Process (hereafter MTGP), that we have implemented in this work. As the name suggests, MTGP considers different tasks, which are essentially different sets of independent training data and their corresponding cross-covariance to perform a simultaneous learning, which in-turn is utilised to make predictions. We begin by introducing the standard Single-Task Gaussian Process and then proceed to the newer MTGP formalism implemented here.

\subsection{Single-Task Gaussian Process}
We introduce the individual/Single-Task Gaussian Process (GP), with a brief description. However, we keep the discussion in this subsection to a minimum as it has been introduced numerous times in the cosmological context (see e.g., \cite{seikel12, Seikel13, Shafieloo12})\footnotemark. GP is essentially a generalisation of the single Gaussian distribution to the probability distributions of a function over a range of independent variables. In principle, this can be any stochastic process but is way simpler in a gaussian scenario and is also more often the case. While GP can be implemented for both \textit{classification} and \textit{regression} problems, for the cosmological concerns in this paper we focus on \textit{regression}. Given a task to be learnt for a set of $n$ independent variables $\textbf{x} = \{x_1,x_2, .., x_{\textrm{n}}\}$ and the corresponding dependent outputs $\textbf{y} = \{y_1,y_2, .., y_{\textrm{n}}\}$, GP provides predictions $\textbf{y}^{\ast}$ at the target inputs $\textbf{x}^{\ast}$ by placing a Gaussian prior on the latent function $f(x)$ that maps $\textbf{x}$ onto $\textbf{y}$. It is in this context that GP is also deemed as the collection of random variables and can be represented as,
\footnotetext{However, for an in-depth discussion, we strongly encourage the reader to refer to \cite{Rasmussen06}. See \cite{McHutchon11} for a discussion of the Gaussian process with noisy inputs, which is more relevant for cosmological purposes. }
\begin{equation}
    \label{eqn:GP}
    f(x_1)=\mathcal{GP}(\mu(x_1),\textrm{cov}[f(x_1),f(x_1)])
\end{equation}
where $\mu(x_1)$, and $\textrm{cov}[f(x_1),f(x_1)]$ are the mean and the variance of the random variable at $x_1$, respectively. The prior assumption on $\mu(x)$ is usually assumed to be 0, and does not play any role in obtaining the reconstructions, as the constant(deterministic) $\mu(x)$ only remains an additive factor for the final predictions. However, it is not the case if one wants to assume an explicit functional form for $\mu(x)$ with additional parameters \cite{Blight75, OHagan78} and {would no longer remain a minimal assumption analysis as the nature of this functional form is now imposed on the reconstructions. Also, note that far away from the regions in which data are available, the reconstructed posterior mean tends to the prior mean}. {However, we will comment more in detail on the assumption of mean towards the end of this section}. Given that one needs to reconstruct the function $f(x)$, we model the covariance between values of this function at different positions as,
%\footnote{{It is also worth mentioning that depending on the strength of the data, even a constant prior mean can influence the final reconstructions (see also \cite{Shafieloo12}), more in particular the derivative reconstructions.}}
\begin{equation}
    \label{eqn:cov}
    \textrm{cov}[f(x),f(x')] = k(x,x')
\end{equation}
where $k$ is an a priori assumed covariance model (\textit{kernel}, in GP terminology) for the particular reconstruction and is appropriately replaced in \Cref{eqn:GP}. We use $\textbf{x}$ to represent a set of inputs values and $x$ for an individual input. Correspondingly, $k(x,x')$ is a single value and $K(\textbf{x},\textbf{x}')$ represents a matrix of dimension $n\times n$, where $n$ is the length of the input vector. In the real-world scenarios often-times the dependent variables are only obtained with uncertainties (\textit{noise}, in the GP terminology) or even possible covariances ($\Sigma$) amongst different measurements. In such cases, the learning is performed with an additional noise covariance term $\Sigma$ along with the modelled $K(\textbf{x},\textbf{x}')$. The standard covariance functions are usually stationary models, where the estimate of the covariance is dependent only on the difference between the positions $|x-x'|$ and not on the actual positions. These models for covariance functions are often referred to as stationary kernels. In this work we implement the standard Gaussian/Squared-Exponential ($SE$) and the Mat\'ern ($M_{\nu}$) class of kernels. The $SE$ kernel is defined as,
\begin{equation}
    \label{eqn:kSE}
    k_{SE}(x,x') = \sigma_f^2 \exp\left(-\frac{|x-x'|^2}{2 l^2}\right),
\end{equation}
where $\sigma_f$ is the function/signal variance (amplitude) and $l$ is the length scale that dictates the ability to model features in the predicted region. These two parameters are often called hyperparameters as they are not the parameters of the function but of the covariance function. In the later descriptions to follow we redefine $\tau = |x-x'|$, and explicitly write $|x-x'|$ only when necessary, as this is consistent with all the kernels implemented here, which are stationary in nature. The $SE$ kernel, however, is a very smooth covariance function and so oftentimes not regarded as the best assumption for reconstructing features. To this aid, the Mat\'ern class of kernels are very useful and the general functional form can be written as,
\begin{equation}
    k_{M_{\nu}}(\tau) = \sigma_f^2 \frac{2^{1-\nu}}{\Gamma(\nu)} \left( \frac{\sqrt{2 \nu}\tau}{l} \right)^{\nu} K_{\nu}\left( \frac{\sqrt{2 \nu}\tau}{l} \right),
\end{equation}
where $K_{\nu}$ is the modified Bessel function of second kind with $\nu$ is strictly a positive parameter and $\Gamma(\nu)$ is the standard Gamma function. $\sigma_f, l > 0$ carry the usual definitions. The modified Bessel function provides an explicit analytic functional form for half-integer values of $\{\nu = 1/2, 3/2, 5/2 .. \}$ and as $\nu \to \infty $ the $ \textrm{M}_{\nu} $ covariance function converges to the $SE$ kernel. Amongst the several possibilities $\nu = 9/2 $ and $ \nu = 7/2$ are of particular interest as they are neither too smooth nor do they predict rapid variations\footnotemark[\value{footnote}]. {We would also like to assert that for the current cosmological data available, the flexibility of the Mat\'ern class kernels is sufficient and there is no immediate need for more complicated kernels, which will become relevant with forthcoming stringent data and for extrapolated predictions.} For instance, the Mat\'ern covariance function for $\nu = 9/2$ is given as,
\footnotetext{Please see \cite{Seikel13} for a good discussion on the relevancy of these issues for the cosmological data. It has also been argued that reconstructions using Mat\'ern kernels with $\nu \geq 7/2$ cannot be distinguished for finite and noisy training samples (see Sec.4.2 of \cite{Rasmussen06} and references therein). }
\begin{equation}
    \label{eqn:kM92}
    k_{M_{9/2}}(\tau) = \sigma_f^2 \exp\left(-\frac{3 \tau}{l}\right)\left[ 1 +\frac{3 \tau }{l} + \frac{27 \tau^2}{7l^2} +\frac{18 \tau^3}{7 l^3} +\frac{27 \tau^4}{35 l^4}\right],
\end{equation}
where $l > 0$. Also, it is worth mentioning that the Mat\'ern kernels are differentiable only to the order $\lfloor \nu \rfloor$ and so the kernels with $ \nu < 5/2$ lose their interest for predicting the derivatives\footnotemark[\value{footnote}] and hence relevant cosmological features, however still relevant for predictions at a single target position. Finally, the hyperparameters ${\Theta} \equiv \{\sigma_f, l\}$ are learned by optimising the log marginal likelihood (see \cite{Rasmussen06} for details) that is defined as,
\begin{equation}
    \label{eqn:LML}
    \mathcal{L}({\Theta}) = -\frac{1}{2}\textbf{\textrm{y}}^{\textrm{T}}K_{\textrm{y}}^{-1}\textbf{\textrm{y}} -\frac{1}{2}\ln|K_\textrm{y}|+\frac{n}{2}\ln(2 \pi)
\end{equation}
where $K_{\textrm{y}} = K(\textbf{x},\textbf{x}') + \Sigma_{n\times n}$ for a set of $n$ observations and assuming mean $\mu = 0$. The reconstructions of the regions can be performed by either learning the hyperparameters from a simple optimisation (global) of \Cref{eqn:LML} or by marginalising over them in a Bayesian way. In fact, the latter procedure often puts GP in the light of a non-parametric method\footnote{In a similar line of reasoning, as we are modelling the covariance but not the latent function ($f(x)$, {which is marginalised upon}) itself, GP is also deemed as model-independent. In a more recent work \cite{Gomez-Valent18}, discussion on similar issues was presented.}. {We emphasise that in the context of cosmology, model-independent implies independent of cosmological model, and the assumptions of kernel models must however be subjected to proper kernel selection to estimate which of them provides a better description of the data (see \Cref{subsec:msel}).} However, it has been argued \cite{Rasmussen06}(also in \cite{Seikel13}), that optimisation provides equally good predictions, while being computationally lighter. The differences in optimisation $vs.$ marginalisation becomes significant when the data are too sparse or if they are too noisy. The latter is, in fact, true with the CC data, as was found in \cite{Gomez-Valent18}. However, if the data are well constrained the probability distributions of the hyperparameters will be similarly peaked\footnote{Although the log marginal likelihood would be peaked at the global maximum, it may also be the case that there exist several local maxima. Appropriate care must be implemented to avoid these local maxima (see Sec 5.4 of \cite{Williams06}). We also provide more discussion on the same in the \Cref{subsec:msel}}. {In this work we perform global optimisation within a very large prior region ($0 < \sigma_f, l < 10$) for either of the hyperparameters (see \Cref{subsec:DataImple})}. Implementing the optimisation method, once the hyperparameters are learned, one can predict the mean and variance of the latent function at the test/target points through,
\begin{equation}
    \label{eqn:MaV}
\begin{split}
    \bar{f}(\textbf{x}^{\ast}) &= K(\textbf{x}^{\ast},\textbf{x}) K_{\textrm{y}}^{-1}\textbf{\textrm{y}} \\
    cov[f(\textbf{x}^{\ast})] &= K(\textbf{x}^{\ast},\textbf{x}^{\ast}) - K(\textbf{x}^{\ast},\textbf{x}) K_{\textrm{y}}^{-1}K(\textbf{x},\textbf{x}^{\ast}).
\end{split}
\end{equation}
One can extend the GP predictions also to the derivatives of the latent functions, however limited by the differentiability of an assumed kernel function. As is shown in \cite{Solak03}, derivative of a certain GP would also be a Gaussian Process, in fact even an integration or combinations of different orders of  derivatives can be approximated as Gaussians within the GP formalism. As described in \cite{Seikel12a}(see also \cite{McHutchon11}), this formalism can be useful to predict relevant dynamics in cosmology. The covariances involving the derivatives are written as,
\begin{equation}
    \label{eqn:dcov}
\begin{split}
    \textrm{cov}[f(x_{1}),f'(x_{2})] &= \dfrac{\partial k(x_{1},x_{2})}{\partial x_{2}}  \\
    \textrm{cov}[f'(x_{1}),f'(x_{2})] &= \dfrac{\partial^{2} k(x_{1},x_{2})}{\partial x_{1}\partial x_{2}}
\end{split}
\end{equation}
where $f'$ represent the derivatives with respect to their corresponding independent variables. Once the covariance between the derivative(s) of the latent function are defined we can obtain the mean and variance for the predictions of function derivatives. The mean of the $i^{th}$ derivative and the covariance between $i^{th}$ and $j^{th}$ derivatives can be written as,
\begin{equation}
    \label{eqn:dMaV}
\begin{split}
    \bar{f^{(i)}}(\textbf{x}^{\ast}) &= K^{(i)}(\textbf{x}^{\ast},\textbf{x}) K_{\textrm{y}}^{-1}\textbf{\textrm{y}} \\
    cov[f^{(i)}(\textbf{x}^{\ast}),f^{(j)}(\textbf{x}^{\ast})] &= K^{(i,j)}(\textbf{x}^{\ast},\textbf{x}^{\ast}) - K^{(i)}(\textbf{x}^{\ast},\textbf{x}) K_{\textrm{y}}^{-1}K^{(j)}(\textbf{x},\textbf{x}^{\ast}),
\end{split}
\end{equation}
respectively. For $i =j$ latter of \Cref{eqn:dMaV} provides the variance of the $i^{th}$ derivative. If the data for derivative functions is available, one can in fact perform a joint analysis, which was by far the only way \cite{Wang17c}, however requiring to assume physical constants in the cosmological context. Subsequently, one can extend this formalism to obtain predictions for combinations of $f(x)$ and its derivatives ($f'(x), f''(x), ..$). The prediction at a given target ($x^{\ast}$) for the combination is sampled from a multivariate normal distribution of the necessary components, appropriately considering the covariance amongst them. This is especially crucial for derivatives of higher order with larger uncertainties, where simple propagation of errors will not fare well.

\subsection{Multi-Task Gaussian Process}
\begin{figure}[h]
    \centering
\includegraphics[width=0.42\textwidth]{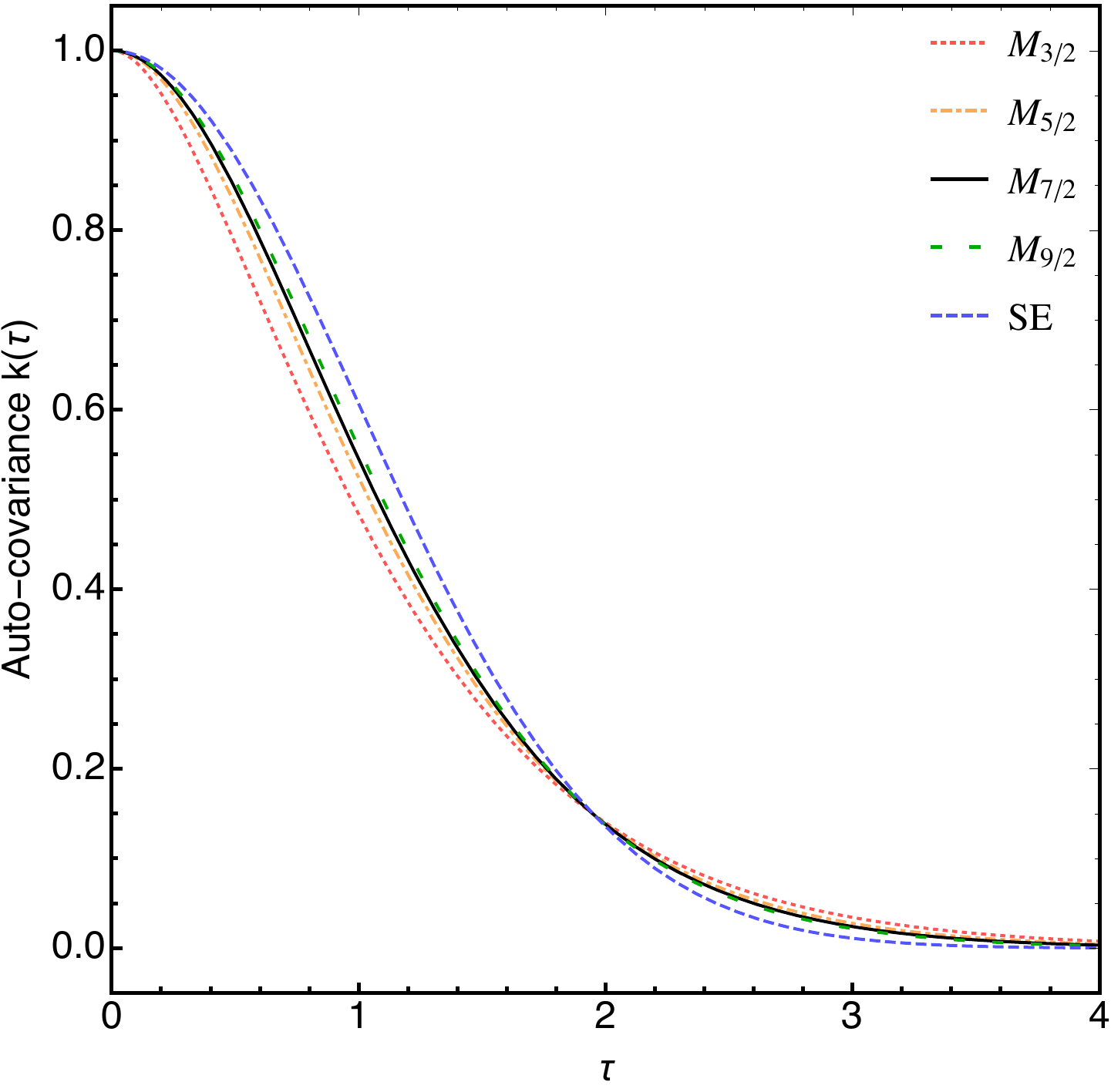}
\hspace{0.15in}
\includegraphics[width=0.42\textwidth]{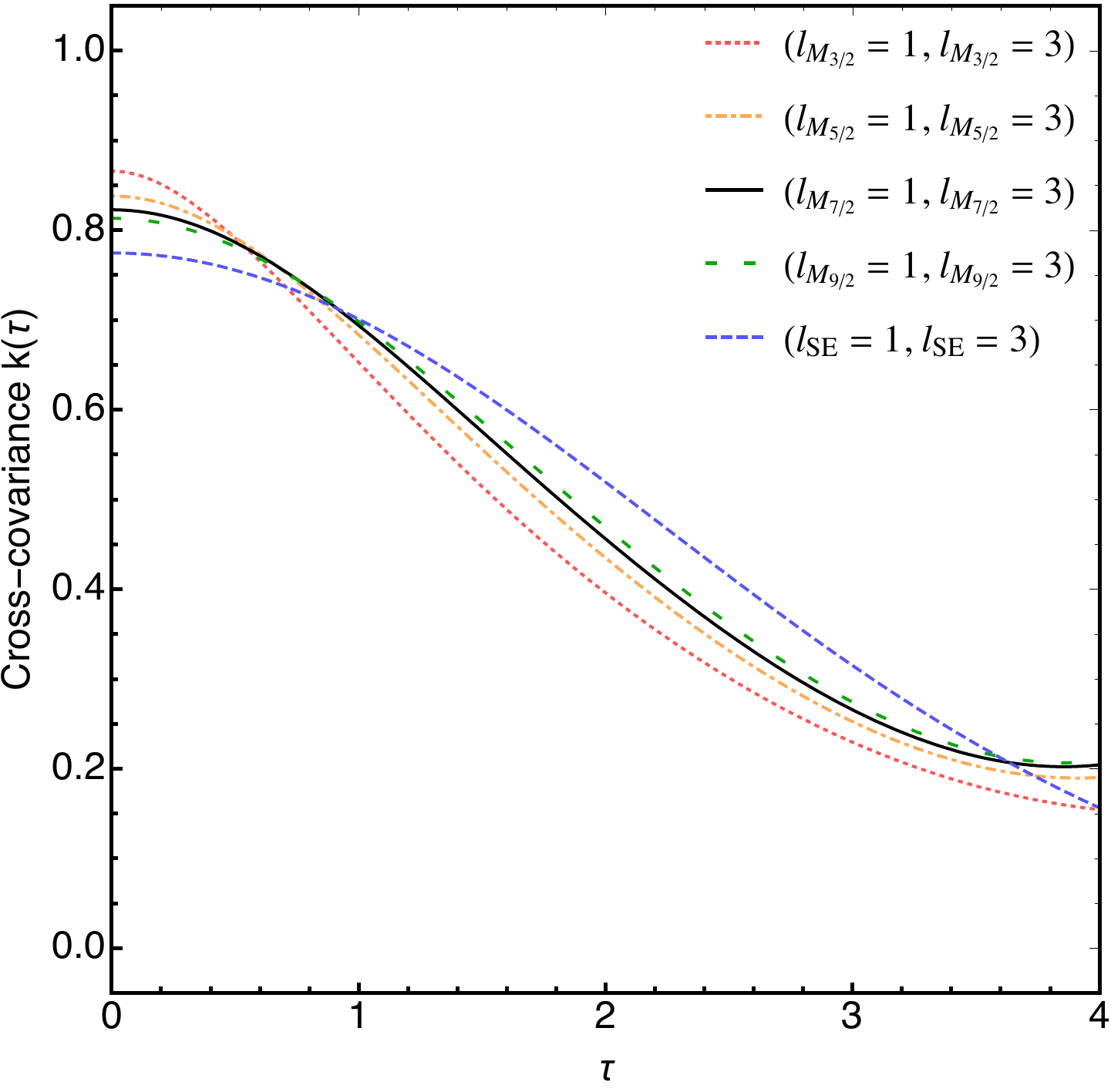}
\caption{We show the auto-covariance (left) of various kernel functions and cross-covariance (right) between same kernel with different length scale hyperparameter ($l$) as a function of $\tau$. In all the cases we have assumed the signal variance hyperparameter $\sigma = 1$.}
    \label{fig:KerComp}
\end{figure}

While the Single-Task GP models the covariance between the values of the same task at different positions, the MTGP allows one perform a simultaneous learning of multiple tasks by modelling the covariance between different tasks at their respective positions \cite{Bonilla08, Bonilla07, Melkumyan11, Vasudevan12}\footnotemark. For this purpose, similar to \Cref{eqn:cov} one can write,
\footnotetext{For an historical overview of Multi-Task Gaussian Process, please refer to \cite{Caruana98, Rasmussen06}}
\begin{equation}
    \label{eqn:covMTGP}
    \textrm{cov}[f_{l}(\textbf{x}_l),f_{m}(\textbf{x}_m)] = k_{lm}(\textbf{x}_l,\textbf{x}_m),
\end{equation}
where $f_l{\textbf{x}_l}, f_m(\textbf{x}_m)$ represent the latent functions of the $l^{th} $ and the $ m^{th}$ task at their corresponding input variables, respectively. The covariance between tasks $l$ and $m$ is modelled by $k_{lm}$. In this work we implement a more general Multi-Kernel Gaussian process, where the formalism has been developed so one can utilise different kernels for different tasks, unlike in the standard Multi-Task scenario which only allows for single kernel across different tasks. Therefore, one can choose to use ${M}_{9/2}$ and $SE$ kernels to model two tasks and their covariance to perform a joint learning. Consider two sets of data $(\textbf{x}_l,\textbf{y}_l)= \{(x_{l_1},y_{l_1}),(x_{l_2},y_{l_2}), .. , (x_{l_{n_{l}}},y_{l_{n_l}}) \}$ and $(\textbf{x}_m,\textbf{y}_m)= \{(x_{m_1},y_{m_1}),(x_{m_2},y_{m_2}), .. , (x_{m_{n_m}},y_{m_{n_m}}) \}$, with $n_l$ and $n_m$ being the number of data points in each set, respectively. The current formalism is based on the fact that the kernel function can be written as a convolution of the underlying basis function \footnotemark, evaluated at the two positions between which the covariance needs to be modelled. Following \cite{Melkumyan11}, one can write the kernel function as,
\footnotetext{The underlying basis function(s) of a kernel models how the strength of covariance/correlation between two inputs decays with increasing separation ($\tau$) between them, see \cite{Rasmussen06} for more details on basis functions.}
\begin{equation}
    \label{eqn:Acov}
    k_{ll}(x,x') \equiv g_{l}(x_l) \otimes g_{l}(x'_l) = \int^{+\infty}_{-\infty}g_{l}(u - x_l) g_{l}(x'_l- u) \textrm{d}u.
\end{equation}
where $g_{l}(x)$ is the basis function of the kernel ($k_{ll}$) defining the auto-covariance between two different data points of the same task. Given that one needs to model the cross-covariance which can be written as the convolution of the basis functions of the two different kernels for two individual tasks,
\begin{equation}
    \label{eqn:Ccov}
    k_{lm}(x,x') \equiv g_{l}(x_l) \otimes g_{m}(x_m) = \int^{+\infty}_{-\infty}g_{l}(u - x_l) g_{m}(x_m- u) \textrm{d}u.
\end{equation}
Here $k_{lm}$ denotes the cross-covariance between two different tasks $l$, $m$ and $g_{l}$, $g_{m}$ represent the basis functions of the two individual kernels, $x_l$, $x_m$ represent the independent variables of their respective tasks. Given the above formalism, the method to obtain the basis functions has been derived in \cite{Melkumyan11}\footnote{See also the technical report ($ACFR-TR-2011-002$) with detailed derivations and explanations, available at \href{http://www.acfr.usyd.edu.au/techreports/}{http://www.acfr.usyd.edu.au/techreports/}.}. The basis function of a given kernel $k(\tau)$ can be written as,

\begin{equation}
    g(\tau)  = \frac{1}{\sqrt[4]{2 \pi}}   \mathcal{F}^{-1}_{s \to \tau} \left[\sqrt{\mathcal{F}_{\tau \to s}\left[k(\tau)\right]}\right],
\end{equation}
where $\mathcal{F}^{-1}$, $\mathcal{F}$ represent the Inverse Fourier and Fourier transform, respectively. Once the basis functions are evaluated one can write the cross-covariance following \Cref{eqn:Ccov}. Finally, we model the auto-covariance for each of the dataset and the corresponding cross-covariance between different datasets as,
\begin{equation}
    \label{eqn:covMT}
\begin{split}
    \textrm{cov}[f_l(x_l),f_m(x_m')] &= k_{lm}(x_l,x_m'),
\end{split}
\end{equation}
where ${l,m} \in { 1, 2, .. N }$ for $N$ number of datasets. This formalism ensures that we obtain a positive semi-definite matrix for the modelled covariance and allows one to perform a joint learning on different datasets/tasks which can now be incorporated in the joint GP.
Given the covariance between the individual tasks, we write the joint covariance matrix as,
\begin{gather}
    \label{eqn:JcovKy}
    K_{\textrm{y}}
    =
    \begin{bmatrix}
        K_{{ll}} & K_{{lm}} \\
        K_{{ml}} & K_{{mm}} \\
    \end{bmatrix}
    +
    \begin{bmatrix}
        \Sigma_{n_l\times n_l} & 0 \\
        0 & \Sigma_{n_m\times n_m} \\
    \end{bmatrix}
\end{gather}
where the cross-covariance between different datasets is modelled by $K_{ml}$ and $\Sigma_{n_i\times n_i}$ represents the corresponding noise observed on the data, with $n_i$ being the number of data points in task $i$. If different tasks are correlated as well, the non-diagonal terms in \Cref{eqn:JcovKy} will be non-zero. This formalism can be easily extended to many datasets and the learning can be performed by optimising \Cref{eqn:LML}, appropriately replacing $K_{\textrm{y}}$ and the data $\textrm{y} \equiv \{\textrm{y}_1,\textrm{y}_2, .. \textrm{y}_N\}$ for the joint analysis of $N$ tasks. Note that also the reconstruction of the predictive regions is performed taking into account the total covariance in \Cref{eqn:JcovKy}. Finally, as an example, we show the cross-covariance terms evaluated for the $SE$ and $M_{\nu}$ kernels. The basis functions of the $SE$ and $M_{9/2}$ kernels can be analytically evaluated and are written as,
\begin{equation}
    \label{eqn:basf}
\begin{split}
        g_{SE}(x) = \sigma_{SE}\left(\frac{2}{\pi l_{SE}^2}\right)^{1/4}\exp{\left(-\frac{x}{l_{SE}}\right)^2},\\
        g_{M_{9/2}}(x) = \frac{48 \sigma_{M}}{\pi} \sqrt{\frac{3}{35  l_{M}^5}} x^2 K_{2}\left(\frac{2 |x|}{l_{M}}\right),
\end{split}
\end{equation}
where $\sigma_{i}, l_{i}$ have the usual meanings and $K_2$ represents the modified Bessel function (with $\nu=2$). Note that the basis function of the $M_{9/2}$ is now given in terms of $K_2$, which cannot be explicitly written in an analytical form and hence all further operations should be carried out numerically. With the basis functions defined, the cross-covariance can be evaluated by performing the convolution as in \Cref{eqn:Ccov}. The cross-covariance between two SE kernels with different hyperparameters can be analytically obtained as,
\begin{equation}
    \label{eqn:ksese}
    k_{SE_1 \times SE_2}(\tau) = \sigma_1 \sigma_2 \sqrt{\frac{2 l_1 l_2 }{l_1^2+l_2^2}} \exp{\left(-\frac{\tau^2}{l_1^2+l_2^2}\right)}.
\end{equation}
where indices 1,2 correspond to the first and the second SE kernels. Note that the $k_{SE_1 \times SE_2}(\tau)$ reduces to the standard $k_{SE}$ (\Cref{eqn:kSE}), if hyperparameters of the two kernels are same. Although, we do not have an analytical functional form for the cross-covariance involving Mat\'ern class kernels, they can be evaluated by numerically performing the convolution. In \Cref{fig:KerComp} we provide some illustrative comparison of the auto-covariance and cross-covariance functions with several combinations of hyperparameters. In the \textit{left} panel we show the standard kernel auto-covariance as a function of $\tau = |x-x'|$. It can be clearly noticed that the Mat\'ern class kernels provide lesser correlation than the $SE$ kernel for smaller $\tau$ and and vice-versa. This in turn allows $M_{\nu}$ kernels to provide more features in the predictive region as the value $\nu$ is reduced. In the \textit{right} panel we show the cross-covariance between same kernels with different length scale hyperparameter for each. Notice that in this case the covariance between $k_{M_{\nu_{1}}\times M_{\nu_{2}}}(\tau)$ is greater for smaller $\nu_{1}$ and $\nu_{2}$, at lower values of $\tau$. This plays a very important role in the determination of posterior at a particular value of $x^{\ast}$ for different tasks in the MTGP formalism. Also, the cross-covariance between the derivative of a first kernel and a second kernel is still stationary but anti-symmetric with respect to $\tau$.

While the improved formalism seems very lucrative for cosmological analysis, there lie also a few cautions one must be aware of, before making cosmological inferences. Therefore, we would like to summarise advantages of this formalism and a few cautions to mind, when using MTGP in a joint analysis.
\newline
\textit{Advantages}:
\begin{itemize}
    \item Complementary datasets can help improve the reconstructions of data with sparse/fewer or noisy data points. This approach will provide better predictions one knows a priori that the two task are related, however with an unknown functional form, e.g., combining $E(z)$ and $H(z)$ datasets. 
    \item Two closely related datasets that are different due to an unknown multiplicative or additive factor can be modelled together, and in fact the unknown factor(s) can be inferred from MTGP instead of having to assume them a priori. {This can also be extended to analyses where the unknown factors are not simply constants but functions, e.g. powers of (1+z), by appropriately adapting the kernels. }
    \item In the current cosmological context, even if a statistical evaluation of the most suited kernel for a particular prediction can be eluding, given the physical nature of data one can infer the better suited kernel through heuristic arguments.
    \item While MTGP provides the freedom to model different tasks with different underlying latent functions, one can also enforce the equivalence and variedly utilise the shared hyperparameters for an extended analysis.
\end{itemize}

\textit{Disadvantages/cautions}:
\begin{itemize}
    \item One should be careful while combining two unrelated datasets, especially if they have very different constraining strength. As the dataset with stronger significance might force its own features on to the reconstruction of less stronger dataset. Such a phenomenon is called \textit{negative transfer} (please refer to \cite{Chai10, Leen12, Lee16} for extensive discussion on this issue). We demonstrate this effect in \Cref{subsec:sys}.

    \item One must be careful in the choice of the covariance functions. As the number of tasks increases, the possible combinations of kernels rise drastically. It would be important to assess the nature of the data and guess a reasonable covariance function. However, this is only of dire consequence if the data are not functionally related.
    
    \item While the data are related, it is still possible to obtain an \textit{over-fitting}, if one of the dataset is much stronger than the other. This can however be easily seen when the stringent posteriors of one dataset over-power the reconstructions of other datasets, in the regions where they are void of data.
\end{itemize}

    Here \textit{over-fitting} and \textit{negative transfer} are very similar phenomenon, which can however be conceptually differentiated. If the datasets are clearly unrelated and the features of one dataset over-power the other to provide incorrect features (nature of the curve) in the reconstruction of second dataset, it is negative transfer. If the datasets are a priori known to be related, and if the stronger dataset over-powers the second dataset to provide stringent reconstruction in the regions where there are no measurements available for the second dataset, this is also negative transfer that is of the over-fitting kind.

{On the assumption of the mean $\mu(x) = 0$, often times when no prior knowledge of the latent function is available, the mean is assumed to be 0 by symmetry, to obtain posteriors of $f(x^{\ast})$ i.e., $y^{\ast}$. Although, the reasoning for the same may not be apparent in the function-space viewpoint elaborated here, they become clearer in an equivalent basis function or weight-space\footnote{Providing a complete description on these two viewpoints of Gaussian Processes is beyond the scope and necessity of this paper, we urge the reader to refer to Section 2.2 of \cite{Rasmussen06} and Section 6.4 of \cite{Bishop06} for more details.} viewpoint \cite{Rasmussen06}. In a weight-space viewpoint the latent function can be written as,
\begin{equation}
        f(\textbf{x}) = \textbf{w}^{\textrm{T}}\Phi(\textbf{x}) ,
\end{equation}
where $\Phi(x)$\footnotemark is a $M$-dimensional vector/set of basis functions and `\textbf{w}' is the corresponding vector of weights. This coincides with the formalism presented here, as the covariance kernel can now be written as $ k(\textbf{x},\textbf{x}') = \Phi(\textbf{x})^{\textrm{T}}\Sigma(\Theta)\Phi(\textbf{x}') $, which is equivalent to the convolution formulae, \Cref{eqn:Ccov} and \Cref{eqn:Acov}. Here $\Sigma(\Theta)$ is the covariance matrix of joint distribution on the weights, equivalently defined as a function of the hyperparameters `$\Theta$' and the kernel itself is a kind of prior on the expected nature of the latent functions, which however includes infinite possibilities. A prior assumption of the mean $\mu(x) = 0$ is in fact equivalent to assuming the mean of the prior on probability of weights $p(\textbf{w},\Theta) \sim \mathcal{N}(\textbf{w}|0,\Sigma(\Theta))$ to be 0 in the weight-space view \cite{Bishop06}. The probability distribution on `$\textbf{w}$' therefore characterizes $f(x)$, as $\bar{f}(\textbf{x}) = \phi(\textbf{x})^{\textrm{T}}\bar{w}(\textbf{x})$. A non-zero prior on the mean of the weights implies a non-minimal assumption\footnotemark[\value{footnote}] analysis as a preferred functional form is a priori implied, from amongst all the infinite possibilities. And so, the prior of $\mu(x) = 0$ is seemingly conservative over any other asymmetric prior to make predictions, unless an accurate prior knowledge of the latent function is available or if the motive is ulterior, such as a residual analysis. Also note that the mean of the process itself is not necessarily the mean of the data. However, if the dataset is stringent enough, different assumptions of prior should not effect the reconstructions within the range of data. Therefore, especially if we do not expect that the data can alleviate the effect of different prior mean assumptions, the most unbiased prior that one can assume is $\mu(x) = 0$. Needless to say, such a constant mean prior implies that the posteriors $y^{\ast}$ far outside the data range will invariably tend to 0 and varied assumptions of mean could alter the results in the extrapolated regions beyond agreement. An asymmetric assumption of prior mean might also tend to provide stricter, but biased reconstructions of the derivatives. While the symmetric $\mu =0$ assumption provides fairly less stringent and hence unbiased, conservative reconstructions.
}

\footnotetext{For example, $\Phi(x) = \{1, x, x^2, x^3 .. , x^{M-1}\}$ is the vector of basis functions for polynomial regression. In fact, this makes GP a generalization over the polynomial regression methods. A non-zero constant prior mean of `c' on a 3-dimensional basis function vector of this kind, imposes a prior belief of the functional form to be $f(x) = c+ cx+cx^2+cx^3$, which is clearly a non-minimal supposition. }

\section{Theory and Data}
\label{sec:Data}
In this section we provide a brief introduction to the theoretical framework of the standard cosmological model and then describe the data, different combinations of the same that we implemented with the MTGP method described in the previous section. In our current work we use the most-recent ``low-redshift'' BAO, CC, and SN datasets. We utilise the expansion rate measurements available from these data, which are however rescaled according to their corresponding observational methods. Finally, we have also presented a brief discussion on the criteria for kernel selection to infer the best description of the data.

\subsection{Theoretical framework}
\label{subsec:TheoFrame}
The first Friedmann equation with all necessary degrees of freedom at ``low-redshifts'' is given by,
 \begin{equation}
\label{eqn:HUE}
H(z)^{2} \equiv {H_0^2 E(z)}^2= {H_0}^2\left[ \Omega_{m}(1+z)^3  + \Omega_{k}(1+z)^2 +\Omega_{DE}f(z)\right],
 \end{equation}
where, $H_{0}$ is the present expansion rate, while $\Omega_{m}$, $ \Omega_{DE}$ and $ \Omega_k $ are the dimensionless density parameters of matter, dark energy (DE) and curvature, respectively. The dynamics of the DE component are determined by $f(z)$, which is written as,
\begin{equation}
    f(z) = \exp\left( 3 \int^{z}_0 \frac{1+w_{DE}(\xi)}{1+\xi} \diff \xi\right),
\end{equation}
where $w_{DE}(z)$ is the equation of state (EOS) parameter of dark energy. The dimensionless density parameters are assumed to obey the cosmic sum rule of $\Omega_{m}+\Omega_{DE}+\Omega_{k}=1$. Predominantly, in this work we assumed the concordance $\Lambda$CDM, with $\Omega_k = 0$ and a constant DE equation of state parameter ($w_{DE} = -1$) with $f(z) = 1$ for the diagnostics implemented here. Through the second Friedmann equation, $ \ddot{a}/a = - 4\pi/3 G\sum_i\rho_i(1+3w_i)$, one can assess the present accelerated state of the universe by estimating the deceleration parameter which is written as,
  \begin{equation}
  \label{eqn:DP}
  q(z) = -a \frac{\ddot{a}}{\dot{a}^{2}} \equiv (1+z)\frac{H'(z)}{H(z)}-1.
  \end{equation}
A negative value of the deceleration parameter implies an accelerated expansion and vice-versa. In addition, one can derive $q_0 = q(z = 0) = \frac{3}{2} \Omega_m - 1$ in the standard $\Lambda$CDM scenario. The transition from deceleration to accelerated expansion is observed as a change in sign of $q(z)$, which is marked as the transition redshift $z_T$. Having significant constraints on the $z_T$ and $q_0$ remain crucial for any cosmological inferences in the current GP framework.

The sound horizon at drag epoch is a relevant cosmological scale for the BAO data that is imprinted in the galaxy-clustering BAO data which is necessary to model the data. The sound horizon $r_d$ at the drag epoch is given as,
   \begin{equation}
     r_d =\int_{z_d}^{\infty} \frac{c_s(z)}{H(z)}dz.
     \label{eq:r_d}
   \end{equation}
where $c_s$ is the sound speed and $z_d$ is the redshift at drag epoch. This physical scale is for rescaling the expansion rate data coming from BAO observations.

In the context of the standard framework, we utilise the $\mathcal{O}_m(z)$ diagnostic \cite{Sahni14, Zunckel08},
\begin{equation}
    \label{eqn:OM}
    \mathcal{O}_m(z) = \frac{(E(z))^2-1}{(1+z)^3-1}.
\end{equation}
If the underlying expansion history $E(z) = H(z)/H_0$ is given by the $\Lambda$CDM, $\mathcal{O}_m(z)$ is essentially a constant and is equal to the matter density $\Omega_m$. Therefore, any deviations from this can be used to infer a dynamic nature of dark energy or an intrinsic dynamic nature of matter itself. The derivative of the $\mathcal{O}'_m(z)$ w.r.t $z$ provides information regarding the possible variations in $\mathcal{O}_m(z)$. Note that $\mathcal{O}'_m(z)$ utilises the information from both $H(z)$ and $H'(z)$ reconstructions, which is the additional information inferred from the GP analysis. Therefore, is more often susceptible to the prior assumption of the GP framework (i.e., kernel) itself. Apart from these well-known diagnostics, we also implement a simple modification of the $\mathcal{O}_m(z)$ by considering the derivative of the first Friedmann equation (\Cref{eqn:HUE}) in the $\Lambda$CDM scenario, which provides,
\begin{equation}
    \label{eqn:OM2}
    \mathcal{O}_m^{(2)}(z) = \frac{2 E(z) E'(z)}{3 (1+z)^2}.
\end{equation}
It is straightforward that the reconstructed $\mathcal{O}_m^{(2)}(z)$ should once again be equal to a constant $\Omega_m$ if the underlying cosmology is $\Lambda$CDM.

\subsection{Data and Implementation}
\label{subsec:DataImple}
In this subsection we summarise the dataset implemented in the current analysis. We have collected datasets which provide information regarding the cosmic expansion history as a function of redshift. The straight-forward relatedness of these datasets allows one to perform the MTGP learning and conduct a joint analysis. The most recent expansion rate data from SN observations plays a very crucial role in our analysis, as it provides significant improvement at lower redshifts.

\textit{Baryon Acoustic Oscillations}: In this work we utilise $H(z) r_{d}$ (hereafter abbreviated as $\textrm{BAO}$) observables provided in \cite{Alam16, Zhao16, Wang16, Bautista17, MasdesBourboux17, Zhao18}. In \cite{Alam16}, using the Sloan Digital Sky Survey(SDSS) III DR-12, these observables were reported for the galaxy-clustering data set at three effective binned redshifts $z = 0.32, 0.57, 0.61$ (hereafter 3z). From the same data set also a tomographic analysis was performed in \cite{Zhao16, Wang16}, providing observables at 9 binned redshifts (9z). Implementing both the 3z and 9z datasets, we find very consistent results within the current methodology and remain to use 3z data set to quote the BAO contribution. More recently, \cite{Zhao18} have reported 4 measurements of the same observables at redshifts $z = 0.98, 1.32, 1.52, 1.94 $. In fact, these 4 measurements provide data in a much needed redshift range of $1 \lesssim z \lesssim 2$, that could play very important role also in the context of model-independent analysis and reconstructions. We also utilised the ``high-redshift'' Lyman-$\alpha$ measurements in \cite{MasdesBourboux17} and \cite{Bautista17} at redshifts $z=2.33$ and $z=2.4$. For all the datasets mentioned here we appropriately use the covariance matrices that have been provided \footnotemark. While the observations are presented at times for $H(z)\left(r_{d,fid}/r_d\right)^{-1}$, we homogenise the dataset to $r_d H(z)$ considering the appropriate $r_{d,fid}$ used in the respective works. $r_d$ clearly being a physical constant set at a much higher redshift ($z\sim 10^3$), the data is interpreted as expansion rate multiplied by a physical constant.
\footnotetext{All values of the mean, dispersion and covariances of $H(z) r_d$ observables for the galaxy clustering BAO data are taken from \href{https://data.sdss.org/sas/dr12/boss/papers/clustering/}{https://data.sdss.org/sas/dr12/boss/papers/clustering/}. }

{\renewcommand{\arraystretch}{1.25}%
\begin{table}[h]
\begin{center}
\caption{ $E(z)$ obtained from the inversion of the $E(z)^{-1}$ data reported in Table 6 of \cite{Riess18}. Note the difference in the estimate of $E(z=1.5)$, from the actual quoted value. We simply adopt the mean value obtained form our inversion of quoted $E(z)^{-1}$ data.}
\label{tab:Eofz}
\footnotesize
\vspace{0.2in}
\begin{tabular}{|c|c|}
\hline
 $z$ &$E(z)$  \\
\hline
\hline
0.07 & 0.997  $\pm$ 0.023   \\

0.20 & 1.111 $\pm$ 0.021   \\

0.35 &  1.127  $\pm$ 0.037   \\

0.55 & 1.366   $\pm$ 0.062   \\

0.9 &  1.524   $\pm$ 0.121  \\

1.5 &  2.924   $\pm$ 0.675  \\
\hline
\end{tabular}
\end{center}
\end{table}
}

\textit{Supernova data:} We have implemented the most recent binned expansion rate data provided in \cite{Riess18}, where 6 $E(z)^{-1}$ (hereafter SN) data with their covariances have been provided for flat cosmologies ($\Omega_k=0$). We invert the $E(z)^{-1}$, data to obtain the $E(z)$ points and corresponding covariance. However, the $E(z=1.5)$ has been reported with an asymmetric uncertainty, while the conversion is unable to account for the same. We simply proceed to utilise the symmetric $E(z=1.5)$ estimate obtained from the inversion, as this data point with its large uncertainty would not be of key importance and also with little significance in the current joint analysis MTGP formalism\footnote{We also verify that the reconstructions show very little to no change, if the $E(z=1.5)$ data point is omitted. Also, \cite{Gomez-Valent18} have commented in similar lines, regarding the $E(z=1.5)$ data point.}. The $E(z)$ data implemented in this work are summarised in \Cref{tab:Eofz}.

{\renewcommand{\arraystretch}{1.25}%
\begin{table}[h]
\begin{center}
\caption{ Cosmic chronometers data obtained using both the BC03 and MaStro models compiled from all the available literature \cite{Simon05, Stern10, Moresco12b, Moresco16a, Moresco15, Zhang14, Ratsimbazafy17} are summarised here.}
\label{tab:Hofz}
\footnotesize
\vspace{0.1in}
\begin{tabular}{|c|c|c|c|}
\hline
 $z$ & \multicolumn{2}{c|}{$H(z)$ [$\textrm{km/s Mpc}^{-1}$]} & Ref. \\
\hline
  & BC03 ($\textrm{CC}$ \& $\textrm{CC}_B$) & MaStro ($\textrm{CC}_M$) & \\
\hline
\hline
0.0708 & 69.0 $\pm$ 19.68 & - & \cite{Zhang14}\\
0.09 & 69.0 $\pm$ 12.0 & - & \cite{Simon05} \\
0.12 & 68.6 $\pm$ 26.2 & - & \cite{Zhang14} \\
0.17 & 83.0 $\pm$ 8.0 & - & \cite{Simon05} \\
0.1791 & 75.0 $\pm$ 4.0 & 81.0 $\pm$ 5.0 & \cite{Moresco12b} \\
0.1993 & 75.0 $\pm$ 5.0 & 81.0 $\pm$ 6.0 & \cite{Moresco12b} \\
0.2 & 72.9 $\pm$ 29.6  & - & \cite{Zhang14} \\
0.27 & 77.0 $\pm$ 14.0 & - & \cite{Simon05} \\
0.28 & 88.8 $\pm$ 36.6  & - & \cite{Zhang14} \\
0.3519 & 83.0 $\pm$ 14.0 & 88.0 $\pm$ 16.0  & \cite{Moresco12b} \\
0.3802 & 83.0 $\pm$ 13.5 & 89.3 $\pm$ 14.1  & \cite{Moresco16a} \\
0.4 & 95.0 $\pm$ 17.0  & - & \cite{Simon05} \\
0.4004 & 77.0 $\pm$ 10.2 & 82.8 $\pm$ 10.6 & \cite{Moresco16a} \\
0.4247 & 87.1 $\pm$ 11.2 & 93.7 $\pm$ 11.7 & \cite{Moresco16a} \\
0.4497 & 92.8 $\pm$ 12.9 & 99.7 $\pm$ 13.4 & \cite{Moresco16a} \\
0.47 & 89.0 $\pm$ 34.0 & - & \cite{Ratsimbazafy17}\\
0.4783 & 80.9 $\pm$ 9.0 & 86.6 $\pm$ 8.7 & \cite{Moresco16a} \\
0.48 & 97.0 $\pm$ 62.0 & - & \cite{Stern10} \\
0.5929 & 104.0 $\pm$ 13.0 & 110.0 $\pm$ 15.0 & \cite{Moresco12b} \\
0.6797 & 92.0 $\pm$ 8.0 & 98.0 $\pm$ 10.0 & \cite{Moresco12b} \\
0.7812 & 105.0 $\pm$ 12.0 & 88.0 $\pm$ 11.0 & \cite{Moresco12b} \\
0.8754 & 125.0 $\pm$ 17.0 & 124.0 $\pm$ 17.0 & \cite{Moresco12b} \\
0.88 & 90.0 $\pm$ 40.0 & - & \cite{Stern10} \\
0.9 & 117.0 $\pm$ 23.0 & - & \cite{Stern10} \\
1.037 & 154.0 $\pm$ 20.0 & 113.0 $\pm$ 15.0 & \cite{Moresco12b} \\
1.3 & 168.0$\pm$ 17.0 & - & \cite{Stern10} \\
1.363 & 160.0$\pm$ 33.6 & 160.0$\pm$ 33.6 & \cite{Moresco15} \\
1.43 & 177.0$\pm$ 18.0 & - & \cite{Stern10} \\
1.53 & 140.0 $\pm$ 14.0 & - & \cite{Stern10} \\
1.75 & 202.0 $\pm$ 40.0 & - & \cite{Stern10} \\
1.965 & 186.5 $\pm$ 50.4 & 186.5 $\pm$ 50.4 & \cite{Moresco15} \\
\hline
\end{tabular}
\end{center}
\end{table}
}

\textit{Cosmic Chronometers:} The measurements of cosmic expansion rate have been estimated using the differential age method suggested in \cite{Jimenez02}, which considers pairs of massive and passively evolving red galaxies at similar redshifts to obtain $\diff z/\diff t$, which are deemed as the \textit{Standardisable Clocks}. In this work, we utilise two different compilations of CC data, differentiated based on the stellar evolution models assumed for obtaining them. A dataset of 31 (CC) uncorrelated data points is compiled from \cite{Simon05, Stern10, Moresco12b, Moresco16a, Moresco15, Zhang14, Ratsimbazafy17} and is the most utilised compilation in literature. Data points in this compilation are obtained utilising the BC03\footnotemark models for the stellar evolution and comprise of larger number of data points as it was the only method implemented in earlier works such as \cite{Simon05, Stern10, Zhang14}. Along side this standard dataset we compile a different set of data that are obtained using the MaStro\footnotemark[\value{footnote}] stellar evolution models. This compilation consists of 15 ($\textrm{CC}_M$) points taken from \cite{Moresco12b, Moresco16a, Moresco15}. Given, the fewer number of data points this compilation does not add significant weight in the standard model-dependent residual analysis. However, in a model-independent analysis, the estimate of $H_0$ as an intercept of the reconstruction of $H(z)$ data can provide significant constraints on the mean with both datasets. Also, complementing with other datasets in our analysis we find suitable arguments to comment on the systematics within the CC dataset. The 15 point compilation of $\textrm{CC}_M$ is compared against the corresponding 15 point compilation obtained using the BC03 ($\textrm{CC}_B$) method. All the CC data yet available in the literature are summarised in \Cref{tab:Hofz}. Notice that the data at redshifts $z=1.363$ and $z= 1.965$ are quoted as same for both the methods, as contribution of systematic error for different choice of the stellar evolution method is included in quadrature within the reported error. Repeating our analysis decoupling this systematic error, we however find compatible results to the ones discussed here. Finally, we refer to the recent analysis of \cite{Moresco18}, that explored the dependence of these data on a possible contamination due to an young underlying component, assessing that for current measurements this effect has been confined by the selection criteria well below the estimated errors.

\footnotetext{This BC03 compilation was implemented in several works such as \cite{Farooq16, Yu17, Gomez-Valent18}, also at times including the BAO measurements assuming an $r_d$. In \cite{Moresco12b, Moresco16a, Moresco15} the estimates from BC03 models were also accompanied with the estimates obtained using MaStro models. }

All three datasets mentioned above are implemented as different tasks in the MTGP formalism. For this purpose, we start by rescaling/rewriting the data with arbitrary fiducial values such as, $H_{0}^{fid} = 100, r_d^{fid} = 148$\footnote{Please note the difference from $r_{d,fid}$, which is the fiducial value assumed in the process of acquiring the measurement in various works and $r_d^{fid}$ is simply a rescaling factor used only in our analysis, for convenience.}. In principle there is no need for this step and it would not effect the results in any way, but we do so to have all the datasets within similar order(s) of magnitude, which we find also convenient in the numerical computations. The $E(z)$ data is not rescaled but we do include a theoretical data point of $E(z = 0) = 1$. This assumption is justified assuming the results are essentially restricted to a comparison for flat cosmologies within standard framework where $H(z) = H_0 E(z)$ and it also helps improve the features in the reconstructions. To summarise, we have three tasks: $E(z)$ (SN), $r_d H(z)(H_0^{fid} r_d^{fid})^{-1}$ (BAO) and $H(z) (H_0^{fid})^{-1}$ (CC). The CC dataset however is replaced for different compilations that are mentioned above. We perform the MTGP on these three datasets considering several combinations of kernels described in \Cref{sec:Method}. Given the known relatedness of the data i.e., all the datasets are essentially providing expansion history, while being rescaled accordingly with physical constants ($H_0$, $r_d H_0$), we do not immediately perform any inter-kernel analysis. We assume the same covariance function for all the three tasks, however with their own hyperparameters and the appropriate cross-covariance. We will comment in detail about this assumption of the kernels and their implications in the \Cref{sec:Results}. Therefore, we have to finally optimise the log Marginal likelihood for six hyperparameters in total, two for each task.

\subsection{Kernel selection}
\label{subsec:msel}
The Single-Task GP formalism so far implemented in the cosmological context has been deprived of a proper kernel selection criteria, to evaluate the best possible model of the covariance. Given that the data are not very stringent, there was however no strong motivation to discuss the same. As is also mentioned in \cite{Gomez-Valent18}, this dependence on the prior choice of the covariance functions enforces a subjectivity, that needs to be properly accessed {to select the kernel model that provides best description of data and hence for making reliable predictions}. As we anticipate the MTGP formalism is capable of providing better reconstructions of the data, the need to discuss kernel selection will become relevant. Note that here kernel selection should not be confused with model selection as is discussed in the context of Gaussian Process literature. The standard idea of model selection in a Gaussian Process regression (GPR) is to select the best possible explanation for the data given the assumed model. In a non-Bayesian approach, as is implemented in this paper, the best explanation for the data is obtained by optimising the log marginal likelihood, $\log[p(\mathcal{D}|\mathcal{M})]$ (equivalent to \Cref{eqn:LML}). Here $\mathcal{D}$ are the observations and $\mathcal{M}(\Theta)$ is the assumed probabilistic model, with parameters $\Theta$. In the Bayesian formalism the posterior distribution of $\Theta$ is given according to the Bayes' rule as,
\begin{equation}
    \label{eqn:BayesRule}
    p(\Theta|\mathcal{D},\mathcal{M}) = \dfrac{p(\mathcal{D}|\Theta,\mathcal{M})p(\Theta|\mathcal{M})}{p(\mathcal{D}|\mathcal{M})},
\end{equation}
where $p(\Theta|\mathcal{M})$ is the prior on the models. Here $p(\mathcal{D}|\mathcal{M})$ is the marginalised likelihood which is also referred to as \textit{evidence} and is given as,
\begin{equation}
    p(\mathcal{D}|\mathcal{M}) = \int p(\mathcal{D}|\Theta,\mathcal{M})p(\Theta|\mathcal{M}) \textrm{d}\Theta.
\end{equation}

Therefore a comparison of the \textit{evidence} for two given different models $\mathcal{M}_A$ and $\mathcal{M}_B$, with equal/uniform priors $p(\mathcal{M}_A) = p(\mathcal{M}_B$), can be inferred using the Bayes factor as,
\begin{equation}
    \frac{p(\mathcal{M}_A|\mathcal{D})}{p(\mathcal{M}_B|\mathcal{D})} =\frac{p(\mathcal{D}|\mathcal{M}_A)}{p(\mathcal{D}|\mathcal{M}_B)},
\end{equation}
where $p(\mathcal{M}_i|\mathcal{D})$ is the posterior for a respective model ($\mathcal{M}_i$) given observations $\mathcal{D}$. The comparison of {evidence} can also be extended to the non-Bayesian formalism to select the model that performs better for the given data. However, this comparison is not necessarily suitable to perform selection amongst different kernels, with smooth data such as in the current cosmological context. While the comparison of the {evidence} for different kernels can be performed, it is usually the case with smooth datasets, that one finds the simplest (in this case the SE) covariance function to be probabilistically best suited. To test the performance of an assumed kernel, one can appropriately split the available data into two separate categories, i.e., the training data ($\sim 75\%$) and test data ($\sim 25\%$). Learning is performed on the training dataset and then the predictions based on the optimised model are compared against the test data by estimating either the \textit{standardised mean squared error} or \textit{negative log probability} \cite{Rasmussen06} ({see also \cite{Melkumyan11} for the examples demonstrated therein}).

In this work we attempted to evaluate for a better suited kernel combination for available cosmological data, but to no avail. We performed 5 different tests by selecting  $\sim 25 \%$ or $\sim 33 \%$ of the total data points as test data, over different regions, to perform extrapolation (test data chosen towards one end of the dataset) and interpolation (test data chosen in the central regions of the dataset) tests. We find a more flexible kernel (e.g., $M_{3/2}$) to be more suitable in 2 cases, while $M_{9/2}$ and $M_{5/2}$ are the best suited in 3 other cases. We conclude that the current compilation of the cosmological data is unable to provide strict preference for a particular kernel combination. In the current work we implement all the kernels except the $M_{3/2}$, which we a priori deem not very suitable for the reconstructions of the derivatives, given its complexity (higher flexibility) to predict more features, differentiability only to first order, and the fact that it could suffer from possible over-fitting due to higher cross-covariance compared to other kernels. However, we do utilise $M_{3/2}$ to compare estimates of cosmological parameters at particular target points such as $H_0$, $q_0$ etc. While the discussion presented here is minimal, a proper kernel selection criteria is necessary if one intends to draw more robust inferences from future data to arrive, in a cosmology independent way. Kernel selection is a much more intricate problem and needs dedicated attention \cite{Abdessalem17}, which we intend to leave for an extended future investigation.

However, given the nature of the cosmological data we utilise an heuristic argument to select from amongst the kernel combinations available. As we expect all the expansion rate data to have the same underlying latent functions, the kernel combination that is capable of predicting the same is clearly a better choice. Therefore, we infer the kernel combination with minimum amount of complexity required to predict equivalent latent functions to provide the best description of data. Increasing the flexibility of the kernels beyond this limit might very well provide same latent functions across datasets, however being over-fitted or predicting additional/unrealistic features. Finally, we choose from amongst the smoother kernels with these criteria to report suitable constraints and comment on the cosmological inferences \textit{, while all the kernel choices remain equally valid in a statistical context of kernel selection}.

\section{Results and Discussion}
\label{sec:Results}
In this section we present the results obtained from the above described formalism \Cref{sec:Method} implemented with the data presented in \Cref{sec:Data}. We present the results for estimation of cosmological parameters, reconstructions and then proceed to report the inferred diagnostics, and the corresponding improvement in comparison to the previous works. Finally, we also comment on the issue of systematics within the CC data.
\subsection{Reconstruction of data and cosmological parameters}
\label{subsec:MainResults}
\begin{figure}[h]
    \centering
\includegraphics[width=0.42\textwidth]{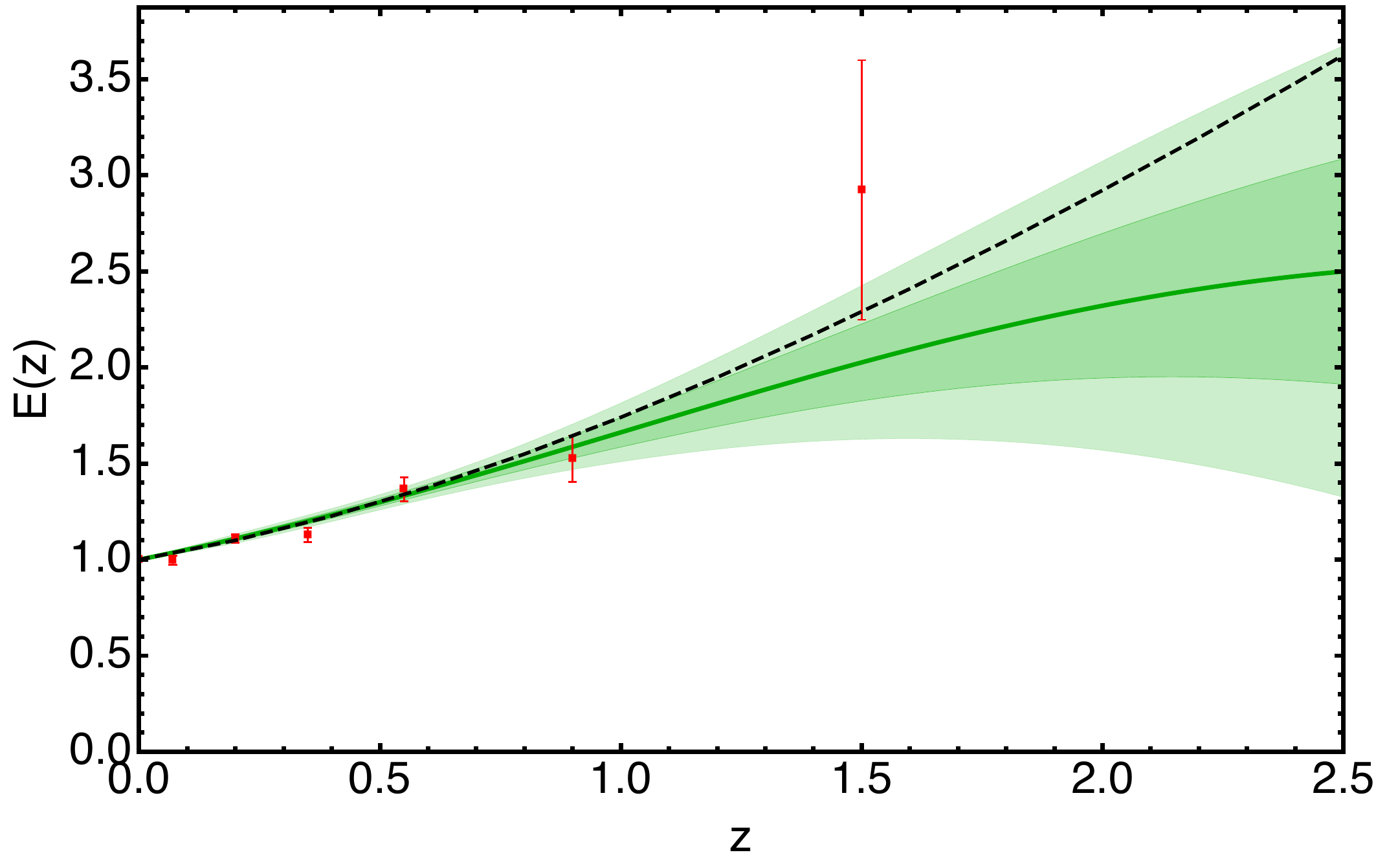}
\hspace{0.15in}
\includegraphics[width=0.42\textwidth]{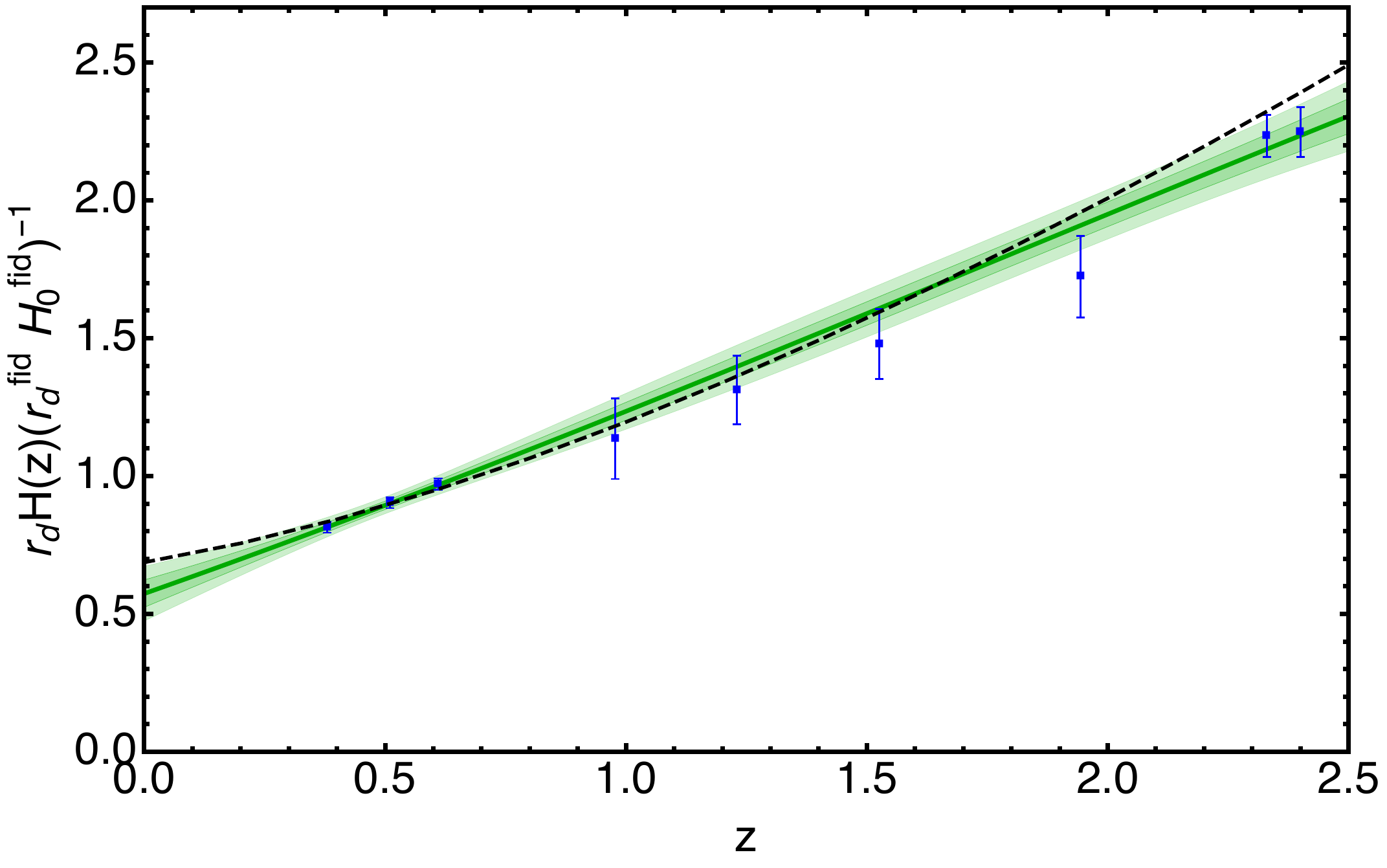}
\vfill
\includegraphics[width=0.42\textwidth]{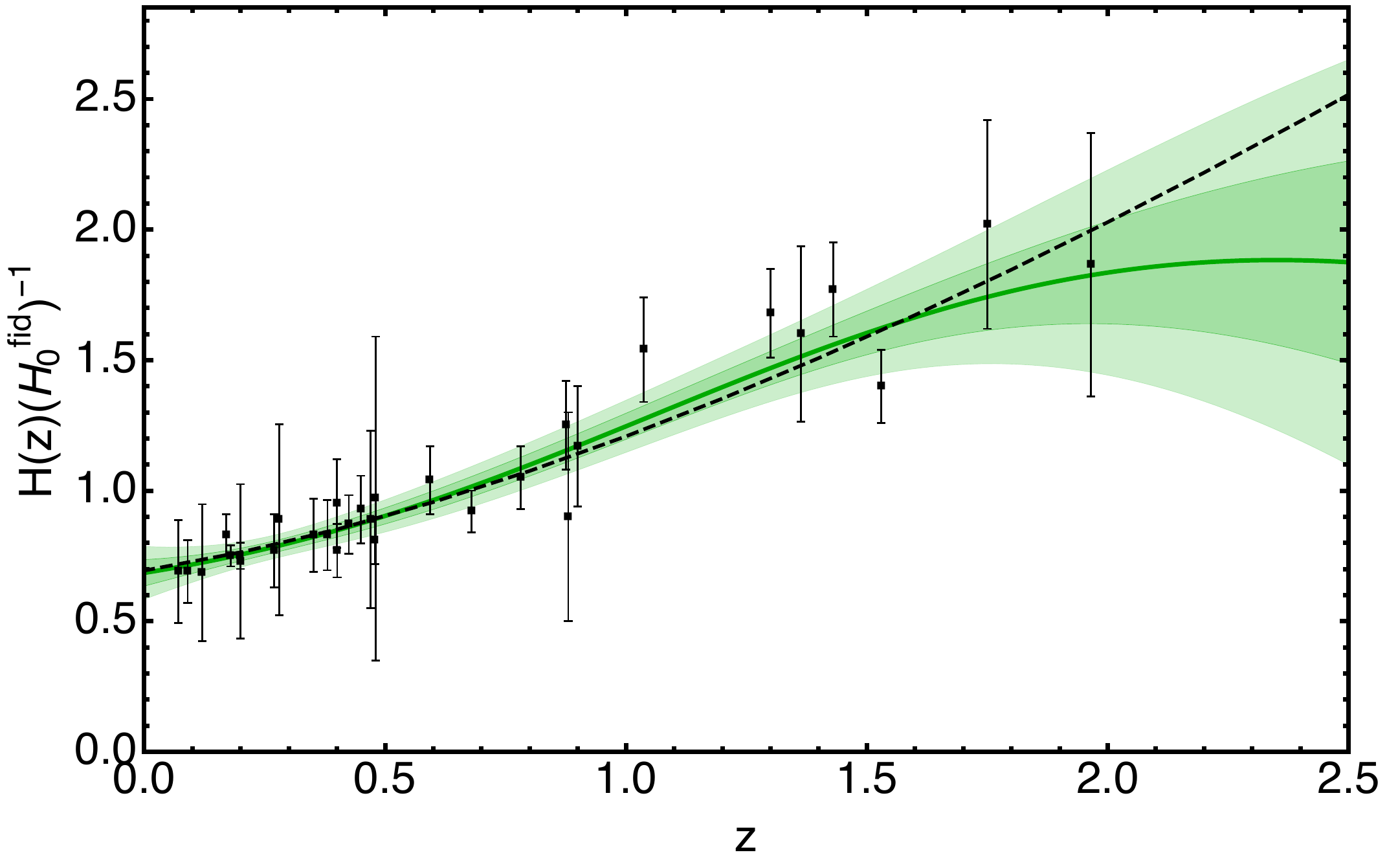}
\hspace{0.15in}
\includegraphics[width=0.42\textwidth]{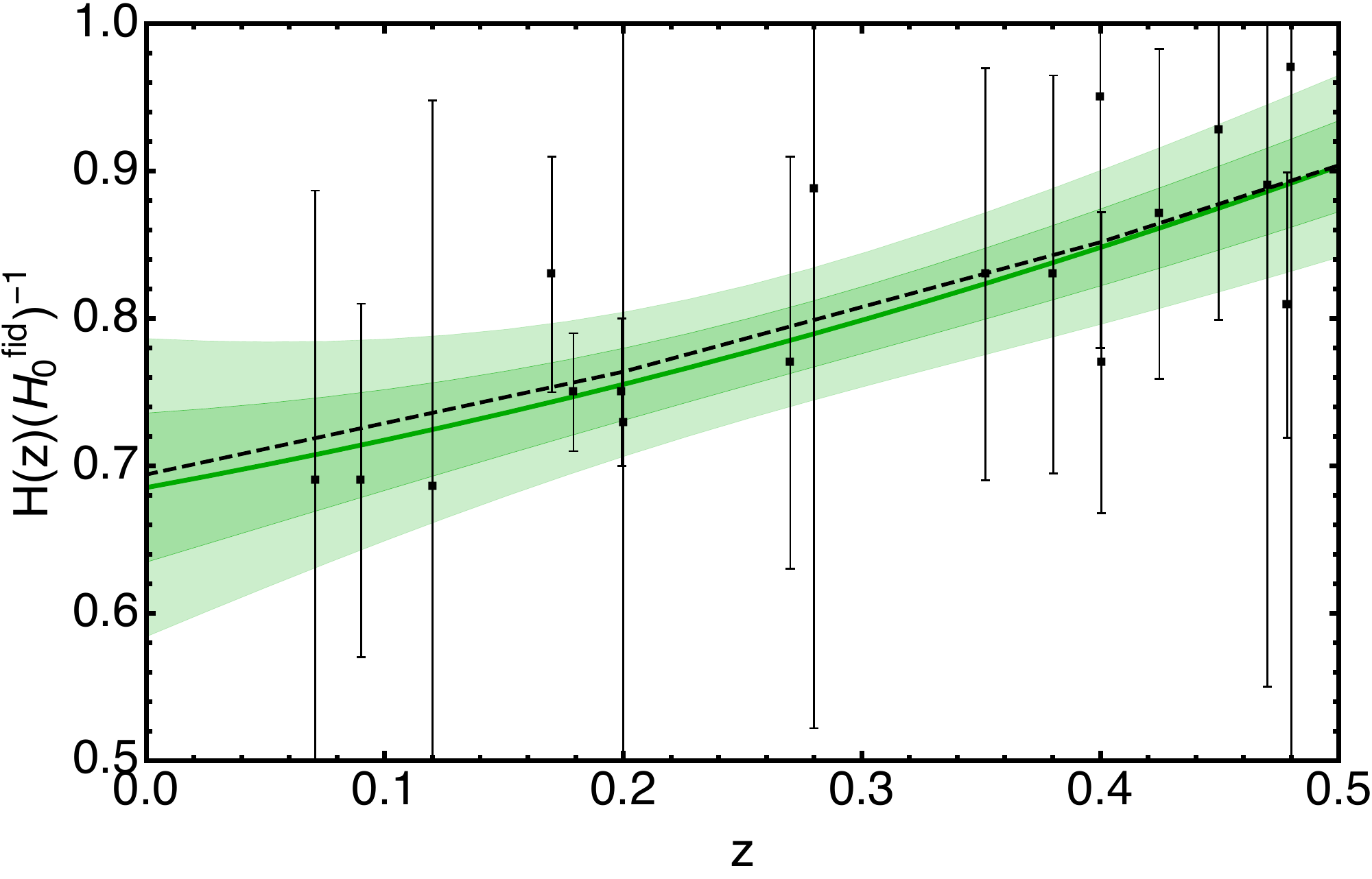}
\caption{We show the reconstructions of all three datasets, SN (\textit{top-left}), BAO (\textit{top-right}) and CC (\textit{bottom-left}) using the individual GP formalism with an assumed $M_{9/2}$ kernel. In the bottom-right panel we show the CC reconstruction enlarged in the redshift range $z < 0.5$ The data have been rescaled with fiducial values $H_0^{fid} = 100$ $\textrm{km/s Mpc}^{-1}$ and $r_d^{fid} = 148 $ Mpc. The dashed line corresponds to best-fit $\Lambda$CDM to the three datasets, with $\Omega_m = 0.284$ and $H_0 = 69.77$ $\textrm{km/s Mpc}^{-1}$, in all the frames (see \Cref{sec:Diagnostics}). {The symmetric prior of $\mu(x) = 0$, with practically unbounded maximum of $\sigma_f \sim 10$ ($68\%$ region) reflects the entire region of the plots and more on either side of the y-axis.}}
    \label{fig:STGPCC}
\end{figure}

We start by reproducing the existing results through individual GPs and then proceed to discuss our results from the MTGP method. In \Cref{fig:STGPCC} we show the individual GP reconstructions of all three datasets considered in this work using the $M_{9/2}$ kernel, as an example. The well-known reconstruction of the CC data is very similar to the one obtained using the $SE$ kernel (see also \cite{Yu17}), which we re-assert. Also, we find very little to no difference from one kernel to another even for the SN and BAO datasets. Reconstruction of the SN data is clearly at a disadvantage for higher redshifts ($z \gtrsim 1.5$) due to the very large uncertainty of the data point at $E(z = 1.5)$. The inclusion or exclusion of this data point makes very little difference as was also mentioned in \cite{Gomez-Valent18} (while they opt to omit this data point we let it remain, with no dire consequences). Similarly, the BAO data is unable to show any features in the reconstructions due to the sparse distribution, and the unavailability of data at redshifts $z \lesssim 0.3$. Note that using Mat\'ern kernels that are necessarily modelled to allow more dynamics/features do not fare immediately well in Single-Task GP. We do not find much difference, except for an increase in the predictive region in the $M_{3/2}$ case compared to $M_{9/2}$. In fact, their utility only becomes significant in our MTGP formalism or when the derivatives are inferred. The estimates for $H_0$ using several kernels from the reconstruction of CC data alone are summarised in \Cref{tab:CCH0main}.

\begin{figure}[h]
    \centering
        \includegraphics[width=1.0\textwidth]{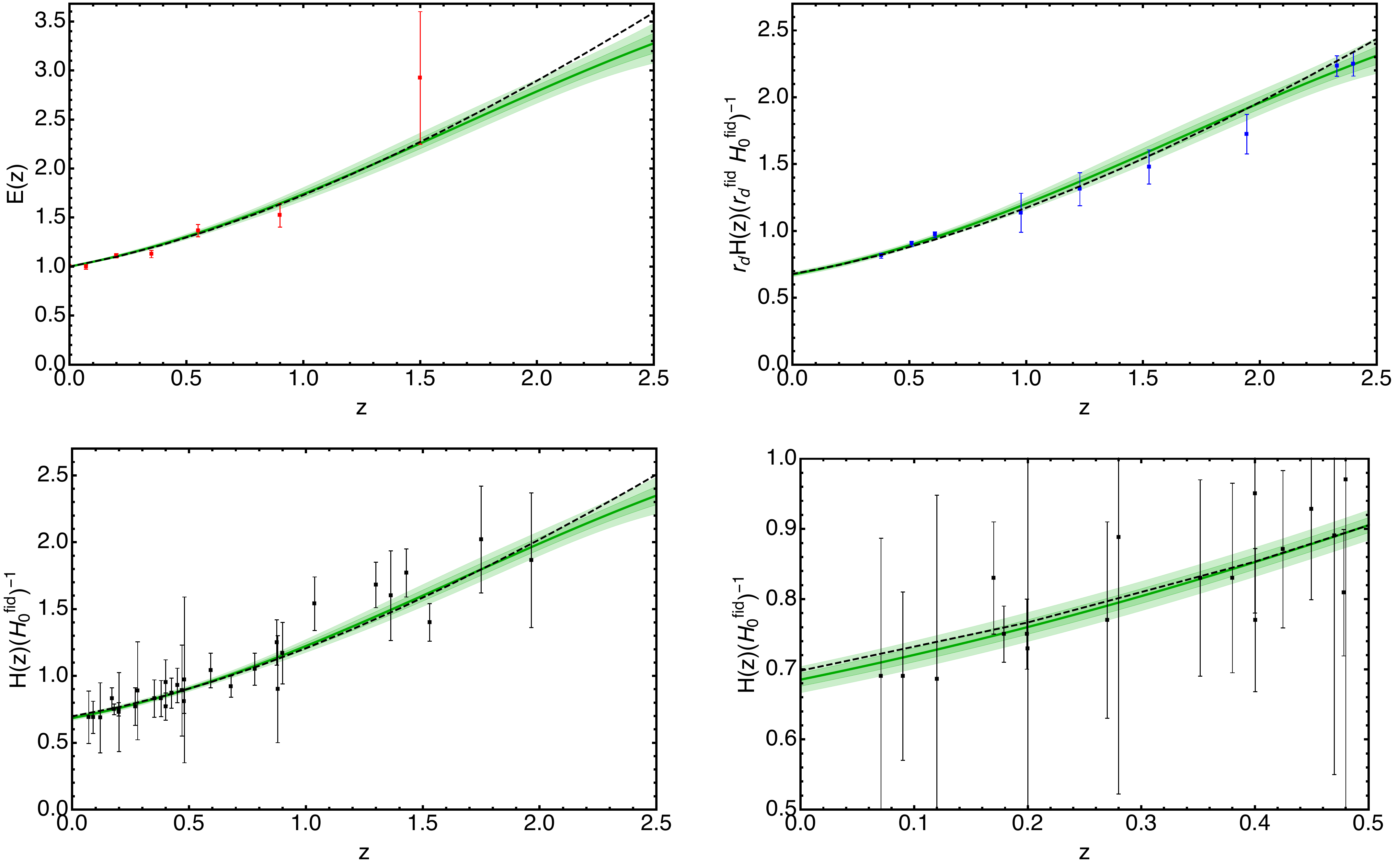}
        \caption{We show the reconstructions of the all the three datasets using the MTGP formalism with an assumed $M_{7/2}$ kernel for each of the dataset. The right-bottom panel shows the enlarged image of the $z<0.5$ region to emphasise the improvement in the intercept of the CC reconstruction. In all three the panels the data are rescaled with $H_0^{fid} = 100$ $\textrm{km/s Mpc}^{-1}$. The dashed line corresponds to the best-fit $\Lambda$CDM with $\Omega_m = 0.284$, $H_0 = 69.77$ $\textrm{km/s Mpc}^{-1}$, to the current data (see \Cref{sec:Diagnostics}). {The symmetric prior of $\mu(x) = 0$, with practically unbounded maximum of $\sigma_f \sim 10$ ($68\%$ region) reflects the entire region of the plots and more on either side of the y-axis.}}
        \label{fig:MTGPM72}
\end{figure}

{Another important feature worth noticing is the lowering of the posterior mean in the SN and CC reconstructions in the left column of \Cref{fig:STGPCC}. As mentioned earlier in \Cref{sec:Method}, this feature could essentially be due to the lack of data in this region and the tendency of reconstructed Gaussian posterior to converge to a 0 (prior) mean with $68\%$ confidence regions given by the optimized $\sigma_f$ hyperparameter, far away from the data. Clearly, this implies that the extrapolated reconstructions are less reliable in individual GPs for making inferences, e.g., $H_0$\footnote{However, it can be noticed in \Cref{fig:STGPCC} that the CC reconstruction does not immediately start tending to the prior $\mu = 0$ as $z \rightarrow 0$.}.} In the MTGP formalism however, these three datasets are modelled to complement each other, thereby providing better reconstructions for all the three datasets (see \Cref{fig:MTGPM72}). {This essentially implies that the overall range of available data is now an union of all three dataset ranges i.e., $0<z\lesssim 2.5$ and as expected the effects of extrapolation in all three data planes are now mitigated within this joint redshift range.}
\newline

\begin{figure}[h]
    \centering
\includegraphics[width=0.45\textwidth]{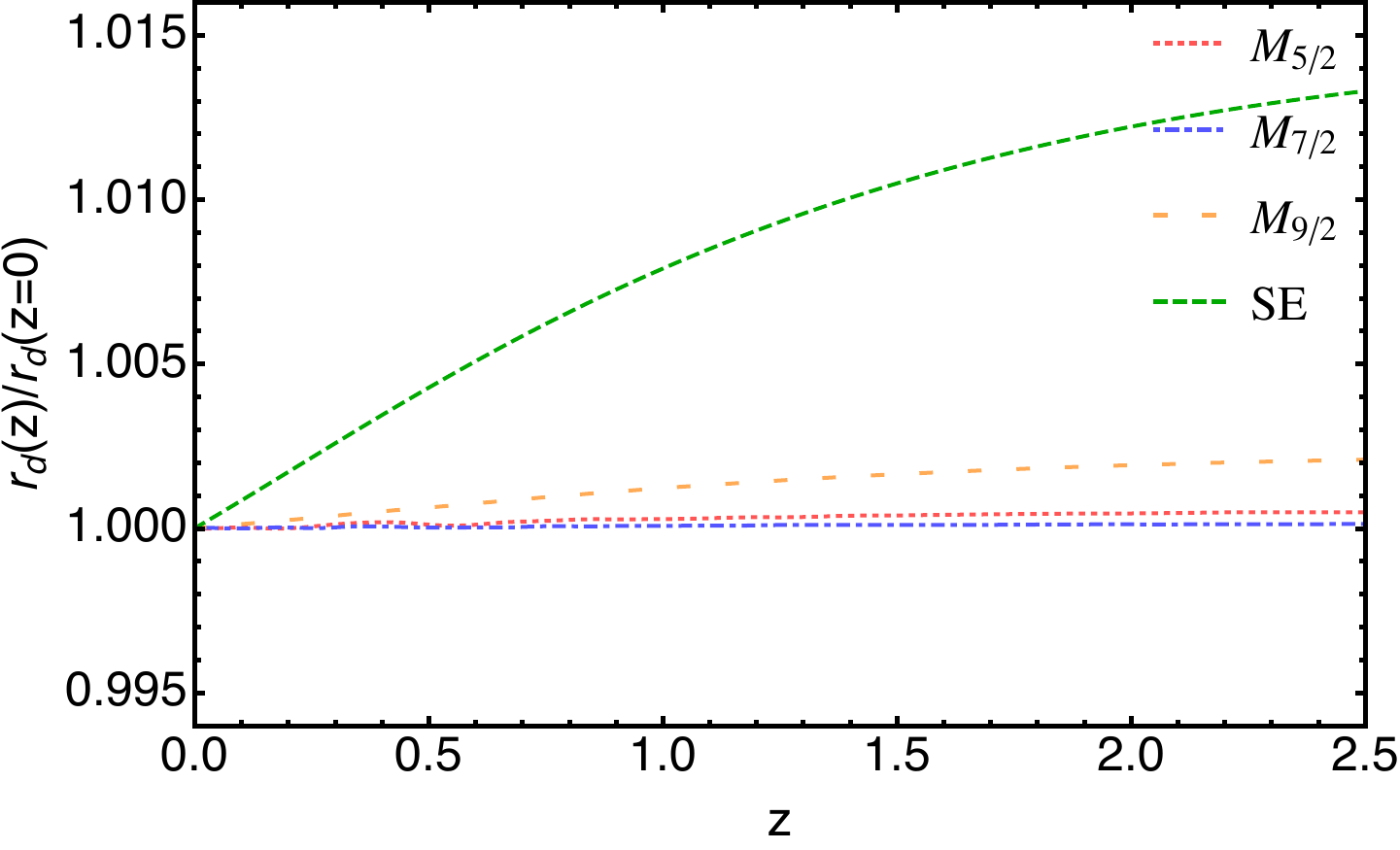}
\hspace{0.15in}
\includegraphics[width=0.44\textwidth]{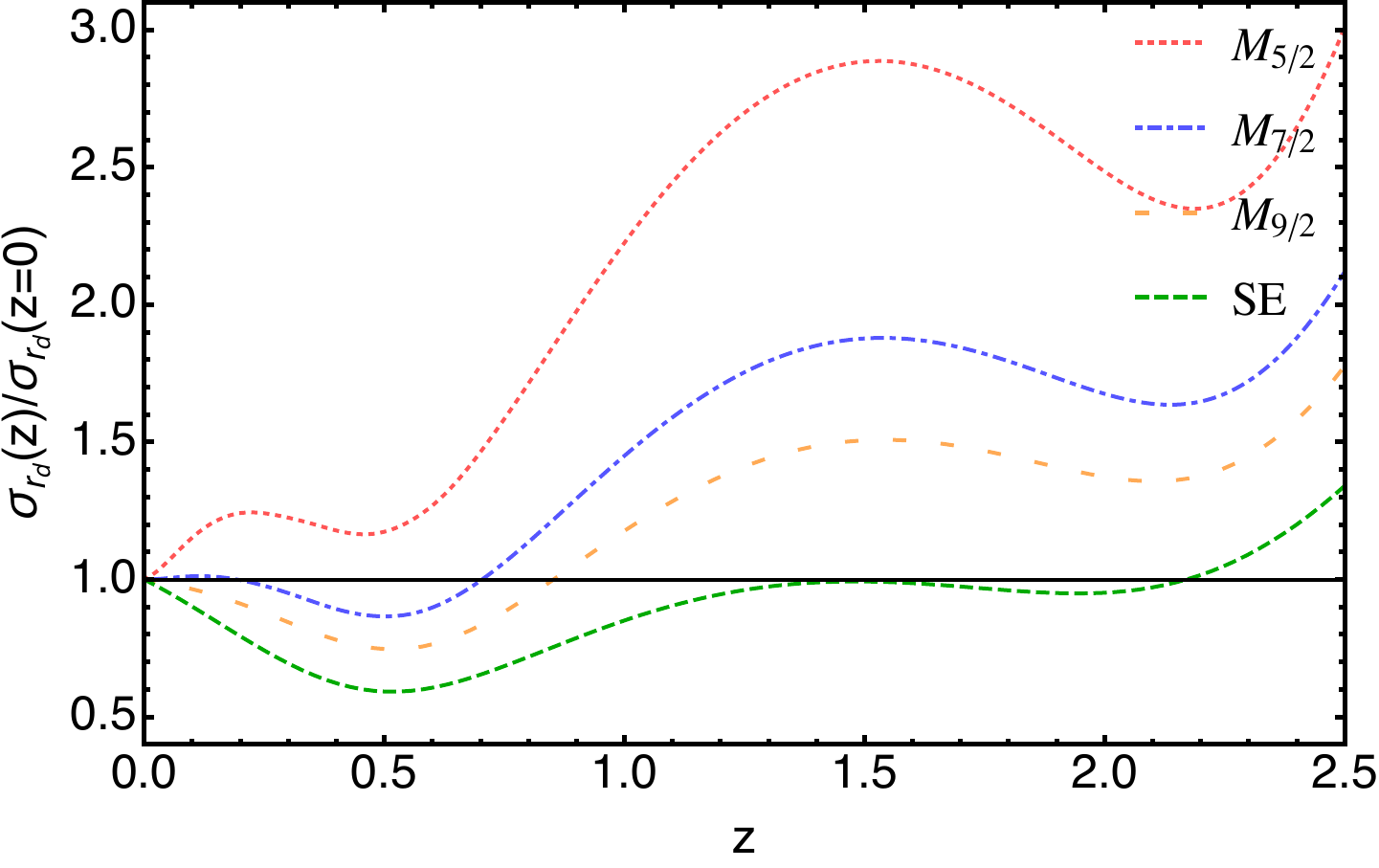}
\caption{In the left panel we show the variation in the $r_d$ estimate normalised to $r_d(z=0)$ at the intercept, over redshift using different kernels. Similarly, in the right panel we show the variation of the uncertainty $\sigma_{r_d}$ in redshift, also normalised to $\sigma_{r_{d}}(z=0)$. }
    \label{fig:rdofz}
\end{figure}

As mentioned earlier (\Cref{subsec:DataImple}), we implement the same kernel across all the data planes considered, as one would expect that the dynamic nature of the reconstructions for these three datasets is essentially the same. However, this does not imply that we a priori assume the underlying latent functions of the three tasks to be unique and hence one could find differences in the reconstructions from one plane to another (see \Cref{fig:rdofz}), as the log marginal likelihood (\Cref{eqn:LML}) is optimised with different length scale hyperparameters for each task. While it might appear inconsistent if the three data planes do not predict unique expansion histories, this freedom provides us with the criteria to evaluate the best kernel combination, that will predict equivalent latent functions, i.e., similar length scale hyperparameters. {In fact, the equivalence of the posteriors validates that the observed features in MTGP analysis are no longer due to extrapolated effects but due to the optimized data correlations.} After obtaining equivalent predictions in all three planes, the results must be read as the final reconstruction of the CC dataset complemented by the BAO at high-redshift and SN data at low redshifts. Therefore we later proceed to study the dynamics based on the CC reconstruction, obtained from MTGP of all three datasets, although the dynamics inferred from all three reconstructions will be same. We test for a strict equivalence between CC and BAO reconstructions, and expect an agreement amongst CC/BAO and SN to be better than $1\sigma$ at higher redshifts (as SN data completely lacks information in this region). We infer $H_0$ as the intercept of CC reconstruction and use CC and BAO reconstructions to obtain $r_d$ as a rescaling constant at the intercept of BAO reconstruction.

{\renewcommand{\arraystretch}{1.2}%
    \begin{table}[h]
    \begin{center}
    \footnotesize
    \begin{tabular}{|c|c|c|c|}
    \hline
     Dataset(s) & Kernel & $H_0$ [$\textrm{km/s Mpc}^{-1}$] & $r_d$ [Mpc]   \\
    \hline
    \hline
    CC & \begin{tabular}{@{}c@{}c@{}c@{}}$SE$ \\ $M_{9/2}$ \\ $M_{7/2}$ \\ $M_{5/2}$\end{tabular} & \begin{tabular}{@{}c@{}c@{}c@{}}$67.44 \pm 4.75$ \\ 68.52 $\pm$ 5.06 \\ 68.73 $\pm$ 5.18 \\ 68.89 $\pm$ 5.43 \end{tabular} & \begin{tabular}{@{}c@{}c@{}c@{}} - \\ - \\ - \\ - \end{tabular} \\
    \hline
    CC+SN+BAO & \begin{tabular}{@{}c@{}c@{}c@{}}$SE$ \\ $M_{9/2}$ \\ $M_{7/2}$ \\ $M_{5/2}$ \end{tabular} & \begin{tabular}{@{}c@{}c@{}c@{}}$67.39 \pm 1.36$ \\ 68.24 $\pm$ 1.09 \\ \textbf{68.52 $\pm$ 0.94} \\ 68.86 $\pm$ 0.67 \end{tabular} & \begin{tabular}{@{}c@{}c@{}c@{}}144.24 $\pm$ 4.38  \\ 145.35 $\pm$ 3.34 \\ \textbf{145.61 $\pm$ 2.82 } \\ 145.78 $\pm$ 2.02 \end{tabular} \\
    \hline
    CC+SN+BAO+R18 & \begin{tabular}{@{}c@{}c@{}c@{}}$SE$ \\ $M_{9/2}$ \\ $M_{7/2}$ \\ $M_{5/2}$\end{tabular} & \begin{tabular}{@{}c@{}c@{}c@{}}$71.58 \pm 0.34$ \\ 71.35 $\pm$ 0.31 \\ 71.40 $\pm$ 0.30 \\ 71.61 $\pm$ 0.37 \end{tabular} & \begin{tabular}{@{}c@{}c@{}c@{}}140.54 $\pm$ 1.28 \\ 141.45 $\pm$ 1.35  \\ 141.29 $\pm$ 1.31 \\ 141.67 $\pm$ 1.39  \end{tabular} \\
    \hline
    \end{tabular}
    \caption{Estimates of $H_0$ and the corresponding statistical $1\sigma$ uncertainty form combinations of the dataset(s) described in \Cref{sec:Data} implemented for several kernels shown in second column. For the joint analysis with the multiple datasets we implement the same kernel for each of them. In the last column we report the $r_d$ estimates and their uncertainties. Highlighted in bold is the best inference for the reasons stated in the text.}
    \label{tab:CCH0main}
    \end{center}
\end{table}
}

In this respect, we would also like to mention that a unique expansion history can be enforced by restricting to the same kernel across all the planes and then imposing that the length scale hyperparameter ($l$) of the three planes should be the same. This assumption in the MTGP formalism implies that the datasets are in fact rescaled by an absolute constant (i.e., $H_0$ (CC) and $H_0 r_d$ (BAO), are the rescaling factors with respect to the $E(z)$ (SN) reconstruction), and is clearly not the right assumption to make a priori for the estimation of physical constants. {This provision within the formalism becomes extremely useful to estimate the differences in the data once the requirement for unique expansion history is enforced, as will be elaborated later (see \Cref{sec:Diagnostics}).} Reconstructions of the three datasets in the MTGP formalism using $M_{7/2}$ kernel are shown in \Cref{fig:MTGPM72}. One can immediately notice a significant improvement in the reconstruction of the individual datasets in comparison to the single task GP shown in \Cref{fig:STGPCC}. The Mat\'ern class kernels with lower $\nu < 7/2$ force the SN reconstruction to be more in agreement with the CC and BAO data reconstructions, which is a consequence of higher cross-covariance between $M_{\nu \leq 5/2}$ kernels for $\tau \to 0$, as shown in right panel of \Cref{fig:KerComp}.
\newline

\textit{Comments on $H_0$ estimates:} The $H_0$ estimate has been discussed time and again in the context of Gaussian Processes, usually prompting values compatible with the ``high-redshift'' CMB estimate (P16). We show the improvement in the $H_0$ estimate and the corresponding uncertainty from standard GP to the MTGP implemented here, with the addition of complementary datasets in \Cref{tab:CCH0main}. In the MTGP formalism with the inclusion of BAO data alone to the CC dataset there is only a modest improvement in the $H_0$ estimate, while the more important contribution is seen at the high redshifts ($2 \leq z \leq 2.5$), where no CC data is available and hence the BAO data aids the CC reconstruction in this redshift range. For example, assuming $M_{9/2}$ kernel we find $H_0 = 64.00 \pm 3.35$ $\textrm{km/s Mpc}^{-1}$ using CC + BAO datasets. In fact, this uncertainty on $H_0$ is of the order of the single task GP conducted in \cite{Yu17}, where BAO (5 points) data were rescaled with an assumed $r_d$ and a Single-Task GP was performed on a total of 36 (31 CC + 5 BAO) data points. We also find a very good agreement in the predicted mean values of the $H_0$ from this work and \cite{Yu17} (please see Table 2. therein). The SN data helps to improve the CC reconstruction at lower redshifts and hence the $H_0$ estimate to a much higher significance. As shown in \Cref{fig:STGPCC}, the SN data is at a severe disadvantage for $z \gtrsim 1$ in the simple GP, while its reconstruction in the MTGP is improved by several orders (see \Cref{fig:MTGPM72}). More recently, \cite{Gomez-Valent18} have utilised the same SN and CC dataset to obtain the $H_0$ in a modified GP formalism. We do assert the consistency in our findings with the results quoted therein. Our formalism provides slightly stringent constraints, while also allowing for a provision to include the BAO data.

As anticipated, we encounter a subtlety of choosing the best possible reconstruction (i.e., choice of the kernel) to quote the $H_0$ estimate. While it is appreciable that the constraint on $H_0$ gets better with more flexible Mat\'ern class kernels ($\nu \leq 5/2$), it is compensated with the overall predictive quality of the reconstruction. For example, using $M_{3/2}$ we find $H_0 = 69.09 \pm 0.38$ $\textrm{km/s Mpc}^{-1}$, which is even more stringent compared to $M_{5/2}$ constraint shown in \Cref{tab:CCH0main}. While the predictive region is over-fitted at lower redshifts due to higher cross-correlation with SN data, at the same time, for the higher redshift regions the predictive quality deteriorates in comparison to the kernels with $\nu > 3/2$. Implementing the heuristic argument for equivalent latent functions introduced in \Cref{subsec:msel} (also elaborated alongside $r_d$ estimates), we argumentatively infer that the assumption of $M_{7/2}$ kernel provides a better description of the data in comparison to other kernel choices for the final constraints on all the cosmologically relevant quantities. Our best estimate of $H_0 = 68.52 \pm 0.94 $ $\textrm{km/s Mpc}^{-1}$ from CC + SN + BAO datasets, is in agreement with the ``high-redshift'' P16 at 1.41$\sigma$ and in a lesser agreement with the local $H_0$ measurement of R18 at 2.60$\sigma$ \footnote{Our estimate is also consistent with several other previous estimates in \cite{Chen11, Lin17a, Cheng15, Yu17, Haridasu17a, Gomez-Valent18}.}. This remains to be a very consistent scenario as we have also reported in \cite{Haridasu17a} (see also references therein and \cite{Lukovic18}, for discussion on $H_0$ estimates from "low-redshift" data), even in a model-dependent analysis, especially if measurements obtained with other stellar population synthesis models are not considered, as we discuss in \Cref{sec:Diagnostics}. Our estimate seems to be consistent with several other low-redshift estimates obtained in \cite{Chen11}. In any case, we find that all the reconstructions assuming each of the $SE$ or Mat\'ern class kernels ($\nu \geq 5/2$) provide very consistent estimates on $H_0$. 
\newline

\textit{Comments on $r_d$ estimates:} In \Cref{tab:CCH0main}, we also show the constraints on $r_d$ obtained as the rescaling factor between CC and BAO reconstructions. As can be seen in \Cref{fig:MTGPM72}, the formalism is able to realise that the CC and BAO data are in fact better described by similar underlying latent functions and hence provide similar length scale hyperparameters. This implies that two reconstructions are well correlated and hence a better description of the data. The reconstructed regions in \Cref{fig:MTGPM72} are obtained after including the covariance between all the datasets. Therefore, the $r_d$ estimates are simply obtained as the ratios of the intercepts in the BAO and CC reconstruction planes without the need to reconsider the covariance. A cross-check for the same would be if we obtain $\sigma_{r_d} = 0$ when considering the covariance twice while obtaining the uncertainty. In fact, when the $SE$ kernel is considered the MTGP is unable to provide a good equivalence for the latent functions of BAO and CC reconstructions, which leads to a non-zero ($0.36$) uncertainty when covariance is considered twice.

We would also like to remind that the estimates of $r_d$ quoted in \Cref{tab:CCH0main} are obtained as the intercept of the BAO data at $z=0$. However, this constraint is supposed to improve at the best-reconstructed redshift of the BAO/CC data. Using the CC + BAO + SN data we find $r_d = 145.61 \pm 2.44 $ Mpc ($M_{7/2}$), $r_d = 145.45 \pm 2.49$ Mpc ($M_{9/2}$) and $144.86 \pm 2.59$ Mpc ($SE$) estimates at redshifts $z = 0.50$, $z = 0.53$ and $z = 0.51$, respectively. Incidentally, all the kernels predict the best reconstruction at similar redshifts, which is not surprising as the better of BAO data are available in the redshift range of $0.38 < z < 0.61$. This trend is no longer the case when more flexible kernels like $M_{5/2}$ and $M_{3/2}$ are used, as they predict the best BAO reconstruction at $z=0$ under the influence of strong cross-covariance with SN data. This, in fact, is a clear indication of over-fitting and these kernels are therefore not inferred as to provide the best description of the data. Needless to say, the estimate of mean $r_d$ would also slightly vary when obtained from different redshifts if the optimised underlying latent functions are not exactly the same. For example, the mean of $r_d$ estimated at different redshifts with $M_{9/2}$ kernel varies in the range of $145.35 < r_d < 145.65$, accounting for an additional systematic uncertainty of $\sigma_{r_d} (sys) = 0.01$ in quadrature. In the left panel of \Cref{fig:rdofz} one can notice that using $M_{7/2}$ kernel we find that the latent functions of CC and BAO are completely equivalent, which makes it a better choice compared to $SE$ and $M_{9/2}$. On the other hand, it remains better than $M_{3/2}$ and $M_{5/2}$ as they provide over-fitted predictions (see right panel of \Cref{fig:rdofz}). All the above-presented arguments show $M_{7/2}$ kernel in the light of being the better choice to make predictions with the current data. Also in the left panel of \Cref{fig:rdofz}, it can be clearly seen how the uncertainty of the $r_d$ predictions follow the availability of data, by providing better reconstructions around $z \sim 0.5$ and $z \sim 2.3$, compared to their neighbouring regions.

Finally, we also include the R18 estimate of $H_0 = 73.48 \pm 1.66 $ $\textrm{km/s Mpc}^{-1}$ \cite{Riess18a} to the joint analysis and as expected find improvement in the constraints on $H_0$. We find extremely consistent final estimates of $H_0$ with all the covariance functions utilised in this work. In fact, this high consistency discourages us to make any inference for the best-suited kernel. However, we do utilise $M_{7/2}$ to quote the results throughout the discussion, without loss of reasoning. The estimated $H_0$ and $r_d$ values are summarised in \Cref{tab:CCH0main}. As expected, with the addition of R18 to the CC + SN + BAO data the errors on the parameters decrease. Given the increase in $H_0$ values we find a consistent decrease in $r_d$ estimates, also retaining very similar reconstructions using all the kernels. Clearly, the value of $r_d H_0$ is conserved while the estimate of $r_d$ gets rescaled to accommodate the increase in the value of $H_0$. Using $M_{7/2}$ kernel, with and without the inclusion of R18 we find $r_d H_0 = 10088 \pm 83$ $\textrm{km/s}$ and $r_d H_0 = 9977 \pm 237$ $\textrm{km/s}$, respectively. While the latter estimate without the inclusion of R18 is extremely consistent with the ``high-redshift'' CMB estimate of $r_d H_0 = 9952 \pm 97$  $\textrm{km/s}$ \cite{Ade16}, the earlier measure is also consistent at $\sim1\sigma$. Given the agreement in $r_d H_0$, the discrepancy remains within breaking the degeneracy between these two parameters. However, we extend this discussion also to a comparison of $\Omega_m$ estimates in \Cref{sec:Diagnostics}.

{\renewcommand{\arraystretch}{1.2}%
    \begin{table}[h]
    \begin{center}
    \footnotesize
    \begin{tabular}{|c|c|c|c|}
    \hline
     CC & SN & BAO & $H_0$ [$\textrm{km/s Mpc}^{-1}$]   \\
    \hline
    \hline
    $SE$ & $M_{9/2}$ & $SE$ & $65.46 \pm 2.60$ \\
    $M_{9/2}$ & $SE$ & $M_{9/2}$ & $65.47 \pm 2.60$ \\
    $M_{9/2}$ & $SE$ & $SE$ & $68.05 \pm 2.89 $ \\
    $SE$ & $M_{9/2}$ & $M_{9/2}$ & $66.69 \pm 2.75 $ \\
    $M_{9/2}$ & $M_{7/2}$ & $M_{7/2}$ & $67.94 \pm 1.94 $ \\
    \hline
    \end{tabular}
    \caption{Estimates of $H_0$ and the corresponding $1\sigma$ uncertainty for different combinations of the kernels assumed for different datasets shown in first three columns.}
    \label{tab:CCH0MKGP}
    \end{center}
\end{table}
}

\textit{Illustration of Multi-Kernel Gaussian Process:} So far we have performed the analysis assuming same kernel for all the datasets. Now we would like to extend the analysis with the provision within current formalism and utilise different kernels for the datasets and comment on its implications. As anticipated, we find that the initial assumption of one kernel for all the datasets remains best, as we try to implement several combinations. For instance the combination of $M_{9/2}$ (SN), $SE$ (BAO), $SE$ (CC) fares poorly compared to the assumption of $SE$ kernel for all datasets. In this case, we find $H_0 = 65.46 \pm 2.6 $ $\textrm{km/s Mpc}^{-1}$. In \Cref{tab:CCH0MKGP} we show a few combinations of kernels assumed in the joint analysis and the corresponding $H_0$ estimates obtained. One can clearly notice that the assumption of different kernels does not improve the results. This can be easily explained based on the strength of cross-covariance that is obtained using a specific combination of kernels and the nature of data. In the Multi-Kernel formalism a combination including $M_{3/2}$ kernel provides stronger covariance for closer data points (i.e., lower $|x-x'|$) compared to $M_{5/2}, .. SE$ and the contrary for far away data points. As we have described earlier the SN data complements the CC data essentially at lower redshifts, and hence giving rise to larger dispersion at the intercept of CC reconstruction for kernel combinations involving $M_{\nu \geq 5/2}$ due to their lower cross-covariance. Although this provision is an added advantage in the current formalism, given the nature of data it does not present any immediate application in the current work. However, it could be of interest in several other implementations such as \cite{Kim13,Dhawan18,Guillochon17}, presenting which is beyond the scope of this paper.

\subsection{Reconstruction of Diagnostics}
\label{sec:Diagnostics}
In this section we present our results for the reconstructions of $q(z)$ and a few diagnostics inferred from the MTGP analysis described in the \Cref{sec:Data}. We find a very significant improvements in the reconstructed regions, moving from single GP performed in \cite{Seikel12a, Zhang16} to MTGP formalism implemented here. In \Cref{tab:CCqzmain} we present the estimates for $q_0$ and the transition redshift $z_T$, corresponding to the joint analysis presented in \Cref{tab:CCH0main}. We find very consistent results for the estimates of $q_0$ and $z_T$ using all the kernels implemented here. However, one can notice that the error on estimated $q_0$ increases with the flexibility of assumed kernel, while $z_T$ is constrained mildly better by the $M_{9/2}$ and $M_{7/2}$ kernels. As expected, using $M_{3/2}$ we find deteriorated constraints of $q_0 = -0.56 \pm 0.12 $ and $z_T = 0.51^{+0.43}_{-0.06}$ reassuring that the assumption of $M_{3/2}$ is not suitable for reconstructions of derivatives. We would like to remind that $M_{3/2}$ kernel might retain its importance for reconstructing other cosmological features such as dynamical dark energy equation of state (EoS) $w(z)$. Very similar trend is observed even when R18 is included as data, in the reconstruction. Consistently, the inclusion of R18 predicts higher values of $z_T$ and lower values of $q_0$, appropriately suggesting lower $\Omega_m$. It is also the case that inclusion of R18 does not improve the constraints on $q_0$ and $z_T$, as can be seen in \Cref{tab:CCqzmain}.

{\renewcommand{\arraystretch}{1.5}%
    \begin{table}[h]
    \begin{center}
    \footnotesize
    \begin{tabular}{|c|c|c|c|}
    \hline
     Dataset(s) & Kernel & $q_0$ & $z_T$    \\
    \hline
    \hline
    CC & \begin{tabular}{@{}c@{}c@{}c@{}}$SE$ \\ $M_{9/2}$ \\ $M_{7/2}$ \\ $M_{5/2}$ \end{tabular} & \begin{tabular}{@{}c@{}c@{}c@{}}$-0.48^{+0.33}_{-0.29}$ \\ $-0.57^{+0.38}_{-0.33}$ \\ $-0.57^{+0.39}_{-0.35}$ \\ $-0.57^{+0.44}_{-0.38}$ \end{tabular} & \begin{tabular}{@{}c@{}c@{}c@{}} $0.56^{+0.22}_{-0.24} $ \\ $ 0.54^{+0.18}_{-0.17}$ \\ $0.54^{+0.18}_{-0.16}$ \\ $0.54^{+0.18}_{-0.14}$ \end{tabular} \\
    \hline
    CC+SN+BAO & \begin{tabular}{@{}c@{}c@{}c@{}}$SE$ \\ $M_{9/2}$ \\ $M_{7/2}$ \\ $M_{5/2}$ \end{tabular} & \begin{tabular}{@{}c@{}c@{}c@{}}$-0.44 \pm 0.04$ \\ $-0.50 \pm 0.05$ \\ $-0.52 \pm 0.06$ \\ $-0.54 \pm 0.07$ \end{tabular} & \begin{tabular}{@{}c@{}c@{}c@{}} $0.63^{+0.12}_{-0.11} $ \\ $ 0.64^{+0.12}_{-0.09}$ \\ $0.64^{+0.12}_{-0.09}$ \\ $0.63^{+0.15}_{-0.10}$ \end{tabular} \\
    \hline
    CC+SN+BAO+R18 & \begin{tabular}{@{}c@{}c@{}c@{}}$SE$ \\ $M_{9/2}$ \\ $M_{7/2}$ \\ $M_{5/2}$ \end{tabular} & \begin{tabular}{@{}c@{}c@{}c@{}}$-0.57 \pm 0.05$ \\ $-0.57 \pm 0.06$ \\ $-0.58 \pm 0.06$ \\ $-0.59 \pm 0.07$ \end{tabular} & \begin{tabular}{@{}c@{}c@{}c@{}} $0.74^{+0.09}_{-0.07} $ \\ $ 0.73^{+0.11}_{-0.09}$ \\ $0.71^{+0.12}_{-0.09}$ \\ $0.69^{+0.17}_{-0.10}$ \end{tabular} \\
    \hline
    \end{tabular}
    \caption{Estimates of $q_0$ and $z_T$ and the corresponding $1\sigma$ uncertainties form combinations of the dataset(s) described in \Cref{sec:Data} implemented for several kernels shown in second column. For the joint analysis with the multiple datasets we implement the same kernel for each of them. The results in this table correspond to the second and third rows of \Cref{tab:CCH0main}.}
    \label{tab:CCqzmain}
    \end{center}
\end{table}
}

We find our estimates of $z_T$ obtained using all the kernels shown in \Cref{tab:CCqzmain} to be extremely consistent with $z_T = 0.64^{+0.11}_{-0.07}$ quoted in \cite{Moresco16a}, which was however obtained using the CC+$H_0$ data alone. Although, the constraint has not improved in comparison, in this analysis we have the advantage of being cosmology-independent (see \cite{Moresco16a} for more discussion on $z_T$ estimates from various analysis). using our formalism we find as yet the best constraints on $z_T$ in a cosmology-independent formalism. As was also summarised in \cite{Sahni08}, a stringent constraint on $z_T$ would provide useful insights for accessing the nature of dark energy. As mentioned above, our results are very much consistent with the $\Lambda$CDM model where we find the constraints to be $H_0 = 69.77 \pm 1.80$ $\textrm{km/s Mpc}^{-1}$, $\Omega_m = 0.284^{+0.018}_{-0.017}$ and $r_d = 143.9^{+4.0}_{-3.9}$ Mpc using CC + SN + BAO. Including the R18 prior in the model-dependent analysis we find $H_0 = 71.80 \pm 1.20$ $\textrm{km/s Mpc}^{-1}$, $\Omega_m = 0.275^{+0.016}_{-0.015}$ and $r_d = 141.1^{+3.4}_{-3.3}$ Mpc. We use these constraints for comparison with the model-independent inferences throughout the paper.

\begin{figure}[h]
    \centering
\includegraphics[width=0.45\textwidth]{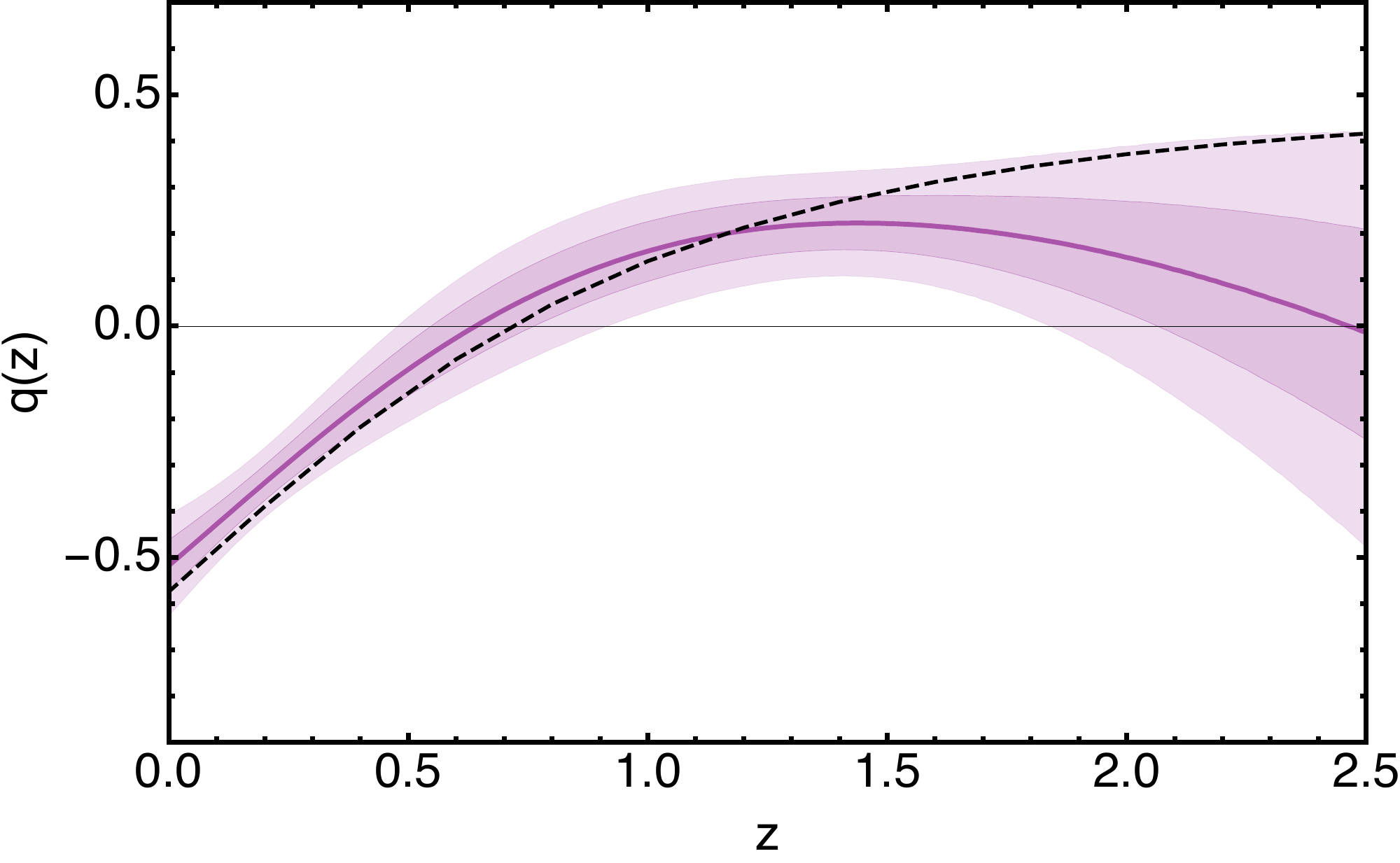}
\hspace{0.15in}
\includegraphics[width=0.45\textwidth]{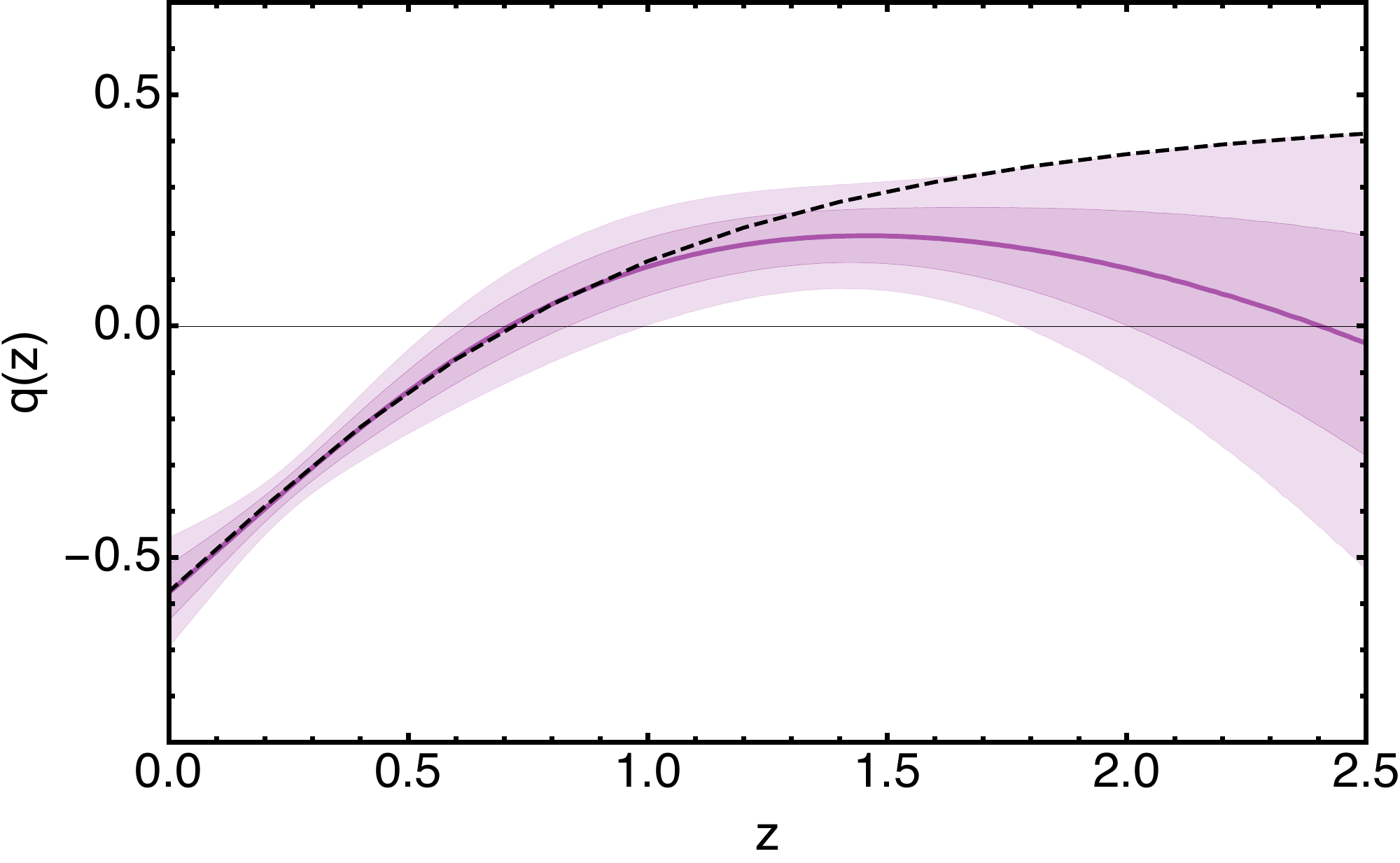}
\caption{We show the reconstructions of $q(z)$ for the datasets CC + SN + BAO (\textit{left}) and CC + SN + BAO + R18 (\textit{right}). All the reconstructions shown here are obtained using $M_{7/2}$ kernel. The dashed line in the left and right panels corresponds to the best-fit $\Lambda$CDM model of $(\Omega_m, H_0) = (0.284, 69.77)$  and $(\Omega_m, H_0) = (0.275, 71.80)$, respectively.}
    \label{fig:qofz}
\end{figure}

\begin{figure}[!ht]
    \centering
\includegraphics[width=0.45\textwidth]{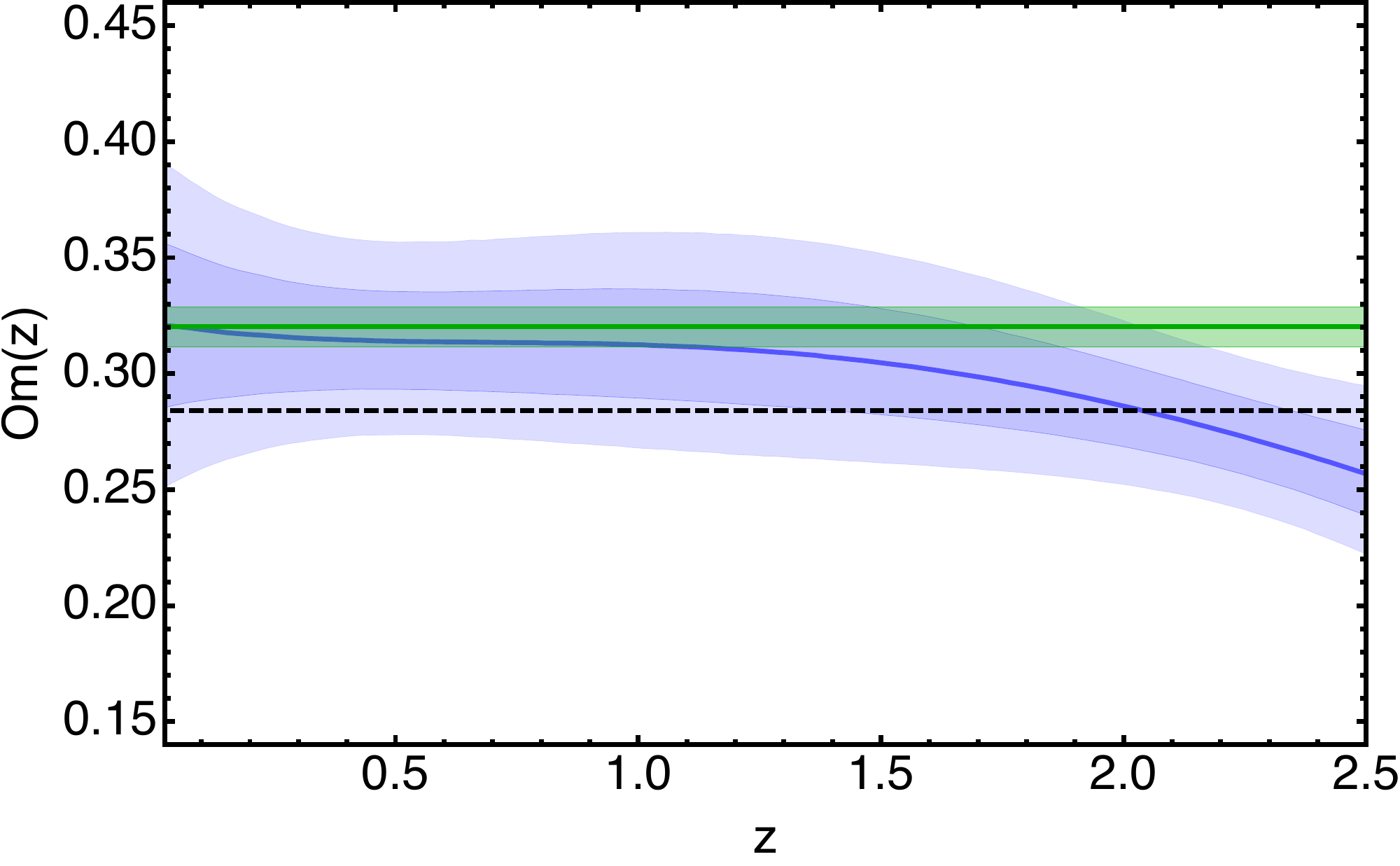}
\hspace{0.15in}
\includegraphics[width=0.45\textwidth]{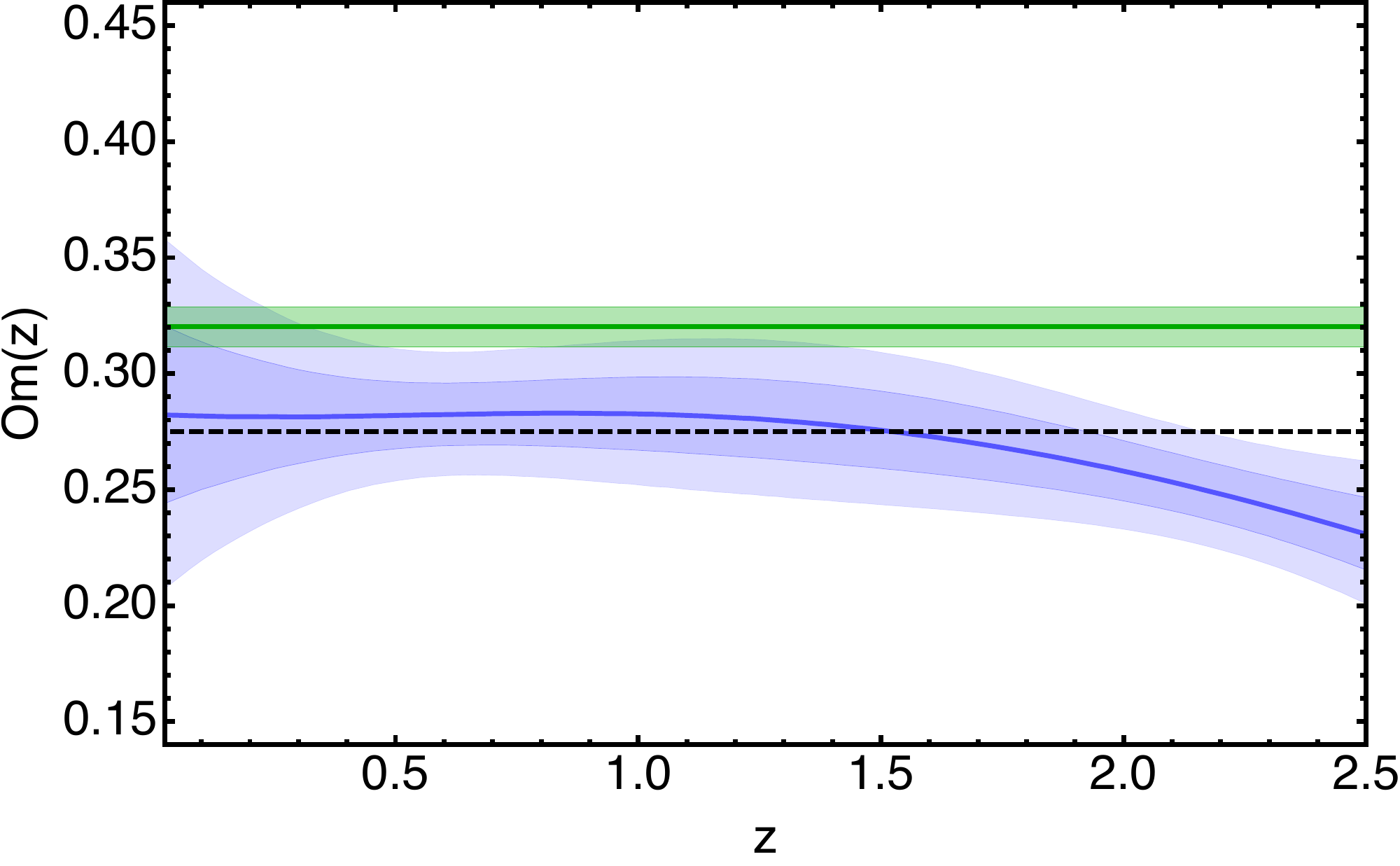}
\vfill
\includegraphics[width=0.45\textwidth]{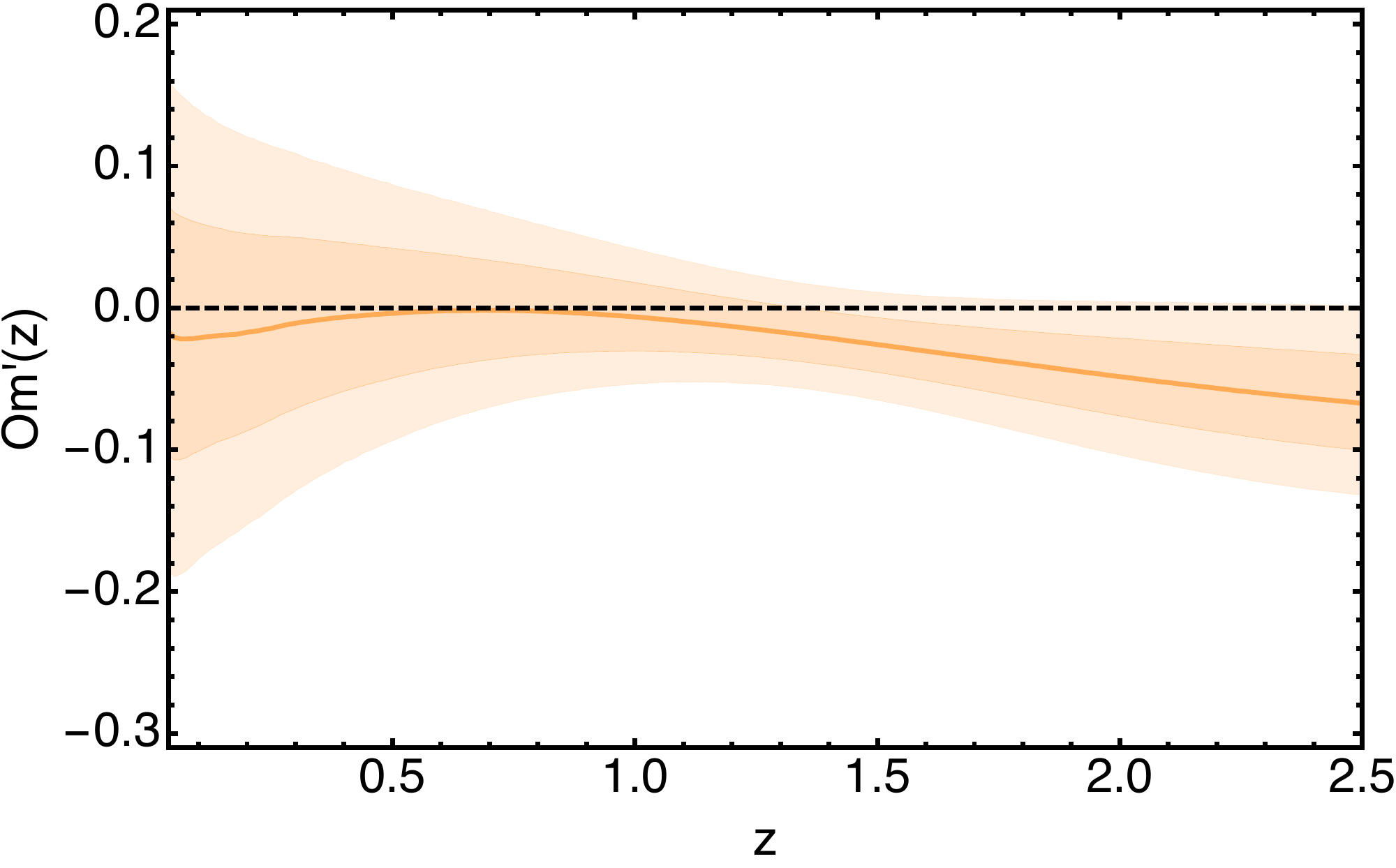}
\hspace{0.15in}
\includegraphics[width=0.45\textwidth]{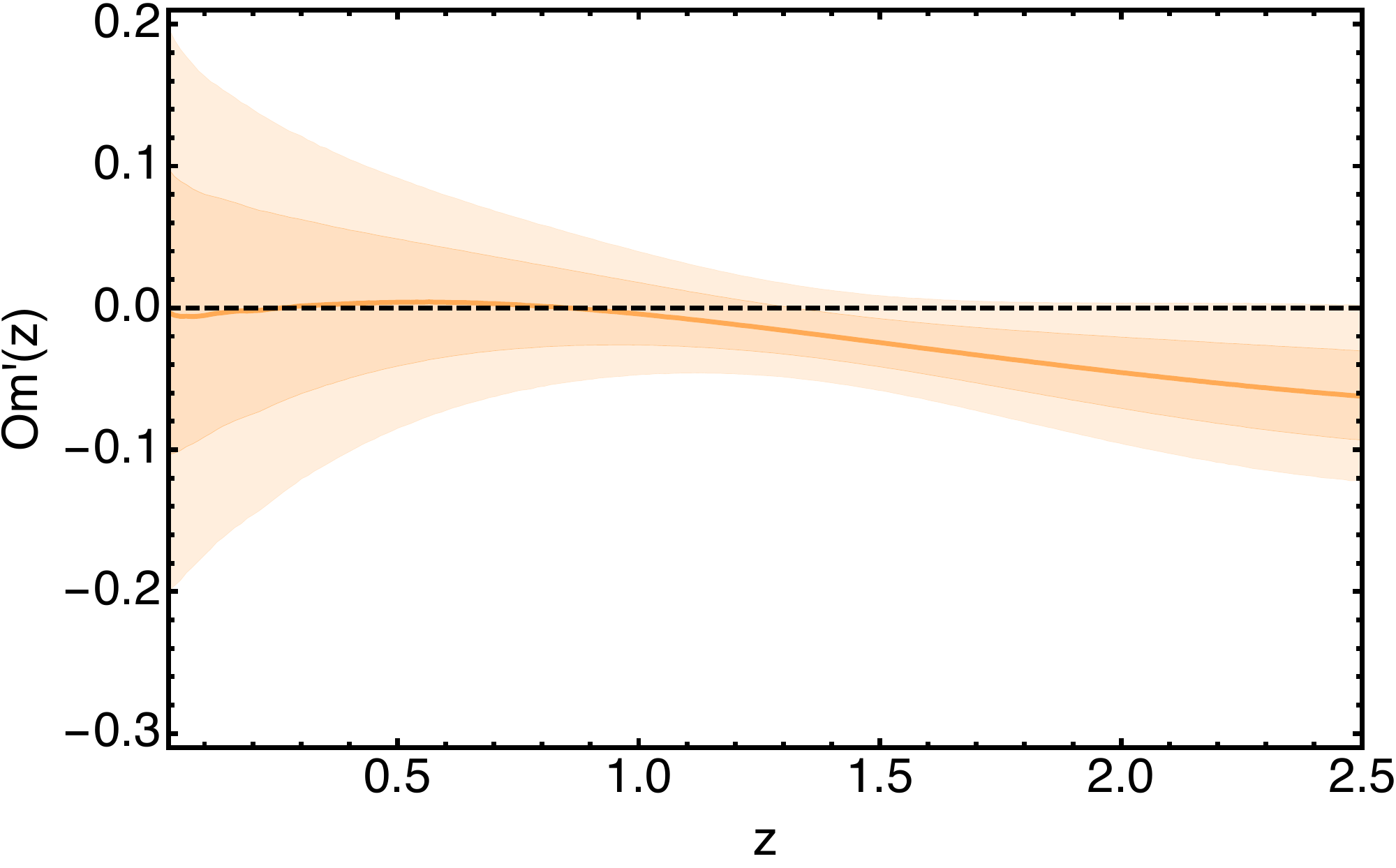}
\vfill
\includegraphics[width=0.45\textwidth]{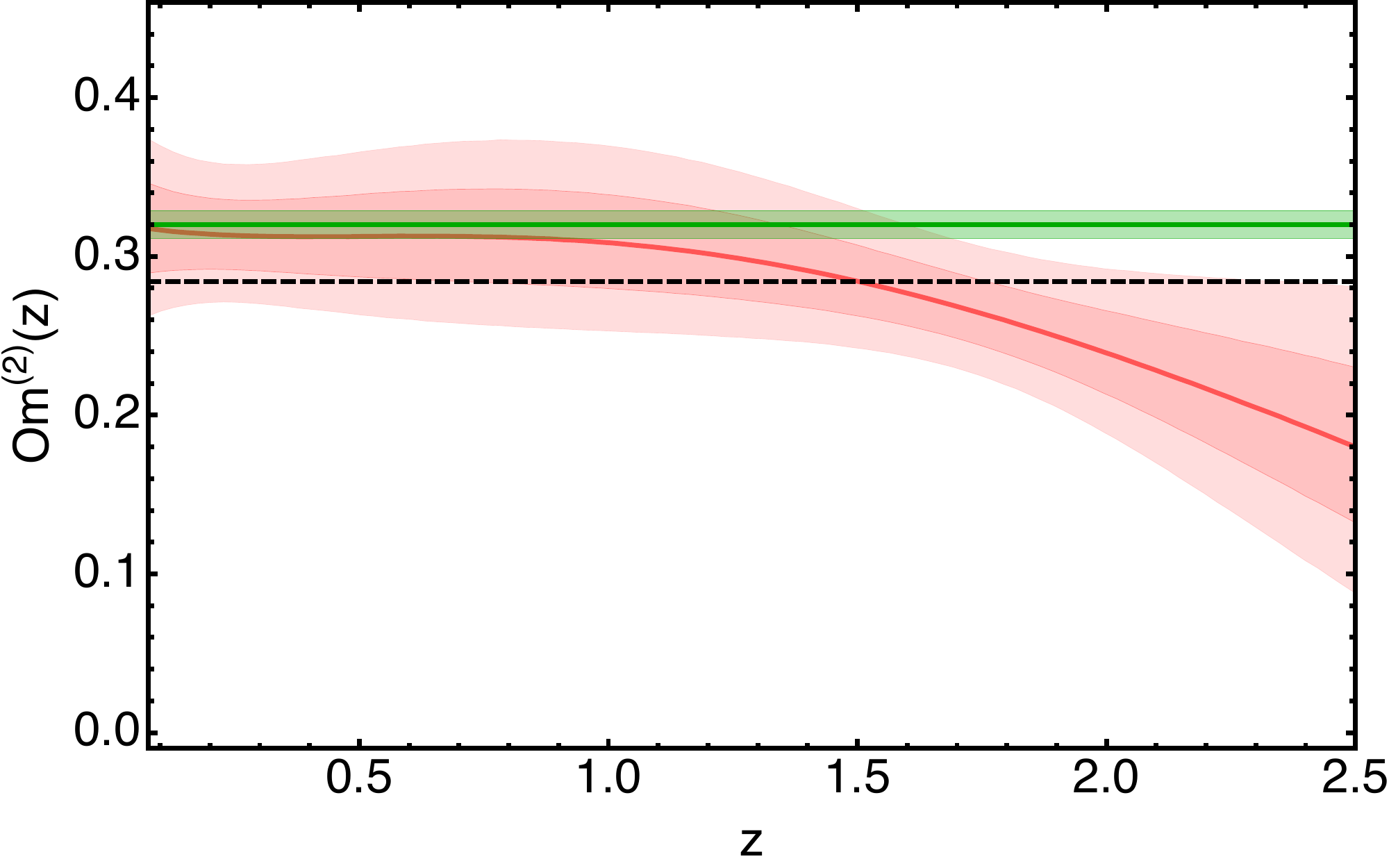}
\hspace{0.15in}
\includegraphics[width=0.45\textwidth]{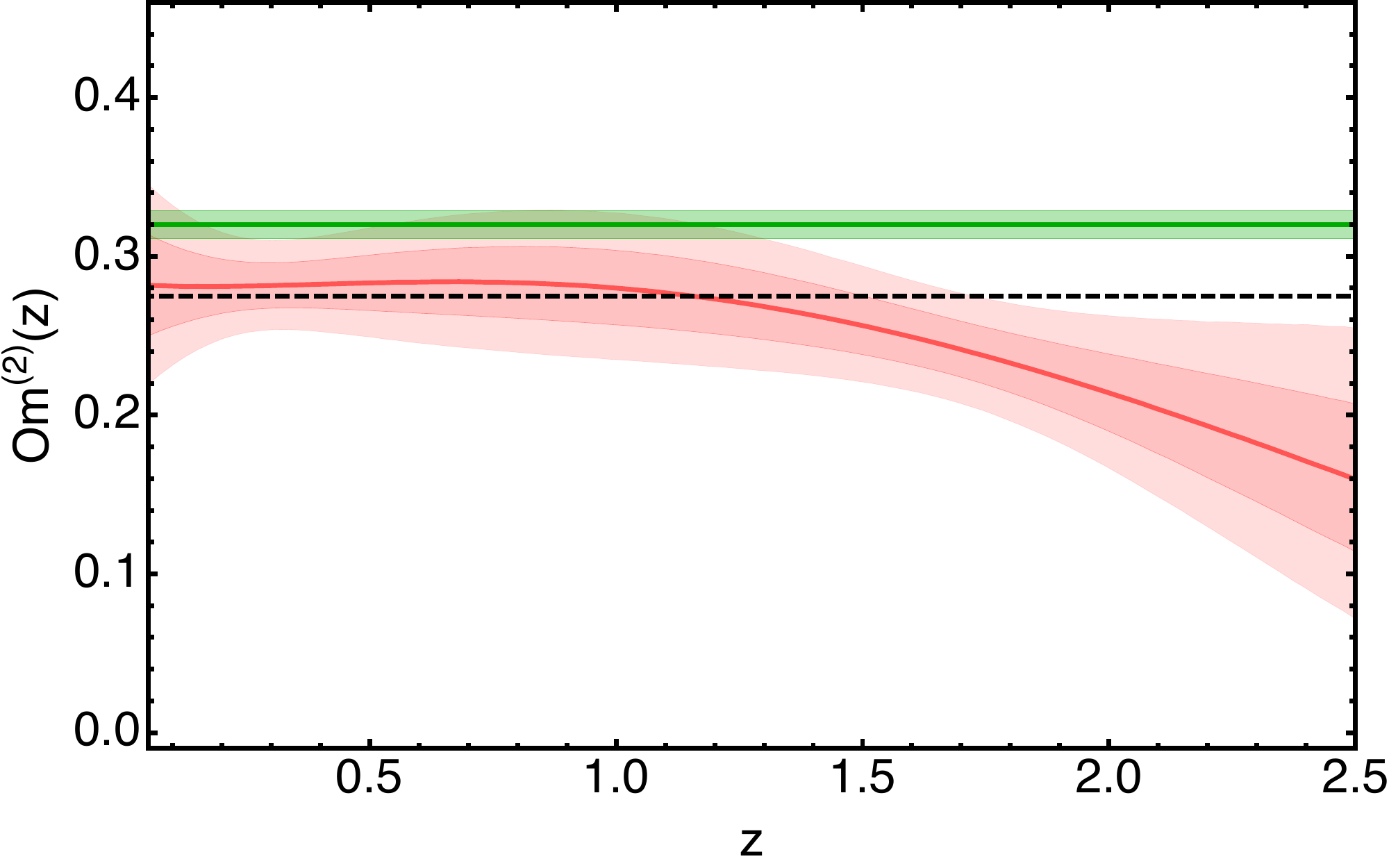}
\caption{We show the reconstructions of $\mathcal{O}m(z)$ (\textit{top}), $\mathcal{O}m'(z)$ (\textit{centre}) and  $\mathcal{O}m^{(2)}(z)$ (\textit{bottom}) for the datasets CC + SN + BAO (\textit{left}) and CC + SN + BAO + R18 (\textit{right}). All the reconstructions shown here are obtained using $M_{7/2}$ kernel. The dashed line in the top and bottom rows corresponds to the best-fit $\Lambda$CDM model of $\Omega_m = 0.284$ (\textit{left}) and $\Omega_m = 0.275$ (\textit{right}) to the current datasets. Similarly, the green region shows the constraint of $\Omega_m = 0.320 \pm 0.0087$ obtained from "high-redshift" CMB fit to $\Lambda$CDM model.}
    \label{fig:Omofz}
\end{figure}

In \Cref{fig:qofz} we show the reconstructions of $q(z)$ with (right) and without (left) inclusion of R18. We find that the $q(z)$ reconstructions are consistent with $\Lambda$CDM deceleration parameter, however only at upper boundary of $2\sigma$ reconstruction. In \Cref{fig:Omofz} we show the $\mathcal{O}m(z)$ (first row) and its derivative $\mathcal{O}m'(z)$ (second row) reconstructions, which are very consistent with a constant $\Omega_m$ and flat $\Omega_{\Lambda} = 1-\Omega_m$ scenario. However, towards redshifts $z \gtrsim 2$ the diagnostics seem to diverge from a constant $\Omega_m \geq 0.3$, while being consistent with an $\Omega_m$ of lower value. In the last row of \Cref{fig:qofz} we show the reconstructions of $\mathcal{O}m^{(2)}(z)$, which shows slightly more deviation from a constant $\Omega_m$ scenario in comparison to the $\mathcal{O}m(z)$ diagnostic. As can be seen from the $\mathcal{O}m(z)$ and $\mathcal{O}m^{(2)}(z)$ reconstructions the "high-redshift" CMB constraints on $\Lambda$CDM do not agree with these reconstructions even within $2\sigma$ at $ z \gtrsim 2$. Clearly this mild discrepancy is amplified when R18 is included in the analysis (see right column of \Cref{fig:Omofz}), although now in terms of $\Omega_m$. The inference of this discrepancy is in agreement with the very recent analysis by \cite{Shafieloo18}, and similar observations have also been made earlier in \cite{Sahni14, Zhao17}. The negative slope of $\mathcal{O}m(z)$ (see also middle row of \Cref{fig:Omofz}) at higher redshifts tentatively indicates a quintessence like behaviour \cite{Sahni08} of freezing kind as it tends to become a constant for $z\lesssim 1.5$, which is contrary to result in \cite{Zhao17} and however, requires an in-depth analysis. It is also noteworthy that the general features of $\mathcal{O}m(z)$ remain unaltered with the inclusion of R18, please see the right column of \Cref{fig:Omofz}.

Slowing down of the cosmic acceleration has been an issue of interest in several works \cite{Shafieloo09, Shafieloo10, Shahalam15, Wang16a, Zhang18, Bonilla18}. In \cite{Shafieloo09} this phenomenon was studied also utilising the $\mathcal{O}m(z)$ diagnostic, which suggests that an increase in $\mathcal{O}m(z)$ at lower redshift can indicate a slowing of acceleration. In \cite{Magana14, Magana17} it has been shown that galaxy cluster gas mass fraction samples and strong lensing datasets show such a slowing down of acceleration. In our current analysis and with the datasets implemented here, we do not find any hints for a slowing down of the cosmic acceleration with any of the covariance functions implemented. Clearly, no strong inferences can be drawn from these reconstructions yet, however we do anticipate better inferences with more stringent data to arrive with \cite{Laureijs11, Collaboration16a, Collaboration16}. In particular, obtaining a few additional CC data points in the midrange redshifts $ 0.5 \leq z \leq 1.5 $ could significantly improve the reconstructions.

Our current formalism suggests that the non-accelerating scenario is highly improbable from the estimated $q(z)$ reconstructions, which remains true for all the assumed covariance functions. Given the discussion led in \cite{Nielsen15, Shariff15, Rubin16, Ringermacher16, Haridasu17, Tutusaus17, Lonappan17a, Dam17, Lin17b} and the analysis presented here, the inference for a present accelerated phase, could in fact be claimed as beyond doubt. Also in this respect, we would like to comment on the inference for non-accelerating power-law models ($a(t) \propto t^{n}$ with $ n \leq 1$) \cite{Dolgov14, Bilicki12, Shafer15, Tutusaus16, Haridasu17, Lonappan17a} and linear-coasting-like models \cite{Mitra14, Bilicki12, John00, Dev02, Melia14a, Melia16, Melia18}, which have been of particular interest in recent works. We find that with all the model-independent flexibility (even with the most flexible $M_{3/2}$) that can be assumed in the MTGP, these models are disfavoured. However, this inference stands only from the joint analysis of all the three datasets, as the analyses of independent datasets are plagued by their own disadvantages, as discussed in next section, and no strict inferences can be made in a model-independent way. More recently, in \cite{Riess18, Tutusaus18}, similar inference for a non-accelerating power-law model was made, that the high-redshift $E(z = 1.5)$ and CMB data disfavours such models.

To summarise, taking into account the predicted equivalence of different datasets, constraints on $q_0$ and $z_T$ presented in the \Cref{tab:CCqzmain}, constraints on $H_0$ and $r_d$ in \Cref{tab:CCH0main} and, predictive quality of the kernels, we infer $M_{7/2}$ kernel as better choice amongst all kernels utilised here, to quote the final constraints on the cosmologically relevant parameters. Clearly this is not a statistical inference, but in the formalism implemented here, the current dataset(s) is(are) unable to allow for a better inference, which as we mentioned earlier shall be a issue of future investigation.

\subsection{Impact of assumptions in data and method}
\label{subsec:sys}
% \newline
\begin{figure}[h]
    \centering
\includegraphics[width=0.45\textwidth]{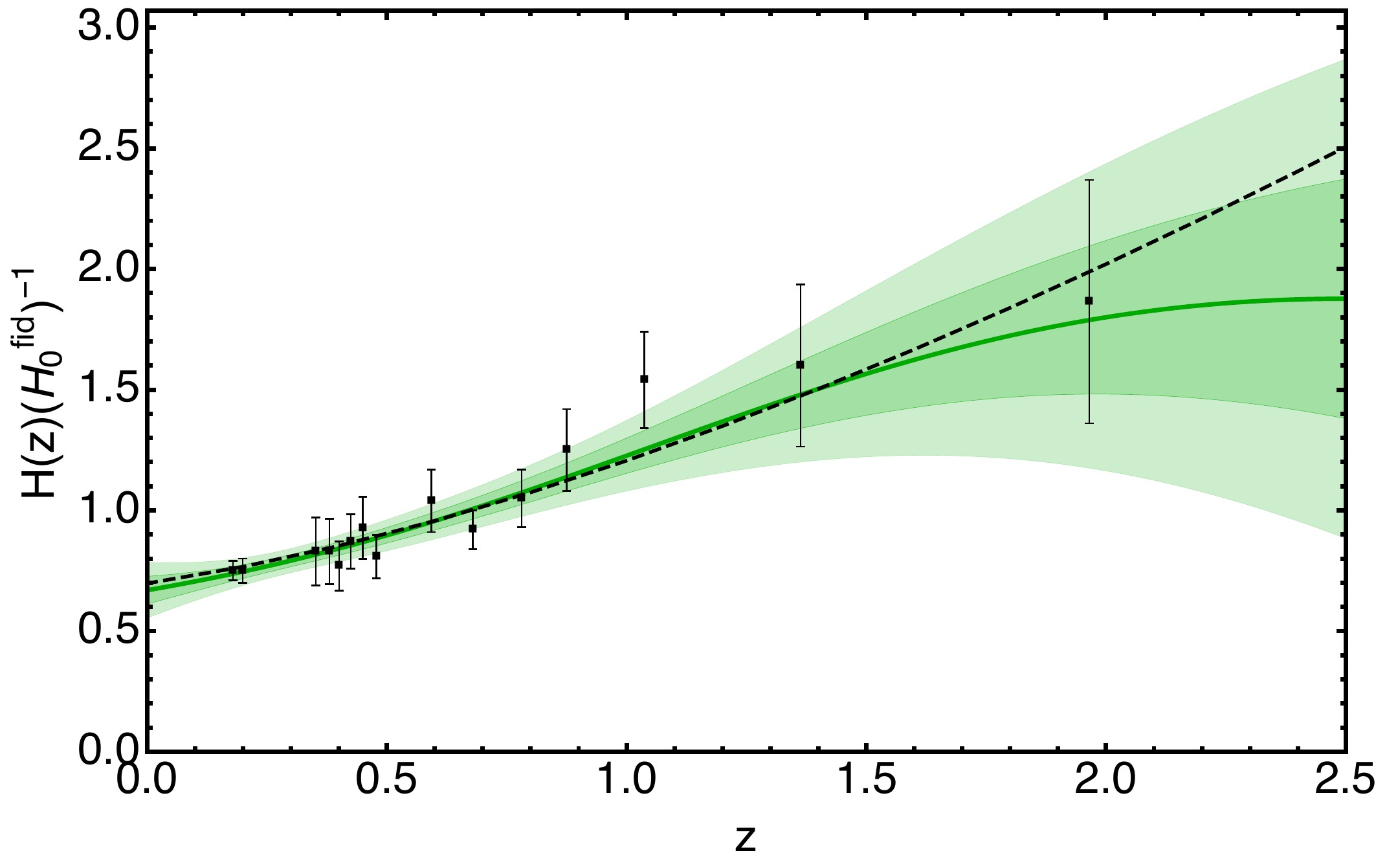}
\hspace{0.15in}
\includegraphics[width=0.45\textwidth]{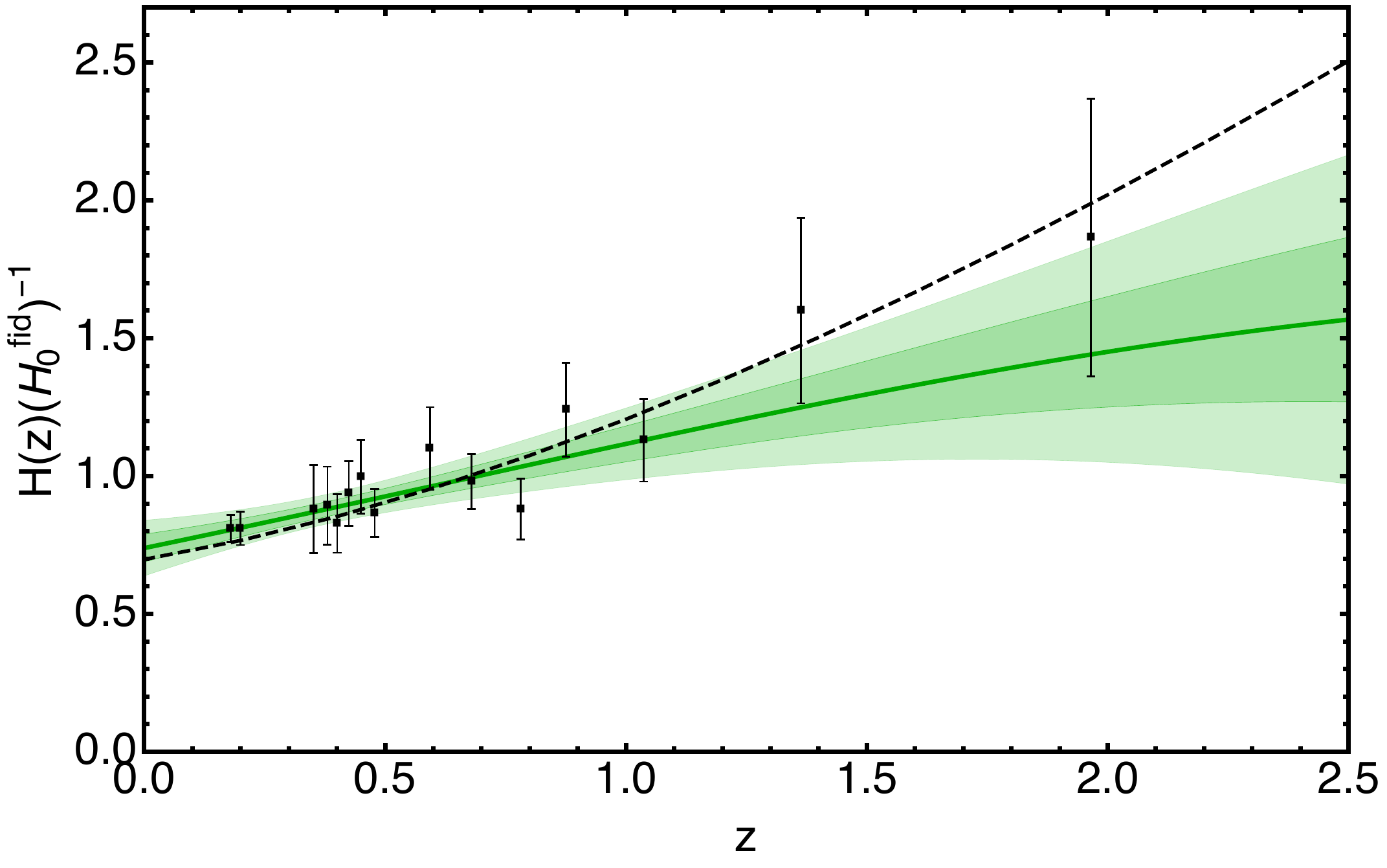}
\vfill
\includegraphics[width=0.45\textwidth]{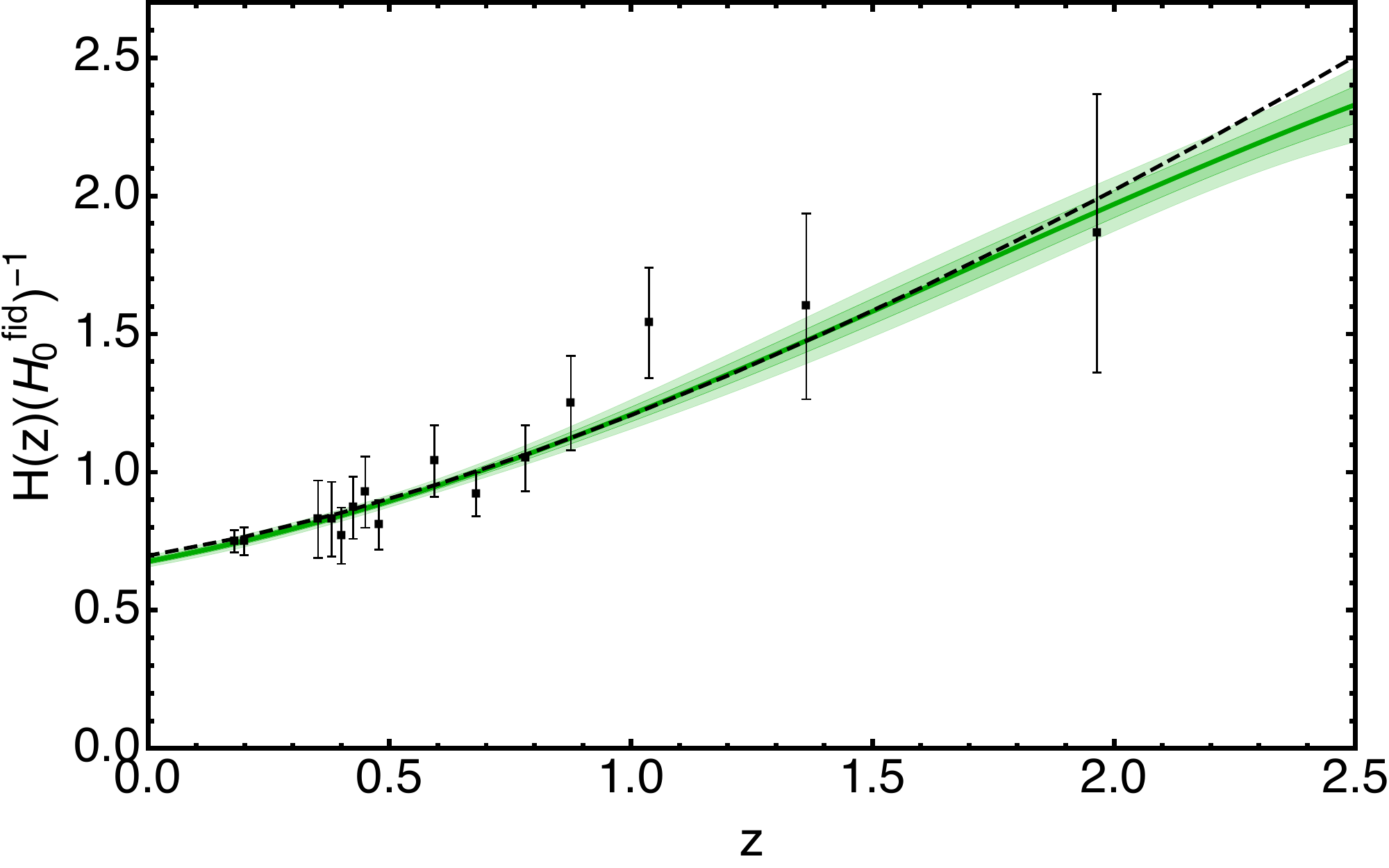}
\hspace{0.15in}
\includegraphics[width=0.45\textwidth]{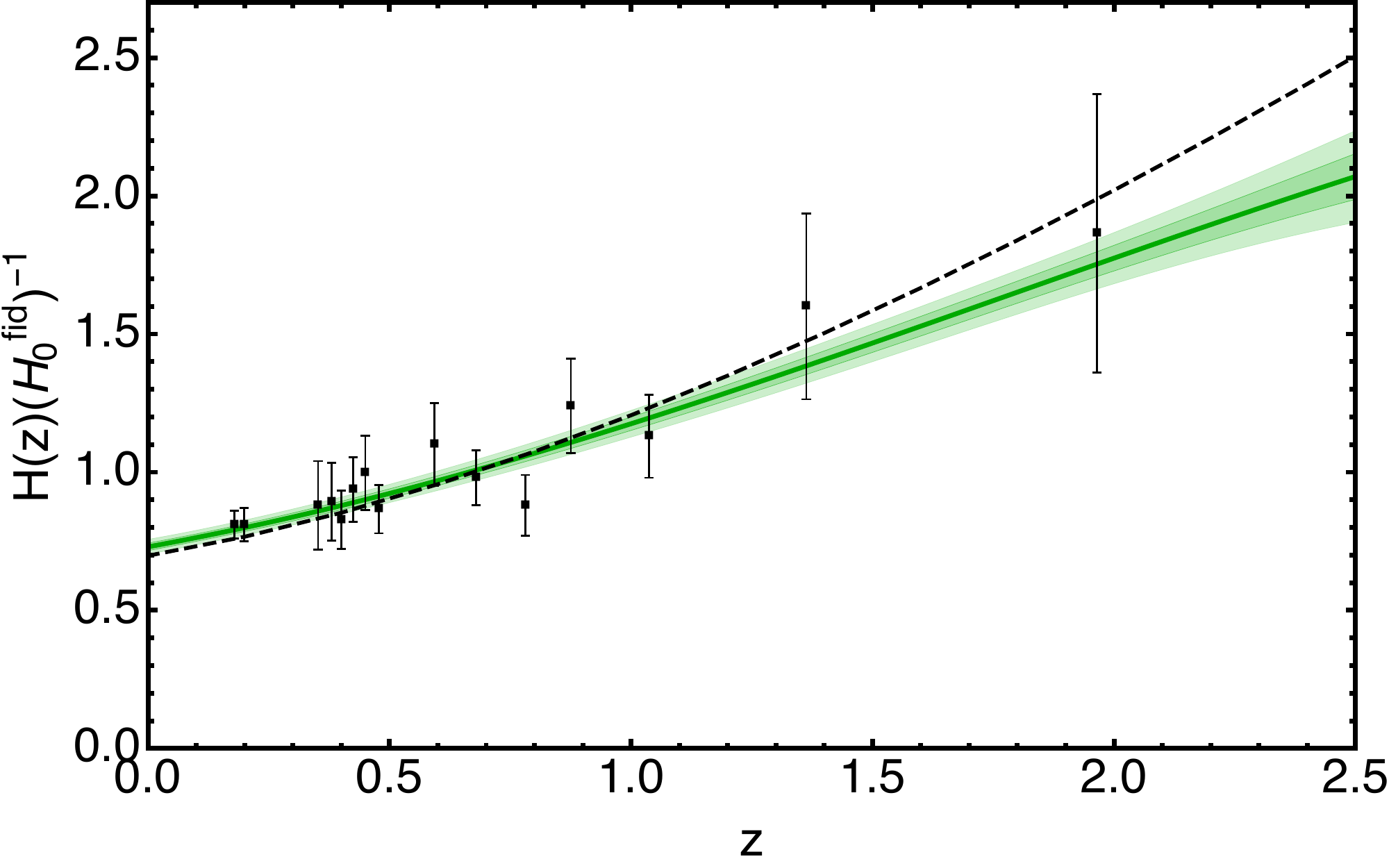}
\vfill
\includegraphics[width=0.45\textwidth]{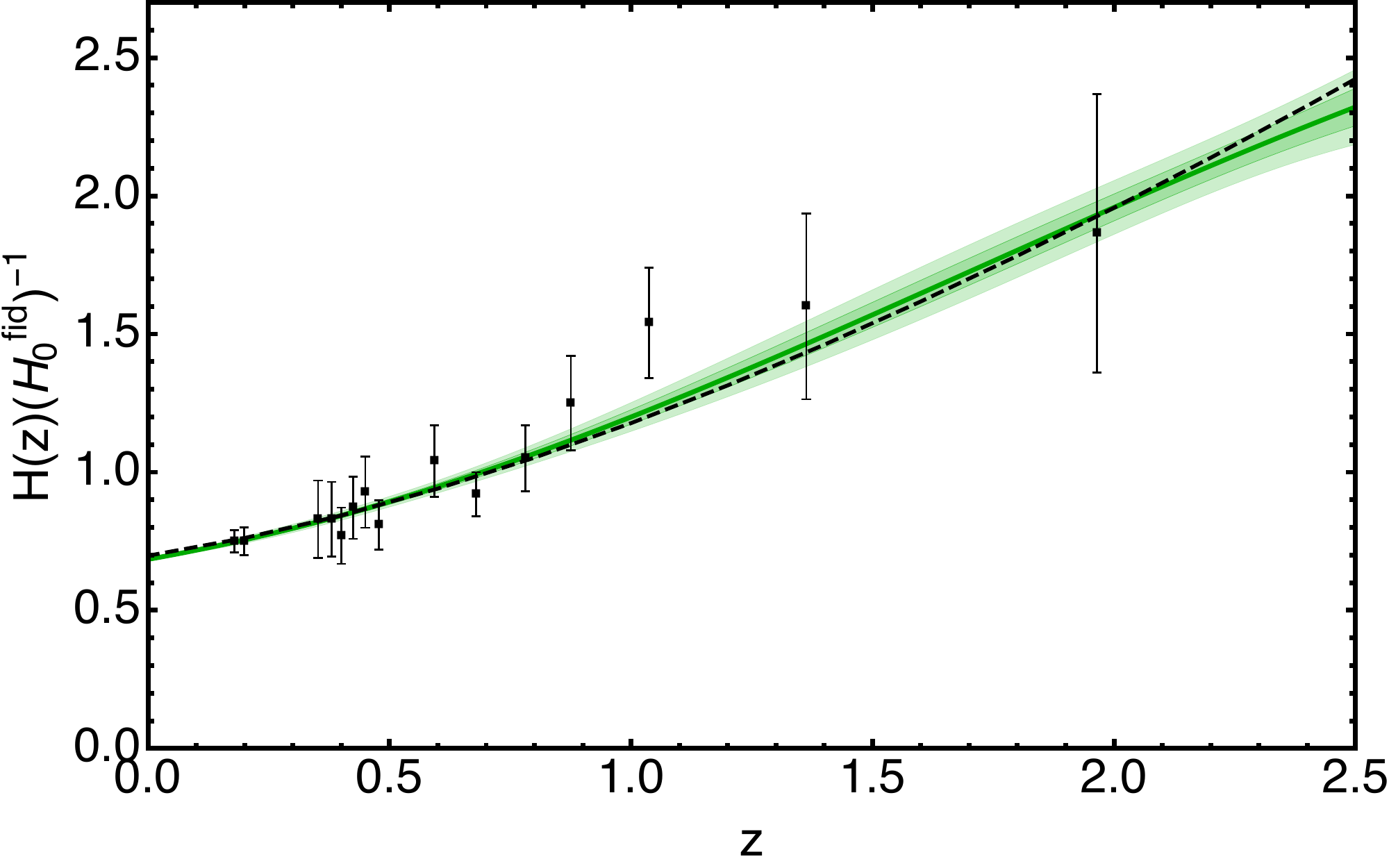}
\hspace{0.15in}
\includegraphics[width=0.45\textwidth]{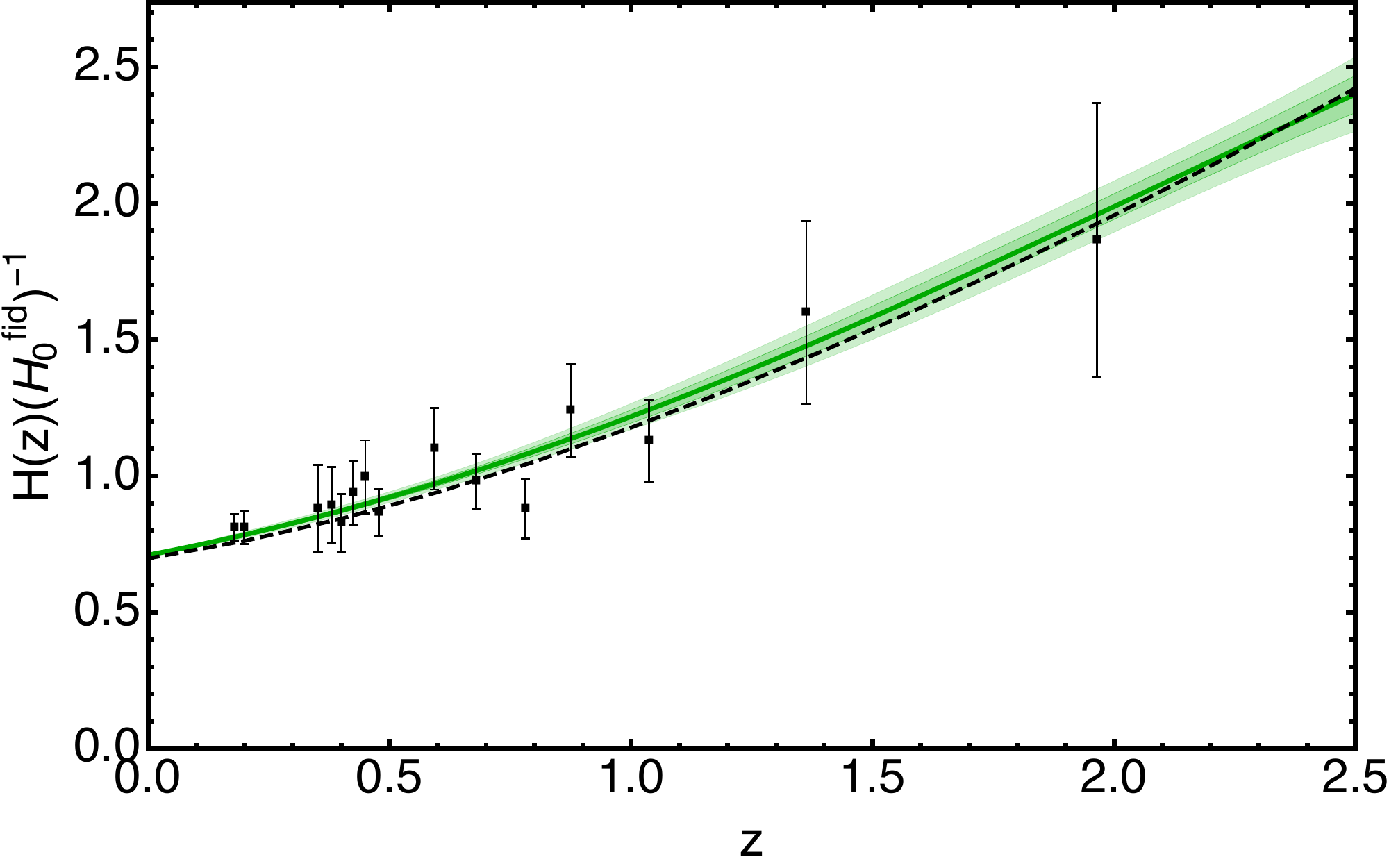}
\caption{We show the reconstructions of the $\textrm{CC}_{B}$ (\textit{left}) and the $\textrm{CC}_{M}$ (\textit{right}) data using the Single-Task GP (\textit{Top}) and MTGP (\textit{centre}) of $\textrm{CC}_{B/M}$ + SN + BAO, assuming $M_{7/2}$ kernel. In the \textit{bottom} row we show the reanalysis of \textit{centre} row, enforcing unique expansion histories across all the datasets. In all the panels data are rescaled with $H_0^{fid} = 100$ $\textrm{km/s Mpc}^{-1}$ and the dashed line corresponds to $\Lambda$CDM with $\Omega_m = 0.284$, $H_0 = 69.77$ $\textrm{km/s Mpc}^{-1}$.}
    \label{fig:STGPCCBM}
\end{figure}

{\renewcommand{\arraystretch}{1.2}%
    \begin{table}[h]
    \begin{center}
    \footnotesize
    \begin{tabular}{|c|c|c|c|}
    \hline
     Dataset(s) &  & $\textrm{CC}_{B}$ & $\textrm{CC}_{M}$   \\
    \hline
      & Kernel & \multicolumn{2}{c|}{$H_0$ [$\textrm{km/s Mpc}^{-1}$]}    \\
    \hline
    \hline
     & \begin{tabular}{@{}c@{}c@{}c@{}}$SE$ \\ $M_{9/2}$ \\ $M_{7/2}$ \\ $M_{5/2}$\end{tabular} & \begin{tabular}{@{}c@{}c@{}c@{}}$64.08 \pm 4.78$ \\ 66.19 $\pm$ 5.47 \\ 66.92 $\pm$ 5.75 \\ 67.56 $\pm$ 6.14 \end{tabular} & \begin{tabular}{@{}c@{}c@{}c@{}}$73.40 \pm 4.90$ \\ 73.68 $\pm$ 5.00 \\ 73.88 $\pm$ 5.04 \\ $74.12 \pm 5.18 $ \end{tabular} \\
    \hline
    +SN & \begin{tabular}{@{}c@{}c@{}c@{}}$SE$ \\ $M_{9/2}$ \\ $M_{7/2}$ \\ $M_{5/2}$\end{tabular} & \begin{tabular}{@{}c@{}c@{}c@{}}$67.97 \pm 1.01$ \\ 68.14 $\pm$ 0.90 \\ 68.23 $\pm$ 0.82 \\ 68.37 $\pm$ 0.68 \end{tabular} & \begin{tabular}{@{}c@{}c@{}c@{}}$71.92 \pm 0.55$ \\ 71.91 $\pm$ 0.58 \\ 71.88 $\pm$ 0.55 \\ 71.83 $\pm$ 0.49 \end{tabular} \\
    \hline
    +BAO & \begin{tabular}{@{}c@{}c@{}c@{}}$SE$ \\ $M_{9/2}$ \\ $M_{7/2}$ \\ $M_{5/2}$\end{tabular} & \begin{tabular}{@{}c@{}c@{}c@{}}$62.38 \pm 3.67$ \\ 62.89 $\pm$ 3.79 \\ 62.57 $\pm$ 3.73 \\ 62.67 $\pm$ 3.75 \end{tabular} & \begin{tabular}{@{}c@{}c@{}c@{}}$71.49 \pm 3.62$ \\ 71.37 $\pm$ 3.68 \\ 71.34 $\pm$ 3.74 \\ 71.32 $\pm$ 3.89 \end{tabular} \\
    \hline
    +SN+BAO & \begin{tabular}{@{}c@{}c@{}c@{}}$SE$ \\ $M_{9/2}$ \\ $M_{7/2}$ \\ $M_{5/2}$\end{tabular} & \begin{tabular}{@{}c@{}c@{}c@{}}$66.44 \pm 1.51$ \\ 67.40 $\pm$ 1.20 \\ 67.72 $\pm$ 1.03 \\ 68.09 $\pm$ 0.73 \end{tabular} & \begin{tabular}{@{}c@{}c@{}c@{}}$73.47 \pm 1.26$ \\ 73.08 $\pm$ 1.32 \\ 72.98 $\pm$ 1.36 \\ 72.84 $\pm$ 1.42 \end{tabular} \\
    \hline
    +SN+BAO+R18 & \begin{tabular}{@{}c@{}c@{}c@{}}$SE$ \\ $M_{9/2}$ \\ $M_{7/2}$ \\ $M_{5/2}$\end{tabular} & \begin{tabular}{@{}c@{}c@{}c@{}}$71.82 \pm 0.63$ \\ $71.72 \pm 0.56$ \\ $71.68 \pm 0.53 $ \\ $71.63 \pm 0.45$ \end{tabular} & \begin{tabular}{@{}c@{}c@{}c@{}}$73.27 \pm 0.73$ \\ $72.87 \pm 0.73$ \\ $73.20 \pm 0.78 $ \\ $72.64 \pm 0.54 $ \end{tabular} \\
    \hline
    \end{tabular}
    \caption{Estimates of $H_0$ and the corresponding $1\sigma$ uncertainty form combinations of the dataset(s) using either the $\textrm{CC}_B$ or $\textrm{CC}_M$, implemented using all the kernels implemented in this work. For the joint analysis with the multiple datasets we implement the same kernel for each of them.}
    \label{tab:CCH0Sys}
    \end{center}
\end{table}
}

As discussed in the preceding sections, one of the key ingredients in the cosmic chronometers measurements to estimate the Hubble parameter is the stellar population synthesis (SPS) model used to calibrate the method. We recall here that a study of the impact of systematic effects due to the assumption of different SPS models has been already provided in \cite{Moresco12a, Moresco16a}. In \cite{Moresco12a}, the constraints on the dimensionless energy density parameters were explored in an open $w_0w_a$CDM model (where the equation of state of dark energy is let free to vary as a function of cosmic time according to the parametrisation given by \cite{Chevallier01, Linder03}), finding no statistical difference when assuming different SPS models. In \cite{ Moresco16a} studying the flat $w_0w_a$CDM model a shift of $\sim$ 5 $\textrm{km/s Mpc}^{-1}$ is found in $H_0$, with M11 models preferring slightly higher values, but with a difference which is compatible with zero, given the estimated errors. This difference, as shown in \cite{Moresco12a, Moresco16a} and also discussed more recently in \cite{Gomez-Valent18}, is due to the fact that expansion rate measurements from M11 models are slightly larger, thus impacting the estimated normalisation of the H(z) relation, i.e. $H_0$.

In parallel to this effect, when different combination of data are explored, they would probe different regions of the parameters space, yielding varied results. Here we present the analysis to evaluate the dependence of the results obtained using both different combinations of data, considering the $\textrm{CC}_{B}$ and $\textrm{CC}_{M}$ compilations provided in \cite{Moresco12a, Moresco16a} separately, to estimate the systematics in our approach both due to data and our MTGP method. For the comparison to be homogeneous, we consider in both cases only the datasets drawn from just these analyses, neglecting other measurements that have been obtained using only BC03 models. The results are shown in \Cref{fig:STGPCCBM}, showing the ST reconstruction of the data in top row with the corresponding $H_0$ values reported in \Cref{tab:CCH0Sys}. We find that $\textrm{CC}_M$ data predicts a higher value of $H_0$, consistent with the estimate from the local model-independent R18 by \cite{Riess18}, while $\textrm{CC}_B$ dataset, as discussed in \Cref{subsec:MainResults}, prefers a value compatible with the estimate from CMB \cite{Collaboration16b}. However, the estimated error bars are large and we confirm that considering the two CC datasets ($\textrm{CC}_M$ and $\textrm{CC}_B$) alone the measurement from the two SPS model are compatible within errors. This result confirms the findings of \cite{Gomez-Valent18}, who also found, with a different approach, a statistically insignificant difference when considering $\textrm{CC}_{M/B}$ datasets on their own.

The large uncertainty on $H_0$ is lowered when a joint analysis is performed, and in this case the differences are more significant. The increase in significance of the difference between the two measurements is mostly due, as demonstrated, by the inclusion of the external datasets of SN and BAO. This analysis indicates a tentative systematic error of $\sim 5.7$ $\textrm{km/s Mpc}^{-1}$, on average. This value is estimated as the average of difference between the mean $H_0$ values in third and fourth column for $\textrm{CC}_{B/M}$ + SN + BAO in \Cref{tab:CCH0Sys}, and also includes the variations due to kernel assumptions. However, unlike $\textrm{CC}_B$ + SN + BAO, we find that the $\textrm{CC}_M$ + SN + BAO combination is unable to provide equivalent expansion histories in all three data planes, even with the utmost flexibility available in the formalism (e.g. see middle row of \Cref{fig:STGPCCBM}). This in turn devolves the purpose/utility of MTGP, to allow the flexibility/freedom and predict equivalence of data. To explore the reasons for the same we produce a spectrum of $H_0$'s coming from several data combinations as are shown in \Cref{tab:CCH0Sys}. This also helps us illustrate the subtleties of MTGP formalism listed in \Cref{sec:Method}.

\begin{figure}[h]
    \centering
\includegraphics[width=0.45\textwidth]{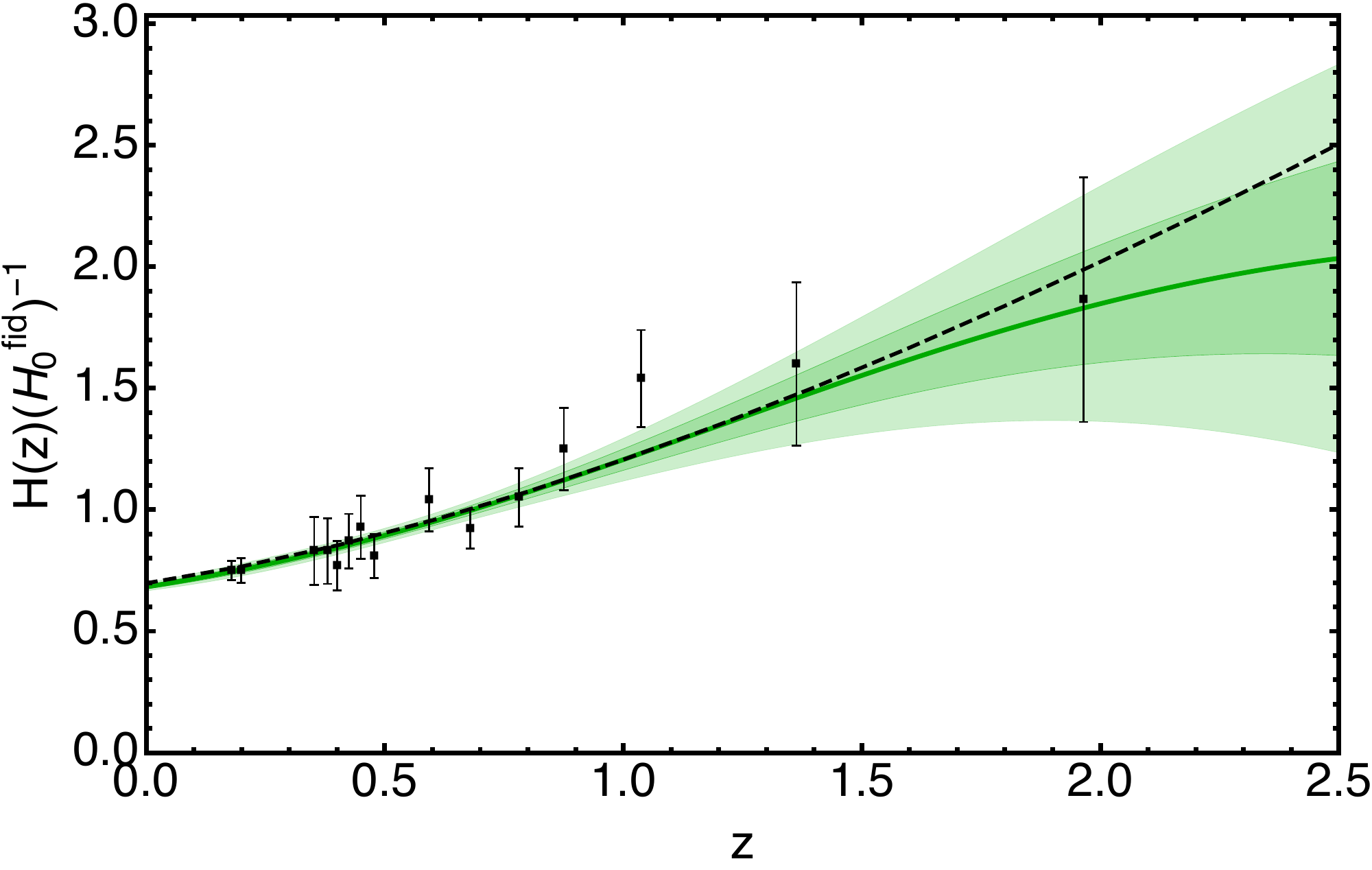}
\hspace{0.15in}
\includegraphics[width=0.45\textwidth]{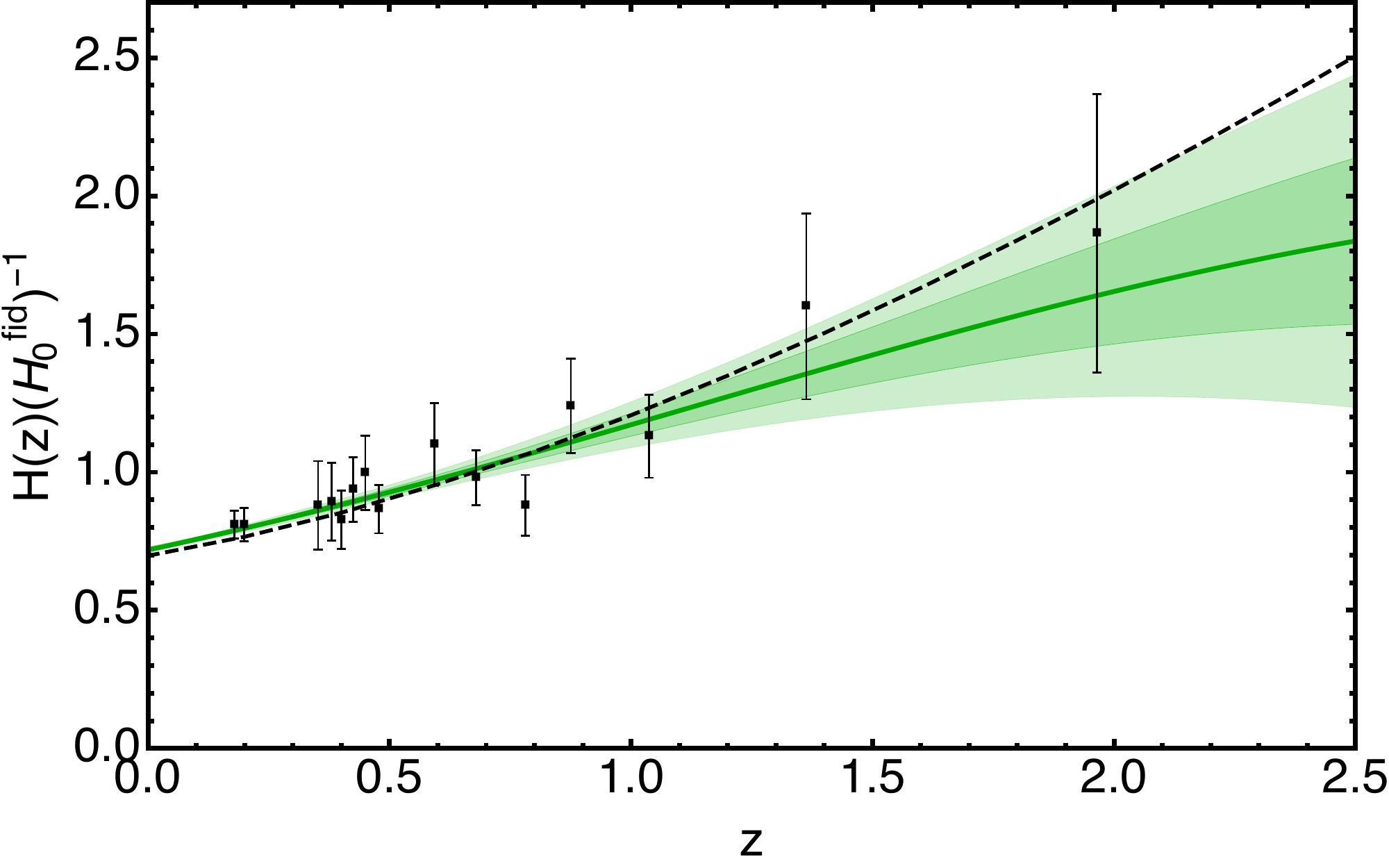}
\vfill
\includegraphics[width=0.45\textwidth]{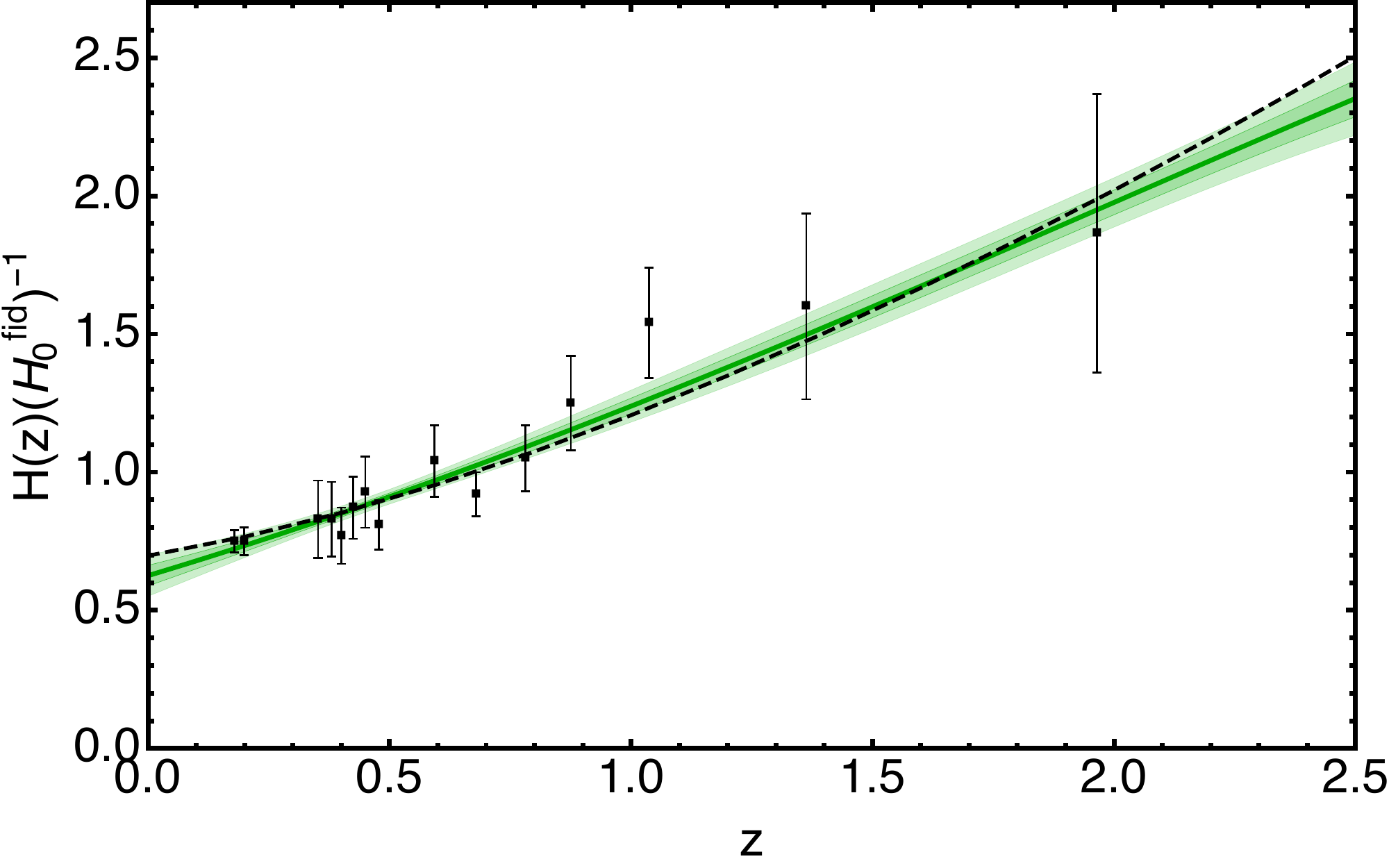}
\hspace{0.15in}
\includegraphics[width=0.45\textwidth]{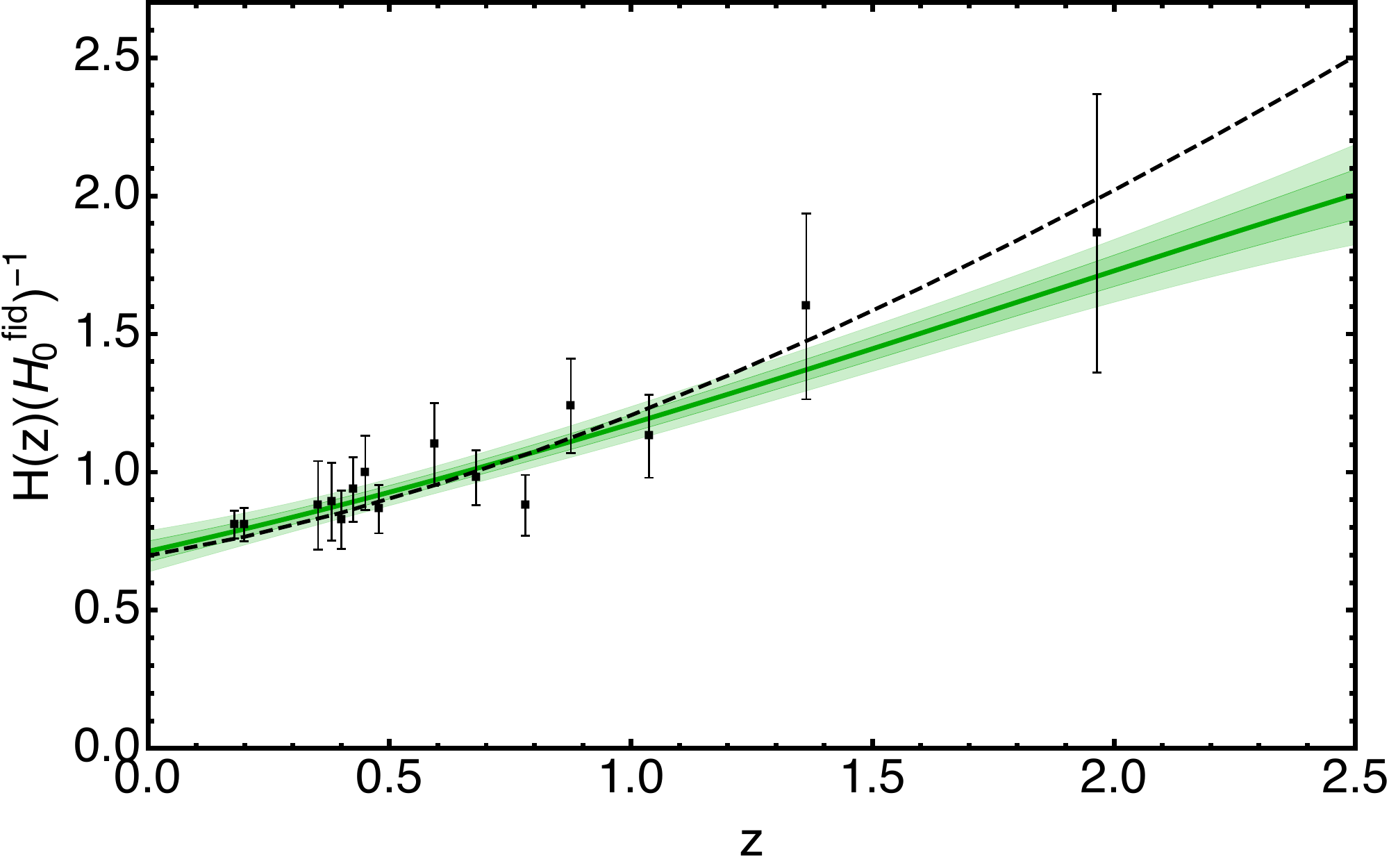}
\caption{We show the reconstructions of the $\textrm{CC}_{B}$(\textit{left})and the $\textrm{CC}_{M}$(\textit{right}) data in combination with SN (\textit{top}) and BAO (\textit{bottom}) using $M_{7/2}$ kernel. In all the panels data are rescaled with $H_0^{fid} = 100$ $\textrm{km/s Mpc}^{-1}$ and the dashed line corresponds to $\Lambda$CDM with $\Omega_m = 0.284$, $H_0 = 69.77$ $\textrm{km/s Mpc}^{-1}$.}
    \label{fig:MTGPCCBM}
\end{figure}

As can be seen in the second row of \Cref{tab:CCH0Sys}, the estimates of uncertainties using CC and SN data alone are much tighter than the uncertainties quoted in \Cref{tab:CCH0main}, with all three datasets. A more comprehensive way to describe these reconstructions is the over-fitting around $z\sim 0$. This over-fitting in fact is much more pronounced in the $\textrm{CC}_M$ than in the $\textrm{CC}_B$, due to the difference in the two data points at $z= 0.781$ and $z= 1.037$. While all the data points in $\textrm{CC}_M$ are higher by $\sim 6$ $\textrm{km/s Mpc}^{-1}$ w.r.t $\textrm{CC}_B$ data points, at these two redshifts the $\textrm{CC}_M$ measurements are lower than the $\textrm{CC}_B$ and this in turn allows the $\textrm{CC}_M$ data to fit more in concert with the SN data, however, giving rise to over-fitting. A clear indication of this is the constant-like behaviour of the $H_0$ estimate at intercept, which we have further confirmed by eliminating the first two data points of the $\textrm{CC}_M$ dataset to find an ever more over-fitted uncertainty of $0.2$. While $\textrm{CC}_B$ also tends to be over-powered by the SN dataset the effect is not as pronounced as is with $\textrm{CC}_M$. In the top row of \Cref{fig:MTGPCCBM}, we show the reconstructions using the combinations of $\textrm{CC}_M$, $\textrm{CC}_B$ with SN data. It is illustrative to see the effect of SN data, which clearly improves the CC reconstruction around $z \sim 0$ than at $z \gtrsim 1.5$. Finally we infer that this effect also takes its toll in the analysis with all three datasets, as this remains to be the clear/only difference why the data combinations with $\textrm{CC}_M$ are unable to predict equivalent expansion histories.

Therefore to estimate the systematics, we aid to the aforementioned provision within MTGP formalism which enforces the equivalence of latent functions by a priori assuming that the length-scale hyperparameters of all three datasets are the same. The SN and BAO data complement $\textrm{CC}_{M/B}$ data and help predicting equivalent expansion histories, hence ensuring that the difference in the intercept is only due to modification from $\textrm{CC}_{B}$ to $\textrm{CC}_{M}$ compilations, while all the other features are primarily driven by SN + BAO datasets (see bottom row of \Cref{fig:STGPCCBM}). As is already mentioned in the beginning of the results section, this provides a constant like rescaling factor(s) amongst datasets. Having established that these values cannot be inferred as physical constants, we are only interested in the difference in the mean estimates from combinations of $\textrm{CC}_M$ to $\textrm{CC}_{B}$. We then infer this difference as the systematic effect arising from the $\textrm{CC}_{B}$ to $\textrm{CC}_{M}$ comparison. Using the $M_{7/2}$ kernel we estimate $\sigma_{H_0} (sys) = 2.51$ $\textrm{km/s Mpc}^{-1}$ as the difference in these constant like intercepts which are $70.97$ and $68.46$ for $\textrm{CC}_{M}$ + SN + BAO and $\textrm{CC}_{B}$ + SN + BAO, respectively. An added advantage in this approach is that the difference remains unaltered (with a maximum variation of $\sim 0.01$) for all assumed kernels, also completely mitigating the variations arising due to different kernel choices. While on the face value the estimates of $H_0$ from $\textrm{CC}_M$ seem discrepant, through more scrutiny we find consistent results with \cite{Yu17, Gomez-Valent18}. However with our slightly larger systematic error, we do not find R18 to be in tension. We refer to a following analysis for a more detailed discussion on the systematic impact on $H(z)$ data due to SPS model assumption in cosmic chronometers. As is known, $r_d$ is estimated as a rescaling factor between BAO and CC and so the systematic in the $H_0$ estimate also propagates to $r_d$. We replicate the same formalism to find a systematic error of $\sigma_{r_d} (sys) = 4.3$ $\textrm{Mpc}$ and $\sigma_{r_d H_0} (sys) = 70 $ km/s, which are also independent of kernel choice. Note that the propagation of the systematics to the $r_d H_0$ estimate are minimal in comparison to the statistical error $\sigma_{r_d H_0} (stat) = 237 $ km/s quoted in \Cref{subsec:MainResults}.
% \newline

In fact when a model-dependent analysis is performed assuming the $\Lambda$CDM model, the discrepancy between the $H_0$ estimates using the $\textrm{CC}_M$ and $\textrm{CC}_B$ compilations separately, reduces in scatter. We find ${H_0} = 68.99 \pm 2.04$ $\textrm{km/s Mpc}^{-1}$ and ${H_0} = 71.60 \pm 2.30$ $\textrm{km/s Mpc}^{-1}$ for $\textrm{CC}_{B}$ + SN + BAO and $\textrm{CC}_{M}$ + SN + BAO, respectively. These values are clearly in no discrepancy among themselves and also very well agree with our best estimate of $H_0 = 68.52^{+0.94 + 2.51 (sys)}_{-0.94} $ $\textrm{km/s Mpc}^{-1}$. We later extend the analysis to estimate the systematic uncertainty even when R18 is included. We find the kernel-independent uncertainties arising from the systematic effect to be $\sigma_{H_0} (sys) = 1.65$ $\textrm{km/s Mpc}^{-1}$ and $\sigma_{r_d} (sys) = 2.63 $ Mpc. We quote the constraints for this combination of data with the additional systematic error for the corresponding $M_{7/2}$ kernel estimates reported in \Cref{tab:CCH0main}.

To illustrate a few more subtleties, in the bottom row of \Cref{fig:MTGPCCBM} we show the features of BAO data being transferred to both the CC datasets. Clearly the $H_0$ estimates are lowered due to the linear nature of BAO reconstruction (see third row of \Cref{tab:CCH0Sys}). Notice the agreement in the mean of $H_0$ estimates from $\textrm{CC}_M$ + SN and $\textrm{CC}_M$ + BAO datasets, while the uncertainty is over-fitted in the prior case. The negative transfer form the BAO data is more pronounced in the $\textrm{CC}_B$ + BAO combination, where the $H_0$ estimate is further lowered to $\sim 62$ $\textrm{km/s Mpc}^{-1}$ in comparison to CC + BAO ($\sim 64$ $\textrm{km/s Mpc}^{-1}$). Given the strength of all the datasets considered in this work, an analysis on the combination of all the three datasets clearly remains necessary for not so trivial reasons, when compared to 2 dataset combinations. {One could even suspect that in an analyses utilising only two datasets \cite{Yu17, Gomez-Valent18}, this could take effect as either of the combinations CC+SN or CC+BAO were implemented, although a straightforward comparison is not available.}

\section{Conclusions and Future prospects}
\label{sec:Conclusions}

In the present work, we have implemented an extension (MTGP) to the standard ``model-independent'' GP formalism utilised so far in the cosmological context. The key improvement in this work is to formulate a methodology to combine different datasets to perform joint analysis in a more cosmology-independent way. To this aid, we have utilised the ``low-redshift'' expansion rate data from SN, BAO and CC observations. Implementing several covariance functions in the MTGP formalism we find an excellent concert amongst them. This, in turn, helps our discussion to override the specificity of choosing a particular covariance function. We have performed several tests, to assess the better kernel assumptions for the cosmological data implemented here, but conclude that it requires a more in-depth investigation and in fact not prospective with the current dataset(s). However, this formalism with its inherent flexibility readily provides us with a heuristic argument to perform a kernel selection, which allows us to infer $M_{7/2}$ as the better kernel choice with the data utilised in this work.

Our model-independent estimates of $H_0 = 68.52^{+0.94 + 2.51 (sys)}_{-0.94} $ $\textrm{km/s Mpc}^{-1}$ is in a good agreement with standard scenario with corresponding $r_d = 145.61^{+2.82}_{ - 2.82 - 4.3 (sys)} $ Mpc. While our statistical uncertainties using CC + SN + BAO are highly competitive, even amongst the existing model-independent analysis, we also included a systematic error specifically utilising the provision in our method and the differences in $\textrm{CC}_{B/M}$ compilations. The statistical estimate obtained here is in agreement with the CMB P16 and local R18 at $1.41\sigma$ and $2.60\sigma$, respectively. Very similar inference was also made with a model-dependent analysis in \cite{Haridasu17a}. When we include the R18 measurement in our analysis we find $H_0 = 71.40^{ + 0.30 + 1.65 (sys)}_{- 0.30 } $ $\textrm{km/s Mpc}^{-1}$ and $r_d = 141.29^{ + 1.31 }_{-1.31-2.63 (sys)}$ $\textrm{Mpc}$. Amongst several advantages of the current formalism, estimation of the systematic error remains to be one. While our statistical uncertainties remain in excellent concert with the recent analyses in \cite{Yu17,Gomez-Valent18}, the systematic uncertainty incurred from our analysis does not show local $H_0$ estimate to be in significant tension at $\sim 1.61 \sigma$. Our constraint on $r_d H_0 = 9977 \pm 237 (stat)$ km/s is extremely consistent with "high-redshift" CMB and the model-dependent $\Lambda$CDM constraints for the datasets implemented here. The effect of systematics in CC data on this value ($\sigma_{r_d H_0} (sys) = 70$ km/s) is minimal compared to the statistical error.

We have also extended the analysis to reconstruct the derivative regions of the data, through which we find a good agreement with the $\Lambda$CDM model up to redshifts $ z \lesssim 2.0 $. We estimate the deceleration parameter $q(z)$, inferring constraints of $q_0 = -0.52 \pm 0.06 $ and $z_T = 0.64_{-0.09}^{+0.12}$, consistent amongst all the kernels implemented here. Utilising the $\mathcal{O}m(z)$ diagnostic we find good agreement with $\Lambda$CDM within $z \lesssim 2.0 $ and a mild discrepancy for $z \gtrsim 2.0$. The inclusion of R18 into the analysis, as expected, magnifies this discrepancy. We find these inferences to be very much in agreement with earlier results in \cite{Shafieloo18, Sahni14, Zhao17}. Work is in progress to extend the current discussion by utilising also the distances and growth rate data, which shall also be implemented to investigate the dark energy EoS in the current model-independent approach. The ability to perform a joint analysis as presented here shall prove extremely useful with more stringent data to arrive \cite{Collaboration16, Collaboration16a, Laureijs11}.

\acknowledgments

B.S.H, V.V.L and N.V acknowledge financial support by ASI Grant No. 2016-24-H.0. M.M acknowledges the grants ASI n.I/023/12/0 ``Attivit\`a relative alla fase B2/C per la missione Euclid'' and PRIN MIUR 2015 ``Cosmology and Fundamental Physics: illuminating the Dark Universe with Euclid''. B.S.H thanks Federico Tosone for useful discussions and comments.

\newpage

  \bibliographystyle{JHEP}
  \bibliography{MIndependent}

\providecommand{\href}[2]{#2}\begingroup\raggedright\begin{thebibliography}{100}

\bibitem{Collaboration16b}
{Planck Collaboration}, N.~Aghanim, M.~Ashdown, J.~Aumont, C.~Baccigalupi,
  M.~Ballardini et~al., \emph{Planck intermediate results. xlvi. reduction of
  large-scale systematic effects in hfi polarization maps and estimation of the
  reionization optical depth},
  \href{https://doi.org/10.1051/0004-6361/201628890}{\emph{\aap} {\bfseries
  596} (Dec., 2016) A107}, [\href{https://arxiv.org/abs/1605.02985}{{\ttfamily
  1605.02985}}].

\bibitem{Alam16}
S.~Alam, M.~Ata, S.~Bailey, F.~Beutler, D.~Bizyaev, J.~A. Blazek et~al.,
  \emph{The clustering of galaxies in the completed sdss-iii baryon oscillation
  spectroscopic survey: cosmological analysis of the dr12 galaxy sample},
  {\emph{arXiv preprint arXiv:1607.03155} (2016) }.

\bibitem{Joudaki17a}
S.~Joudaki, A.~Mead, C.~Blake, A.~Choi, J.~de~Jong, T.~Erben et~al.,
  \emph{Kids-450: testing extensions to the standard cosmological model},
  \href{https://doi.org/10.1093/mnras/stx998}{\emph{\mnras} {\bfseries 471}
  (Oct., 2017) 1259--1279}, [\href{https://arxiv.org/abs/1610.04606}{{\ttfamily
  1610.04606}}].

\bibitem{Riess18}
A.~G. {Riess}, S.~A. {Rodney}, D.~M. {Scolnic}, D.~L. {Shafer}, L.-G.
  {Strolger}, H.~C. {Ferguson} et~al., \emph{{Type Ia Supernova Distances at
  Redshift > 1.5 from the Hubble Space Telescope Multi-cycle Treasury Programs:
  The Early Expansion Rate}},
  \href{https://doi.org/10.3847/1538-4357/aaa5a9}{\emph{\apj} {\bfseries 853}
  (Feb., 2018) 126}, [\href{https://arxiv.org/abs/1710.00844}{{\ttfamily
  1710.00844}}].

\bibitem{Mortonson11}
M.~J. Mortonson, W.~Hu and D.~Huterer, \emph{Simultaneous falsification of
  {$\Lambda$}cdm and quintessence with massive, distant clusters},
  \href{https://doi.org/10.1103/PhysRevD.83.023015}{\emph{\prd} {\bfseries 83}
  (Jan., 2011) 023015}, [\href{https://arxiv.org/abs/1011.0004}{{\ttfamily
  1011.0004}}].

\bibitem{Feng16}
C.-J. {Feng} and X.-Z. {Li}, \emph{{Probing the Expansion History of the
  Universe by Model-independent Reconstruction from Supernovae and Gamma-Ray
  Burst Measurements}},
  \href{https://doi.org/10.3847/0004-637X/821/1/30}{\emph{\apj} {\bfseries 821}
  (Apr., 2016) 30}, [\href{https://arxiv.org/abs/1604.01930}{{\ttfamily
  1604.01930}}].

\bibitem{Zhao17}
G.-B. Zhao, M.~Raveri, L.~Pogosian, Y.~Wang, R.~G. Crittenden, W.~J. Handley
  et~al., \emph{Dynamical dark energy in light of the latest observations},
  \href{https://doi.org/10.1038/s41550-017-0216-z}{\emph{Nature Astronomy}
  {\bfseries 1} (Sept., 2017) 627--632},
  [\href{https://arxiv.org/abs/1701.08165}{{\ttfamily 1701.08165}}].

\bibitem{Miranda17}
V.~Miranda and C.~Dvorkin, \emph{Model-independent predictions for smooth
  cosmic acceleration scenarios}, {\emph{ArXiv e-prints} (Dec., 2017) },
  [\href{https://arxiv.org/abs/1712.04289}{{\ttfamily 1712.04289}}].

\bibitem{Mortonson10b}
M.~J. Mortonson and W.~Hu, \emph{Observational limits on patchy reionization:
  Implications for b modes},
  \href{https://doi.org/10.1103/PhysRevD.81.067302}{\emph{\prd} {\bfseries 81}
  (Mar., 2010) 067302}, [\href{https://arxiv.org/abs/1001.4803}{{\ttfamily
  1001.4803}}].

\bibitem{Mortonson09a}
M.~J. Mortonson, W.~Hu and D.~Huterer, \emph{Falsifying paradigms for cosmic
  acceleration}, \href{https://doi.org/10.1103/PhysRevD.79.023004}{\emph{\prd}
  {\bfseries 79} (Jan., 2009) 023004},
  [\href{https://arxiv.org/abs/0810.1744}{{\ttfamily 0810.1744}}].

\bibitem{Vanderveld12}
R.~A. Vanderveld, M.~J. Mortonson, W.~Hu and T.~Eifler, \emph{Testing dark
  energy paradigms with weak gravitational lensing},
  \href{https://doi.org/10.1103/PhysRevD.85.103518}{\emph{\prd} {\bfseries 85}
  (May, 2012) 103518}, [\href{https://arxiv.org/abs/1203.3195}{{\ttfamily
  1203.3195}}].

\bibitem{Shafieloo12a}
A.~Shafieloo, \emph{Crossing statistic: reconstructing the expansion history of
  the universe},
  \href{https://doi.org/10.1088/1475-7516/2012/08/002}{\emph{\jcap} {\bfseries
  8} (Aug., 2012) 002}, [\href{https://arxiv.org/abs/1204.1109}{{\ttfamily
  1204.1109}}].

\bibitem{Shafieloo06}
A.~Shafieloo, U.~Alam, V.~Sahni and A.~A. Starobinsky, \emph{Smoothing
  supernova data to reconstruct the expansion history of the universe and its
  age}, \href{https://doi.org/10.1111/j.1365-2966.2005.09911.x}{\emph{\mnras}
  {\bfseries 366} (Mar., 2006) 1081--1095},
  [\href{https://arxiv.org/abs/astro-ph/0505329}{{\ttfamily
  astro-ph/0505329}}].

\bibitem{LHuillier17}
B.~L'Huillier, A.~Shafieloo and H.~Kim, \emph{Model-independent cosmological
  constraints from growth and expansion}, {\emph{ArXiv e-prints} (Dec., 2017)
  }, [\href{https://arxiv.org/abs/1712.04865}{{\ttfamily 1712.04865}}].

\bibitem{Cattoen08}
C.~{Catto{\"e}n} and M.~{Visser}, \emph{{Cosmographic Hubble fits to the
  supernova data}},
  \href{https://doi.org/10.1103/PhysRevD.78.063501}{\emph{\prd} {\bfseries 78}
  (Sept., 2008) 063501}, [\href{https://arxiv.org/abs/0809.0537}{{\ttfamily
  0809.0537}}].

\bibitem{Moertsell09}
E.~M{\"o}rtsell and C.~Clarkson, \emph{Model independent constraints on the
  cosmological expansion rate},
  \href{https://doi.org/10.1088/1475-7516/2009/01/044}{\emph{\jcap} {\bfseries
  1} (Jan., 2009) 044}, [\href{https://arxiv.org/abs/0811.0981}{{\ttfamily
  0811.0981}}].

\bibitem{Gruber14}
C.~Gruber and O.~Luongo, \emph{Cosmographic analysis of the equation of state
  of the universe through pad{\'e} approximations},
  \href{https://doi.org/10.1103/PhysRevD.89.103506}{\emph{\prd} {\bfseries 89}
  (May, 2014) 103506}, [\href{https://arxiv.org/abs/1309.3215}{{\ttfamily
  1309.3215}}].

\bibitem{Gomez-Valent18}
A.~G{\'o}mez-Valent and L.~Amendola, \emph{{$H_0$ from cosmic chronometers and
  Type Ia supernovae, with Gaussian Processes and the novel Weighted Polynomial
  Regression method}}, {\emph{ArXiv e-prints} (Feb., 2018) },
  [\href{https://arxiv.org/abs/1802.01505}{{\ttfamily 1802.01505}}].

\bibitem{Capozziello18}
S.~Capozziello, R.~D'Agostino and O.~Luongo, \emph{Cosmographic analysis with
  chebyshev polynomials},
  \href{https://doi.org/10.1093/mnras/sty422}{\emph{\mnras} {\bfseries 476}
  (May, 2018) 3924--3938}, [\href{https://arxiv.org/abs/1712.04380}{{\ttfamily
  1712.04380}}].

\bibitem{seikel12}
M.~Seikel, C.~Clarkson and M.~Smith, \emph{Reconstruction of dark energy and
  expansion dynamics using gaussian processes},
  \href{https://doi.org/10.1088/1475-7516/2012/06/036}{\emph{\jcap} {\bfseries
  6} (June, 2012) 036}, [\href{https://arxiv.org/abs/1204.2832}{{\ttfamily
  1204.2832}}].

\bibitem{Seikel13}
M.~Seikel and C.~Clarkson, \emph{Optimising gaussian processes for
  reconstructing dark energy dynamics from supernovae}, {\emph{ArXiv e-prints}
  (Nov., 2013) }, [\href{https://arxiv.org/abs/1311.6678}{{\ttfamily
  1311.6678}}].

\bibitem{Shafieloo12}
A.~Shafieloo, A.~G. Kim and E.~V. Linder, \emph{Gaussian process cosmography},
  \href{https://doi.org/10.1103/PhysRevD.85.123530}{\emph{\prd} {\bfseries 85}
  (June, 2012) 123530}, [\href{https://arxiv.org/abs/1204.2272}{{\ttfamily
  1204.2272}}].

\bibitem{Vitenti15}
S.~D.~P. Vitenti and M.~Penna-Lima, \emph{A general reconstruction of the
  recent expansion history of the universe},
  \href{https://doi.org/10.1088/1475-7516/2015/09/045}{\emph{\jcap} {\bfseries
  9} (Sept., 2015) 045}, [\href{https://arxiv.org/abs/1505.01883}{{\ttfamily
  1505.01883}}].

\bibitem{Seikel13a}
M.~Seikel, C.~Clarkson and M.~Smith, ``Gapp: Gaussian processes in python.''
  Astrophysics Source Code Library, Mar., 2013.

\bibitem{Zhang16}
M.-J. Zhang and J.-Q. Xia, \emph{Test of the cosmic evolution using gaussian
  processes}, \href{https://doi.org/10.1088/1475-7516/2016/12/005}{\emph{\jcap}
  {\bfseries 12} (Dec., 2016) 005},
  [\href{https://arxiv.org/abs/1606.04398}{{\ttfamily 1606.04398}}].

\bibitem{Bilicki12}
M.~{Bilicki} and M.~{Seikel}, \emph{{We do not live in the R$_{h}$ = ct
  universe}},
  \href{https://doi.org/10.1111/j.1365-2966.2012.21575.x}{\emph{\mnras}
  {\bfseries 425} (Sept., 2012) 1664--1668},
  [\href{https://arxiv.org/abs/1206.5130}{{\ttfamily 1206.5130}}].

\bibitem{Busti14a}
V.~C. Busti, C.~Clarkson and M.~Seikel, \emph{The value of h $_{0}$ from
  gaussian processes},  in \emph{Statistical Challenges in 21st Century
  Cosmology} (A.~{Heavens}, J.-L. {Starck} and A.~{Krone-Martins}, eds.),
  vol.~306 of \emph{IAU Symposium}, pp.~25--27, May, 2014,
  \href{https://arxiv.org/abs/1407.5227}{{\ttfamily 1407.5227}},
  \href{https://doi.org/10.1017/S1743921314013751}{DOI}.

\bibitem{Busti14}
V.~C. Busti, C.~Clarkson and M.~Seikel, \emph{Evidence for a lower value for
  h$_{0}$ from cosmic chronometers data?},
  \href{https://doi.org/10.1093/mnrasl/slu035}{\emph{\mnras} {\bfseries 441}
  (June, 2014) L11--L15}, [\href{https://arxiv.org/abs/1402.5429}{{\ttfamily
  1402.5429}}].

\bibitem{Yu17}
H.~Yu, B.~Ratra and F.-Y. Wang, \emph{Hubble parameter and baryon acoustic
  oscillation measurement constraints on the hubble constant, the deviation
  from the spatially-flat $\lambda$cdm model, the deceleration-acceleration
  transition redshift, and spatial curvature}, {\emph{ArXiv e-prints} (Nov.,
  2017) }, [\href{https://arxiv.org/abs/1711.03437}{{\ttfamily 1711.03437}}].

\bibitem{Wei17}
J.-J. {Wei} and X.-F. {Wu}, \emph{{An Improved Method to Measure the Cosmic
  Curvature}}, \href{https://doi.org/10.3847/1538-4357/aa674b}{\emph{\apj}
  {\bfseries 838} (Apr., 2017) 160},
  [\href{https://arxiv.org/abs/1611.00904}{{\ttfamily 1611.00904}}].

\bibitem{Seikel12a}
M.~Seikel, S.~Yahya, R.~Maartens and C.~Clarkson, \emph{Using h(z) data as a
  probe of the concordance model},
  \href{https://doi.org/10.1103/PhysRevD.86.083001}{\emph{\prd} {\bfseries 86}
  (Oct., 2012) 083001}, [\href{https://arxiv.org/abs/1205.3431}{{\ttfamily
  1205.3431}}].

\bibitem{Sahni14}
V.~{Sahni}, A.~{Shafieloo} and A.~A. {Starobinsky}, \emph{{Model-independent
  Evidence for Dark Energy Evolution from Baryon Acoustic Oscillations}},
  \href{https://doi.org/10.1088/2041-8205/793/2/L40}{\emph{\apjl} {\bfseries
  793} (Oct., 2014) L40}, [\href{https://arxiv.org/abs/1406.2209}{{\ttfamily
  1406.2209}}].

\bibitem{Nair14}
R.~{Nair}, S.~{Jhingan} and D.~{Jain}, \emph{{Exploring scalar field dynamics
  with Gaussian processes}},
  \href{https://doi.org/10.1088/1475-7516/2014/01/005}{\emph{\jcap} {\bfseries
  1} (Jan., 2014) 005}, [\href{https://arxiv.org/abs/1306.0606}{{\ttfamily
  1306.0606}}].

\bibitem{Holsclaw10}
T.~Holsclaw, U.~Alam, B.~Sans{\'o}, H.~Lee, K.~Heitmann, S.~Habib et~al.,
  \emph{Nonparametric dark energy reconstruction from supernova data},
  \href{https://doi.org/10.1103/PhysRevLett.105.241302}{\emph{Physical Review
  Letters} {\bfseries 105} (Dec., 2010) 241302},
  [\href{https://arxiv.org/abs/1011.3079}{{\ttfamily 1011.3079}}].

\bibitem{Holsclaw11}
T.~Holsclaw, U.~Alam, B.~Sans{\'o}, H.~Lee, K.~Heitmann, S.~Habib et~al.,
  \emph{Nonparametric reconstruction of the dark energy equation of state from
  diverse data sets},
  \href{https://doi.org/10.1103/PhysRevD.84.083501}{\emph{\prd} {\bfseries 84}
  (Oct., 2011) 083501}, [\href{https://arxiv.org/abs/1104.2041}{{\ttfamily
  1104.2041}}].

\bibitem{Yennapureddy17}
M.~K. Yennapureddy and F.~Melia, \emph{Reconstruction of the hii galaxy hubble
  diagram using gaussian processes},
  \href{https://doi.org/10.1088/1475-7516/2017/11/029}{\emph{\jcap} {\bfseries
  11} (Nov., 2017) 029}, [\href{https://arxiv.org/abs/1711.03454}{{\ttfamily
  1711.03454}}].

\bibitem{Melia18}
F.~Melia and M.~K. Yennapureddy, \emph{Model selection using cosmic
  chronometers with gaussian processes}, {\emph{ArXiv e-prints} (Feb., 2018) },
  [\href{https://arxiv.org/abs/1802.02255}{{\ttfamily 1802.02255}}].

\bibitem{MartaPinho18}
A.~Marta~Pinho, S.~Casas and L.~Amendola, \emph{Model-independent
  reconstruction of the linear anisotropic stress $\eta$}, {\emph{ArXiv
  e-prints} (Apr., 2018) }, [\href{https://arxiv.org/abs/1805.00027}{{\ttfamily
  1805.00027}}].

\bibitem{Jimenez02}
R.~{Jimenez} and A.~{Loeb}, \emph{{Constraining Cosmological Parameters Based
  on Relative Galaxy Ages}}, \href{https://doi.org/10.1086/340549}{\emph{\apj}
  {\bfseries 573} (July, 2002) 37--42},
  [\href{https://arxiv.org/abs/astro-ph/0106145}{{\ttfamily
  astro-ph/0106145}}].

\bibitem{Moresco16a}
M.~Moresco, L.~Pozzetti, A.~Cimatti, R.~Jimenez, C.~Maraston, L.~Verde et~al.,
  \emph{A 6% measurement of the hubble parameter at $z\sim0.45$: direct
  evidence of the epoch of cosmic re-acceleration},
  \href{https://doi.org/10.1088/1475-7516/2016/05/014}{\emph{\jcap} (2016) },
  [\href{https://arxiv.org/abs/1601.01701v2}{{\ttfamily 1601.01701v2}}].

\bibitem{Eisenstein05}
D.~J. {Eisenstein}, I.~{Zehavi}, D.~W. {Hogg}, R.~{Scoccimarro}, M.~R.
  {Blanton}, R.~C. {Nichol} et~al., \emph{{Detection of the Baryon Acoustic
  Peak in the Large-Scale Correlation Function of SDSS Luminous Red Galaxies}},
  \href{https://doi.org/10.1086/466512}{\emph{\apj} {\bfseries 633} (Nov.,
  2005) 560--574}, [\href{https://arxiv.org/abs/astro-ph/0501171}{{\ttfamily
  astro-ph/0501171}}].

\bibitem{Caruana98}
R.~Caruana, \emph{Multitask learning},  in \emph{Learning to learn},
  pp.~95--133.
\newblock Springer, 1998.

\bibitem{Bonilla07}
E.~V. Bonilla, F.~V. Agakov and C.~K. Williams, \emph{Kernel multi-task
  learning using task-specific features},  in \emph{Artificial Intelligence and
  Statistics}, pp.~43--50, 2007.

\bibitem{Bonilla08}
E.~V. Bonilla, K.~M. Chai and C.~Williams, \emph{Multi-task gaussian process
  prediction},  in \emph{Advances in neural information processing systems},
  pp.~153--160, 2008.

\bibitem{Riess18a}
A.~G. Riess, S.~Casertano, W.~Yuan, L.~Macri, J.~Anderson, J.~W. MacKenty
  et~al., \emph{New parallaxes of galactic cepheids from spatially scanning the
  hubble space telescope: Implications for the hubble constant},
  \href{https://doi.org/10.3847/1538-4357/aaadb7}{\emph{\apj} {\bfseries 855}
  (Mar., 2018) 136}, [\href{https://arxiv.org/abs/1801.01120}{{\ttfamily
  1801.01120}}].

\bibitem{Abdessalem17}
A.~B. Abdessalem, N.~Dervilis, D.~J. Wagg and K.~Worden, \emph{Automatic kernel
  selection for gaussian processes regression with approximate bayesian
  computation and sequential monte carlo}, {\emph{Frontiers in Built
  Environment} {\bfseries 3} (2017) 52}.

\bibitem{Leon06}
J.~P. de~Leon, \emph{An analytical model for the transition from decelerated to
  accelerated cosmic expansion},
  \href{https://doi.org/10.1142/S0218271806008929}{\emph{International Journal
  of Modern Physics D} {\bfseries 15} (2006) 1237--1257},
  [\href{https://arxiv.org/abs/gr-qc/0511150}{{\ttfamily gr-qc/0511150}}].

\bibitem{Lima12}
J.~A.~S. Lima, J.~F. Jesus, R.~C. Santos and M.~S.~S. Gill, \emph{Is the
  transition redshift a new cosmological number?}, {\emph{ArXiv e-prints} (May,
  2012) }, [\href{https://arxiv.org/abs/1205.4688}{{\ttfamily 1205.4688}}].

\bibitem{Sahni08}
V.~Sahni, A.~Shafieloo and A.~A. Starobinsky, \emph{Two new diagnostics of dark
  energy}, \href{https://doi.org/10.1103/PhysRevD.78.103502}{\emph{\prd}
  {\bfseries 78} (Nov., 2008) 103502},
  [\href{https://arxiv.org/abs/0807.3548}{{\ttfamily 0807.3548}}].

\bibitem{Zunckel08}
C.~Zunckel and C.~Clarkson, \emph{Consistency tests for the cosmological
  constant},
  \href{https://doi.org/10.1103/PhysRevLett.101.181301}{\emph{Physical Review
  Letters} {\bfseries 101} (Oct., 2008) 181301},
  [\href{https://arxiv.org/abs/0807.4304}{{\ttfamily 0807.4304}}].

\bibitem{Moresco12b}
M.~{Moresco}, A.~{Cimatti}, R.~{Jimenez}, L.~{Pozzetti}, G.~{Zamorani},
  M.~{Bolzonella} et~al., \emph{{Improved constraints on the expansion rate of
  the Universe up to z \~{} 1.1 from the spectroscopic evolution of cosmic
  chronometers}},
  \href{https://doi.org/10.1088/1475-7516/2012/08/006}{\emph{\jcap} {\bfseries
  8} (Aug., 2012) 006}, [\href{https://arxiv.org/abs/1201.3609}{{\ttfamily
  1201.3609}}].

\bibitem{Moresco15}
M.~Moresco, \emph{Raising the bar: new constraints on the hubble parameter with
  cosmic chronometers at z $\sim$ 2},
  \href{https://doi.org/10.1093/mnrasl/slv037}{\emph{\mnras} {\bfseries 450}
  (June, 2015) L16--L20}, [\href{https://arxiv.org/abs/1503.01116}{{\ttfamily
  1503.01116}}].

\bibitem{Rasmussen06}
C.~E. Rasmussen and C.~K. Williams, \emph{Gaussian processes for machine
  learning. 2006}, {\emph{The MIT Press, Cambridge, MA, USA} {\bfseries 38}
  (2006) 715--719}.

\bibitem{McHutchon11}
A.~McHutchon and C.~E. Rasmussen, \emph{Gaussian process training with input
  noise},  in \emph{Advances in Neural Information Processing Systems},
  pp.~1341--1349, 2011.

\bibitem{Blight75}
B.~Blight and L.~Ott, \emph{A bayesian approach to model inadequacy for
  polynomial regression}, {\emph{Biometrika} {\bfseries 62} (1975) 79--88}.

\bibitem{OHagan78}
A.~O'Hagan and J.~Kingman, \emph{Curve fitting and optimal design for
  prediction}, {\emph{Journal of the Royal Statistical Society. Series B
  (Methodological)} (1978) 1--42}.

\bibitem{Williams06}
C.~K. Williams and C.~E. Rasmussen, \emph{Gaussian processes for machine
  learning}, {\emph{the MIT Press} {\bfseries 2} (2006) 4}.

\bibitem{Solak03}
E.~Solak, R.~Murray-Smith, W.~E. Leithead, D.~J. Leith and C.~E. Rasmussen,
  \emph{Derivative observations in gaussian process models of dynamic systems},
   in \emph{Advances in neural information processing systems}, pp.~1057--1064,
  2003.

\bibitem{Wang17c}
D.~Wang and X.-H. Meng, \emph{Improved constraints on the dark energy equation
  of state using gaussian processes},
  \href{https://doi.org/10.1103/PhysRevD.95.023508}{\emph{\prd} {\bfseries 95}
  (Jan., 2017) 023508}, [\href{https://arxiv.org/abs/1708.07750}{{\ttfamily
  1708.07750}}].

\bibitem{Melkumyan11}
A.~Melkumyan and F.~Ramos, \emph{Multi-kernel gaussian processes},  in
  \emph{IJCAI Proceedings-International Joint Conference on Artificial
  Intelligence}, vol.~22, p.~1408, 2011.

\bibitem{Vasudevan12}
S.~Vasudevan, \emph{Data fusion with gaussian processes}, {\emph{Robotics and
  Autonomous Systems} {\bfseries 60} (2012) 1528--1544}.

\bibitem{Chai10}
K.~M. Chai, \emph{Multi-task learning with gaussian processes}, .

\bibitem{Leen12}
G.~Leen, J.~Peltonen and S.~Kaski, \emph{Focused multi-task learning in a
  gaussian process framework}, {\emph{Machine learning} {\bfseries 89} (2012)
  157--182}.

\bibitem{Lee16}
G.~Lee, E.~Yang and S.~Hwang, \emph{Asymmetric multi-task learning based on
  task relatedness and loss},  in \emph{International Conference on Machine
  Learning}, pp.~230--238, 2016.

\bibitem{Bishop06}
C.~M. Bishop, \emph{Pattern Recognition and Machine Learning}.
\newblock Springer, 2006.

\bibitem{Zhao16}
G.-B. Zhao, Y.~Wang, S.~Saito, D.~Wang, A.~J. Ross, F.~Beutler et~al.,
  \emph{The clustering of galaxies in the completed sdss-iii baryon oscillation
  spectroscopic survey: tomographic bao analysis of dr12 combined sample in
  fourier space}, {\emph{Monthly Notices of the Royal Astronomical Society}
  {\bfseries 466} (2016) 762--779}.

\bibitem{Wang16}
J.~Wang, F.~Wang, K.~Cheng and Z.~Dai, \emph{Measuring dark energy with the
  eiso--ep correlation of gamma-ray bursts using model-independent methods},
  {\emph{\aap} {\bfseries 585} (2016) A68}.

\bibitem{Bautista17}
J.~E. Bautista, N.~G. Busca, J.~Guy, J.~Rich, M.~Blomqvist, H.~du~Mas~des
  Bourboux et~al., \emph{Measurement of baryon acoustic oscillation
  correlations at z = 2.3 with sdss dr12 ly{$\alpha$}-forests},
  \href{https://doi.org/10.1051/0004-6361/201730533}{\emph{\aap} {\bfseries
  603} (June, 2017) A12}, [\href{https://arxiv.org/abs/1702.00176}{{\ttfamily
  1702.00176}}].

\bibitem{MasdesBourboux17}
H.~du~Mas~des Bourboux, J.-M.~L. Goff, M.~Blomqvist, N.~G. Busca, J.~Guy,
  J.~Rich et~al., \emph{Baryon acoustic oscillations from the complete sdss-iii
  ly$\alpha$-quasar cross-correlation function at $z=2.4$},
  \href{https://arxiv.org/abs/1708.02225v1}{{\ttfamily 1708.02225v1}}.

\bibitem{Zhao18}
G.-B. Zhao, Y.~Wang, S.~Saito, H.~Gil-Mar{\'{\i}}n, W.~J. Percival, D.~Wang
  et~al., \emph{The clustering of the sdss-iv extended baryon oscillation
  spectroscopic survey dr14 quasar sample: a tomographic measurement of cosmic
  structure growth and expansion rate based on optimal redshift weights},
  {\emph{ArXiv e-prints} (Jan., 2018) },
  [\href{https://arxiv.org/abs/1801.03043}{{\ttfamily 1801.03043}}].

\bibitem{Simon05}
J.~Simon, L.~Verde and R.~Jimenez, \emph{Constraints on the redshift dependence
  of the dark energy potential},
  \href{https://doi.org/10.1103/PhysRevD.71.123001}{\emph{\prd} {\bfseries 71}
  (June, 2005) 123001},
  [\href{https://arxiv.org/abs/astro-ph/0412269}{{\ttfamily
  astro-ph/0412269}}].

\bibitem{Stern10}
D.~Stern, R.~Jimenez, L.~Verde, S.~A. Stanford and M.~Kamionkowski,
  \emph{Cosmic chronometers: Constraining the equation of state of dark energy.
  ii. a spectroscopic catalog of red galaxies in galaxy clusters},
  \href{https://doi.org/10.1088/0067-0049/188/1/280}{\emph{\apjs} {\bfseries
  188} (May, 2010) 280--289},
  [\href{https://arxiv.org/abs/0907.3152}{{\ttfamily 0907.3152}}].

\bibitem{Zhang14}
C.~Zhang, H.~Zhang, S.~Yuan, S.~Liu, T.-J. Zhang and Y.-C. Sun, \emph{Four new
  observational h(z) data from luminous red galaxies in the sloan digital sky
  survey data release seven},
  \href{https://doi.org/10.1088/1674-4527/14/10/002}{\emph{Research in
  Astronomy and Astrophysics} {\bfseries 14} (Oct., 2014) 1221--1233},
  [\href{https://arxiv.org/abs/1207.4541}{{\ttfamily 1207.4541}}].

\bibitem{Ratsimbazafy17}
A.~L. Ratsimbazafy, S.~I. Loubser, S.~M. Crawford, C.~M. Cress, B.~A. Bassett,
  R.~C. Nichol et~al., \emph{Age-dating luminous red galaxies observed with the
  southern african large telescope},
  \href{https://doi.org/10.1093/mnras/stx301}{\emph{\mnras} {\bfseries 467}
  (May, 2017) 3239--3254}, [\href{https://arxiv.org/abs/1702.00418}{{\ttfamily
  1702.00418}}].

\bibitem{Moresco18}
M.~Moresco, R.~Jimenez, L.~Verde, L.~Pozzetti, A.~Cimatti and A.~Citro,
  \emph{Setting the stage for cosmic chronometers i. minimizing frosting with
  an optimized selection of cosmic chronometers}, {\emph{ArXiv e-prints} (Apr.,
  2018) }, [\href{https://arxiv.org/abs/1804.05864}{{\ttfamily 1804.05864}}].

\bibitem{Farooq16}
O.~Farooq, F.~R. Madiyar, S.~Crandall and B.~Ratra, \emph{{Hubble parameter
  measurement constraints on the redshift of the deceleration-acceleration
  transition, dynamical dark energy, and space curvature}},
  \href{https://doi.org/10.3847/1538-4357/835/1/26}{\emph{\apj} {\bfseries 835}
  (2017) 26}, [\href{https://arxiv.org/abs/1607.03537}{{\ttfamily
  1607.03537}}].

\bibitem{Chen11}
G.~{Chen} and B.~{Ratra}, \emph{{Median Statistics and the Hubble Constant}},
  \href{https://doi.org/10.1086/662131}{\emph{\pasp} {\bfseries 123} (Sept.,
  2011) 1127--1132}, [\href{https://arxiv.org/abs/1105.5206}{{\ttfamily
  1105.5206}}].

\bibitem{Lin17a}
W.~Lin and M.~Ishak, \emph{Cosmological discordances. ii. hubble constant,
  planck and large-scale-structure data sets},
  \href{https://doi.org/10.1103/PhysRevD.96.083532}{\emph{\prd} {\bfseries 96}
  (Oct., 2017) 083532}, [\href{https://arxiv.org/abs/1708.09813}{{\ttfamily
  1708.09813}}].

\bibitem{Cheng15}
C.~{Cheng} and Q.~{Huang}, \emph{{An accurate determination of the Hubble
  constant from baryon acoustic oscillation datasets}},
  \href{https://doi.org/10.1007/s11433-015-5684-5}{\emph{Science China Physics,
  Mechanics, and Astronomy} {\bfseries 58} (Sept., 2015) 095684},
  [\href{https://arxiv.org/abs/1409.6119}{{\ttfamily 1409.6119}}].

\bibitem{Haridasu17a}
B.~S. Haridasu, V.~V. Lukovi\'c and N.~Vittorio, \emph{Isotropic vs.
  anisotropic components of bao data: a tool for model selection}, {\emph{ArXiv
  e-prints} (Nov., 2017) },
  [\href{https://arxiv.org/abs/1711.03929v2}{{\ttfamily 1711.03929v2}}].

\bibitem{Lukovic18}
V.~V. Lukovi{\'c}, B.~S. Haridasu and N.~Vittorio, \emph{Cosmological
  constraints from low-redshift data}, {\emph{ArXiv e-prints} (Jan., 2018) },
  [\href{https://arxiv.org/abs/1801.05765}{{\ttfamily 1801.05765}}].

\bibitem{Ade16}
P.~Ade, N.~Aghanim, M.~Arnaud, M.~Ashdown, J.~Aumont, C.~Baccigalupi et~al.,
  \emph{Planck 2015 results-xiii. cosmological parameters}, {\emph{\aap}
  {\bfseries 594} (2016) A13}.

\bibitem{Kim13}
A.~G. {Kim}, R.~C. {Thomas}, G.~{Aldering}, P.~{Antilogus}, C.~{Aragon},
  S.~{Bailey} et~al., \emph{{Standardizing Type Ia Supernova Absolute
  Magnitudes Using Gaussian Process Data Regression}},
  \href{https://doi.org/10.1088/0004-637X/766/2/84}{\emph{\apj} {\bfseries 766}
  (Apr., 2013) 84}, [\href{https://arxiv.org/abs/1302.2925}{{\ttfamily
  1302.2925}}].

\bibitem{Dhawan18}
S.~{Dhawan}, S.~W. {Jha} and B.~{Leibundgut}, \emph{{Measuring the Hubble
  constant with Type Ia supernovae as near-infrared standard candles}},
  \href{https://doi.org/10.1051/0004-6361/201731501}{\emph{\aap} {\bfseries
  609} (Jan., 2018) A72}, [\href{https://arxiv.org/abs/1707.00715}{{\ttfamily
  1707.00715}}].

\bibitem{Guillochon17}
J.~{Guillochon}, M.~{Nicholl}, V.~A. {Villar}, B.~{Mockler}, G.~{Narayan},
  K.~S. {Mandel} et~al., \emph{{MOSFiT: Modular Open-Source Fitter for
  Transients}}, {\emph{ArXiv e-prints} (Oct., 2017) },
  [\href{https://arxiv.org/abs/1710.02145}{{\ttfamily 1710.02145}}].

\bibitem{Shafieloo18}
A.~{Shafieloo}, B.~{L'Huillier} and A.~A. {Starobinsky}, \emph{{Falsifying
  $\Lambda$CDM: Model-independent tests of the concordance model with eBOSS
  DR14Q and Pantheon}}, {\emph{ArXiv e-prints} (Apr., 2018) },
  [\href{https://arxiv.org/abs/1804.04320}{{\ttfamily 1804.04320}}].

\bibitem{Shafieloo09}
A.~Shafieloo, V.~Sahni and A.~A. Starobinsky, \emph{Is cosmic acceleration
  slowing down?}, \href{https://doi.org/10.1103/PhysRevD.80.101301}{\emph{\prd}
  {\bfseries 80} (Nov., 2009) 101301},
  [\href{https://arxiv.org/abs/0903.5141}{{\ttfamily 0903.5141}}].

\bibitem{Shafieloo10}
A.~Shafieloo, V.~Sahni and A.~A. Starobinsky, \emph{Tentative evidence for
  slowing down of cosmic acceleration from recent small redshift supernovae and
  bao data},  in \emph{American Institute of Physics Conference Series} (J.-M.
  {Alimi} and A.~{Fu{\"o}zfa}, eds.), vol.~1241 of \emph{American Institute of
  Physics Conference Series}, pp.~294--302, June, 2010,
  \href{https://doi.org/10.1063/1.3462648}{DOI}.

\bibitem{Shahalam15}
M.~Shahalam, S.~Sami and A.~Agarwal, \emph{Om diagnostic applied to scalar
  field models and slowing down of cosmic acceleration},
  \href{https://doi.org/10.1093/mnras/stv083}{\emph{\mnras} {\bfseries 448}
  (Apr., 2015) 2948--2959}, [\href{https://arxiv.org/abs/1501.04047}{{\ttfamily
  1501.04047}}].

\bibitem{Wang16a}
S.~{Wang}, Y.~{Hu}, M.~{Li} and N.~{Li}, \emph{{A Comprehensive Investigation
  on the Slowing Down of Cosmic Acceleration}},
  \href{https://doi.org/10.3847/0004-637X/821/1/60}{\emph{\apj} {\bfseries 821}
  (Apr., 2016) 60}, [\href{https://arxiv.org/abs/1509.03461}{{\ttfamily
  1509.03461}}].

\bibitem{Zhang18}
M.-J. Zhang and J.-Q. Xia, \emph{Physical condition for the slowing down of
  cosmic acceleration},
  \href{https://doi.org/10.1016/j.nuclphysb.2018.02.020}{\emph{Nuclear Physics
  B} {\bfseries 929} (Apr., 2018) 438--451},
  [\href{https://arxiv.org/abs/1701.04973}{{\ttfamily 1701.04973}}].

\bibitem{Bonilla18}
A.~Bonilla and J.~Castillo, \emph{Constraints on dark energy models from galaxy
  clusters and gravitational lensing data},
  \href{https://doi.org/10.3390/universe4010021}{\emph{Universe} {\bfseries 4}
  (Jan., 2018) 21}, [\href{https://arxiv.org/abs/1711.09291}{{\ttfamily
  1711.09291}}].

\bibitem{Magana14}
J.~Maga{\~n}a, V.~H. C{\'a}rdenas and V.~Motta, \emph{Cosmic slowing down of
  acceleration for several dark energy parametrizations},
  \href{https://doi.org/10.1088/1475-7516/2014/10/017}{\emph{\jcap} {\bfseries
  10} (Oct., 2014) 017}, [\href{https://arxiv.org/abs/1407.1632}{{\ttfamily
  1407.1632}}].

\bibitem{Magana17}
J.~Maga{\~n}a, V.~Motta, V.~H. C{\'a}rdenas and G.~Fo{\"e}x, \emph{Testing
  cosmic acceleration for w(z) parametrizations using f$_{gas}$ measurements in
  galaxy clusters}, \href{https://doi.org/10.1093/mnras/stx750}{\emph{\mnras}
  {\bfseries 469} (July, 2017) 47--61},
  [\href{https://arxiv.org/abs/1703.08521}{{\ttfamily 1703.08521}}].

\bibitem{Laureijs11}
R.~Laureijs, J.~Amiaux, S.~Arduini, J.~. Augu{\`e}res, J.~Brinchmann, R.~Cole
  et~al., \emph{Euclid definition study report}, {\emph{ArXiv e-prints} (Oct.,
  2011) }, [\href{https://arxiv.org/abs/1110.3193}{{\ttfamily 1110.3193}}].

\bibitem{Collaboration16a}
D.~Collaboration, A.~Aghamousa, J.~Aguilar, S.~Ahlen, S.~Alam, L.~E. Allen
  et~al., \emph{The desi experiment part ii: Instrument design}, {\emph{ArXiv
  e-prints} (Oct., 2016) }, [\href{https://arxiv.org/abs/1611.00037}{{\ttfamily
  1611.00037}}].

\bibitem{Collaboration16}
D.~Collaboration, A.~Aghamousa, J.~Aguilar, S.~Ahlen, S.~Alam, L.~E. Allen
  et~al., \emph{The desi experiment part i: Science,targeting, and survey
  design}, {\emph{ArXiv e-prints} (Oct., 2016) },
  [\href{https://arxiv.org/abs/1611.00036}{{\ttfamily 1611.00036}}].

\bibitem{Nielsen15}
J.~T. Nielsen, A.~Guffanti and S.~Sarkar, \emph{Marginal evidence for cosmic
  acceleration from type ia supernovae},
  \href{https://doi.org/10.1038/srep35596}{\emph{Scientific Reports} {\bfseries
  6} (Oct., 2016) 35596}, [\href{https://arxiv.org/abs/1506.01354}{{\ttfamily
  1506.01354}}].

\bibitem{Shariff15}
H.~{Shariff}, X.~{Jiao}, R.~{Trotta} and D.~A. {van Dyk}, \emph{{BAHAMAS: New
  Analysis of Type Ia Supernovae Reveals Inconsistencies with Standard
  Cosmology}}, \href{https://doi.org/10.3847/0004-637X/827/1/1}{\emph{\apj}
  {\bfseries 827} (Aug., 2016) 1},
  [\href{https://arxiv.org/abs/1510.05954}{{\ttfamily 1510.05954}}].

\bibitem{Rubin16}
D.~{Rubin} and B.~{Hayden}, \emph{{Is the Expansion of the Universe
  Accelerating? All Signs Point to Yes}},
  \href{https://doi.org/10.3847/2041-8213/833/2/L30}{\emph{\apjl} {\bfseries
  833} (Dec., 2016) L30}, [\href{https://arxiv.org/abs/1610.08972}{{\ttfamily
  1610.08972}}].

\bibitem{Ringermacher16}
H.~I. {Ringermacher} and L.~R. {Mead}, \emph{{In Defense of an Accelerating
  Universe: Model Insensitivity of the Hubble Diagram}}, {\emph{ArXiv e-prints}
  (Nov., 2016) }, [\href{https://arxiv.org/abs/1611.00999}{{\ttfamily
  1611.00999}}].

\bibitem{Haridasu17}
B.~S. Haridasu, V.~V. Lukovi{\'c}, R.~D'Agostino and N.~Vittorio, \emph{Strong
  evidence for an accelerating universe},
  \href{https://doi.org/10.1051/0004-6361/201730469}{\emph{\aap} {\bfseries
  600} (Apr., 2017) L1}, [\href{https://arxiv.org/abs/1702.08244}{{\ttfamily
  1702.08244}}].

\bibitem{Tutusaus17}
I.~Tutusaus, B.~Lamine, A.~Dupays and A.~Blanchard, \emph{Is cosmic
  acceleration proven by local cosmological probes?},
  \href{https://doi.org/10.1051/0004-6361/201630289}{\emph{\aap} {\bfseries
  602} (June, 2017) A73}, [\href{https://arxiv.org/abs/1706.05036}{{\ttfamily
  1706.05036}}].

\bibitem{Lonappan17a}
A.~I. Lonappan, S.~Kumar, Ruchika, B.~R. Dinda and A.~A. Sen, \emph{Bayesian
  evidences for dark energy models in light of current obsevational data},
  {\emph{ArXiv e-prints} (July, 2017) },
  [\href{https://arxiv.org/abs/1707.00603}{{\ttfamily 1707.00603}}].

\bibitem{Dam17}
L.~H. Dam, A.~Heinesen and D.~L. Wiltshire, \emph{Apparent cosmic acceleration
  from type ia supernovae},
  \href{https://doi.org/10.1093/mnras/stx1858}{\emph{\mnras} {\bfseries 472}
  (Nov., 2017) 835--851}, [\href{https://arxiv.org/abs/1706.07236}{{\ttfamily
  1706.07236}}].

\bibitem{Lin17b}
H.-N. {Lin}, X.~{Li} and Y.~{Sang}, \emph{{The local probes strongly favor an
  accelerating universe}}, {\emph{ArXiv e-prints} (Nov., 2017) },
  [\href{https://arxiv.org/abs/1711.05025}{{\ttfamily 1711.05025}}].

\bibitem{Dolgov14}
A.~{Dolgov}, V.~{Halenka} and I.~{Tkachev}, \emph{{Power-law cosmology, SN Ia,
  and BAO}}, \href{https://doi.org/10.1088/1475-7516/2014/10/047}{\emph{\jcap}
  {\bfseries 10} (Oct., 2014) 047},
  [\href{https://arxiv.org/abs/1406.2445}{{\ttfamily 1406.2445}}].

\bibitem{Shafer15}
D.~L. Shafer, \emph{Robust model comparison disfavors power law cosmology},
  \href{https://doi.org/10.1103/PhysRevD.91.103516}{\emph{\prd} {\bfseries 91}
  (May, 2015) 103516}, [\href{https://arxiv.org/abs/1502.05416}{{\ttfamily
  1502.05416}}].

\bibitem{Tutusaus16}
I.~Tutusaus, B.~Lamine, A.~Blanchard, A.~Dupays, Y.~Zolnierowski,
  J.~Cohen-Tanugi et~al., \emph{Power law cosmology model comparison with cmb
  scale information},
  \href{https://doi.org/10.1103/PhysRevD.94.103511}{\emph{\prd} {\bfseries 94}
  (Nov., 2016) 103511}, [\href{https://arxiv.org/abs/1610.03371}{{\ttfamily
  1610.03371}}].

\bibitem{Mitra14}
A.~{Mitra}, \emph{{Why the Big Bang Model does not allow inflationary and
  cyclic cosmologies though mathematically one can obtain any model with
  favourable assumptions}},
  \href{https://doi.org/10.1016/j.newast.2013.12.002}{\emph{\na} {\bfseries 30}
  (July, 2014) 46--50}.

\bibitem{John00}
M.~V. {John} and K.~B. {Joseph}, \emph{{Generalized Chen-Wu type cosmological
  model}}, \href{https://doi.org/10.1103/PhysRevD.61.087304}{\emph{\prd}
  {\bfseries 61} (Apr., 2000) 087304},
  [\href{https://arxiv.org/abs/gr-qc/9912069}{{\ttfamily gr-qc/9912069}}].

\bibitem{Dev02}
A.~Dev, M.~Safonova, D.~Jain and D.~Lohiya, \emph{Cosmological tests for a
  linear coasting cosmology},
  \href{https://doi.org/10.1016/S0370-2693(02)02814-9}{\emph{Physics Letters B}
  {\bfseries 548} (Nov., 2002) 12--18},
  [\href{https://arxiv.org/abs/astro-ph/0204150}{{\ttfamily
  astro-ph/0204150}}].

\bibitem{Melia14a}
F.~{Melia}, \emph{{On recent claims concerning the R$_{h}$ = ct Universe}},
  \href{https://doi.org/10.1093/mnras/stu2181}{\emph{\mnras} {\bfseries 446}
  (Jan., 2015) 1191--1194}, [\href{https://arxiv.org/abs/1406.4918}{{\ttfamily
  1406.4918}}].

\bibitem{Melia16}
F.~{Melia}, \emph{{The linear growth of structure in the R$_{h}$ = ct
  universe}}, \href{https://doi.org/10.1093/mnras/stw2493}{\emph{\mnras}
  {\bfseries 464} (Jan., 2017) 1966--1976},
  [\href{https://arxiv.org/abs/1609.08576}{{\ttfamily 1609.08576}}].

\bibitem{Tutusaus18}
I.~Tutusaus, B.~Lamine and A.~Blanchard, \emph{Model-independent cosmic
  acceleration and type ia supernovae intrinsic luminosity redshift
  dependence}, {\emph{ArXiv e-prints} (Mar., 2018) },
  [\href{https://arxiv.org/abs/1803.06197}{{\ttfamily 1803.06197}}].

\bibitem{Moresco12a}
M.~{Moresco}, L.~{Verde}, L.~{Pozzetti}, R.~{Jimenez} and A.~{Cimatti},
  \emph{{New constraints on cosmological parameters and neutrino properties
  using the expansion rate of the Universe to z \~{} 1.75}},
  \href{https://doi.org/10.1088/1475-7516/2012/07/053}{\emph{\jcap} {\bfseries
  7} (July, 2012) 053}, [\href{https://arxiv.org/abs/1201.6658}{{\ttfamily
  1201.6658}}].

\bibitem{Chevallier01}
M.~Chevallier and D.~Polarski, \emph{Accelerating universes with scaling dark
  matter}, \href{https://doi.org/10.1142/S0218271801000822}{\emph{International
  Journal of Modern Physics D} {\bfseries 10} (2001) 213--223},
  [\href{https://arxiv.org/abs/gr-qc/0009008}{{\ttfamily gr-qc/0009008}}].

\bibitem{Linder03}
E.~V. Linder, \emph{Exploring the expansion history of the universe},
  \href{https://doi.org/10.1103/PhysRevLett.90.091301}{\emph{Physical Review
  Letters} {\bfseries 90} (Mar., 2003) 091301},
  [\href{https://arxiv.org/abs/astro-ph/0208512}{{\ttfamily
  astro-ph/0208512}}].

\end{thebibliography}\endgroup

\end{document}